\documentclass{jfm}

\graphicspath{{figures/}} 
\usepackage{natbib} 
\usepackage[breaklinks,colorlinks = true,linkcolor = blue,urlcolor=blue,citecolor=blue]{hyperref}
\usepackage{newtxtext,newtxmath}
\usepackage{bm}
\usepackage{amsmath}
\usepackage{tikz}
\usepackage{pgfplots,tikz-3dplot}
\usepackage{ar}
\usepackage{subfigure}
\usepackage[]{lineno}
\usepackage{longtable}

\newcommand{\aver}[1]{ \! \left\langle {#1} \right \rangle \!}

\title[Dynamics of the velocity fluctuations in sedimenting suspensions of rigid fibres]{Dynamics of the velocity fluctuations in sedimenting suspensions of rigid fibres}

\author[A. Chiarini, E. Gallorini and M. E. Rosti]{Alessandro Chiarini
\aff{1,}\aff{2}\corresp{\email{alessandro.chiarini@polimi.it}}, 
Emanuele Gallorini
\aff{2,}\aff{3}\corresp{\email{emanuele.gallorini@polimi.it}},
and 
Marco Edoardo Rosti
\aff{1}\corresp{\email{marco.rosti@oist.jp}}
} \affiliation{
\aff{1}Complex Fluids and Flows Unit, Okinawa Institute of Science and Technology Graduate University, 1919-1 Tancha, Onna-son, Okinawa 904-0495, Japan 
\aff{2}  Dipartimento di Scienze e Tecnologie Aerospaziali, Politecnico di Milano, via La Masa 34, 20156 Milano, Italy
\aff{3} Univ. Lille, CNRS, ONERA, Arts et Metiers Institute of Technology, Centrale Lille, UMR 9014 - LMFL - Laboratoire de Mécanique des Fluides de Lille - Kampé de Fériet, F-59000 Lille, France}

\date{\today}
\begin{document}

\maketitle

\begin{abstract} 
We use direct numerical simulations to investigate fluid-solid interactions in suspensions of rigid fibres settling under gravity in a quiescent fluid. The solid-to-fluid density ratio is $\mathcal{O}(100)$, while the Galileo number ($Ga$) and fibre concentration ($n\ell_f^3$) are varied over the ranges $Ga \in [180, 900]$ and $n\ell_f^3 \in [0.36, 23.15]$; $\ell_f$ denotes the fibre length and $n$ the number density.
At high $Ga$ and/or low $n\ell_f^3$, fibres cluster into gravity-aligned streamers with elevated concentrations and enhanced settling velocities, disrupting the flow homogeneity. 
As $Ga$ increases and/or $n\ell_f^3$ decreases, the fluid-phase kinetic energy rises and the energy spectrum broadens, reflecting enhanced small-scale activity. The flow anisotropy is assessed by decomposing the energy spectrum into components aligned with and transverse to gravity. Vertical fluctuations are primarily driven by fluid-solid interactions, while transverse ones are maintained by pressure-strain effects that promote isotropy.
With increasing $Ga$, nonlinear interactions become more prominent, producing a net forward energy cascade toward smaller scales, punctuated by localised backscatter events. 
Analysis of the local velocity gradient tensor reveals distinct flow topologies: at low $Ga$, the flow is dominated by axisymmetric compression and two-dimensional straining; at high $Ga$, regions of high fibre concentration are governed by two-dimensional strain, while voids are associated with axisymmetric extension. The fluid motion is predominantly extensional rather than rotational.
\end{abstract} 


\maketitle

\section{Introduction}

The sedimentation of solid objects, such as spherical particles and fibres, is a long-standing problem that has been extensively studied over the years \citep{davis-acrivos-1985,guazzelli-morris-2011}. It occurs in a wide range of natural phenomena, including sediment transport in rivers, the formation of raindrops, and the dispersion of atmospheric dust. Sedimentation also plays a crucial role in numerous industrial processes, where it is employed for liquid clarification, particle recovery, and the separation of materials based on density or size; it is also central to paper manufacturing. Despite its long history in both practical applications and fundamental research, many aspects of sedimenting suspensions remain poorly understood. The problem is inherently complex and depends on several key parameters, including the particle-to-fluid density ratio $\rho_p/\rho$ (where the subscript ``$p$'' refers to a particle of generic shape), the Galileo number $Ga = ( |\rho_p/\rho - 1| \ell_p^3 g )^{1/2} / \nu$, which quantifies the ratio between buoyancy and viscous forces, and the volume fraction of the suspension $\Phi_V = V_p / V_t$, where $V_p$ and $V_t$ denote the volumes of the solid phase and the entire suspension, respectively. In the above, $g$ is the magnitude of the gravitational acceleration, $\ell_p$ is the characteristic size of the particle, and $\nu$ is the kinematic viscosity of the fluid.

In general, the dynamics of the carrier phase depends on the structure of the suspension and the spatial relative distribution of the particles. In turn, the spatial organisation of the suspension is strongly influenced by the dynamics of the fluid-phase fluctuations. The sedimentation of nonspherical particles, which is the focus of the present work, is an even more difficult task:  the structure of the suspension and the organisation of the fluid velocity fluctuations also depend on the particles orientation.

\subsection{Spherical particles}

The monodisperse suspension of spherical particles is the most studied sedimentation system. In the absence of hydrodynamics interactions, i.e., in the extremely dilute case with $\Phi_V \rightarrow 0$, spherical particles sediment at the Stokes velocity $\bm{u}_p = \bm{u}_{St} \equiv (\rho_p-\rho) \ell_p^2/(18 \nu) \bm{g}$ \citep{batchelor-1967}, where in this case $\ell_p$ is the sphere diameter and $\bm{g}$ is the acceleration due to the gravity. As the suspension concentration $\Phi_V$ increases, the backreaction of the solids on the fluid phase becomes significant, modulating the flow and affecting the settling velocity. 
In semi-dilute and dense regimes, early studies \citep{davis-acrivos-1985} showed that the mean settling velocity decreases due to the upward fluid flux induced by the displaced liquid. This effect is commonly described by a hindrance function, $f(\Phi_V)$, such that the mean particle settling velocity satisfies $\langle \bm{u}_p \rangle_p = f(\Phi_V)\bm{u}_{St}$, where $\langle \cdot \rangle_p$ denotes an average over the particles.
In non-dilute suspensions at sufficiently high Reynolds numbers, the structure is strongly influenced by wake interactions between particle pairs. Early studies examined flow past fixed spheres arranged in specific configurations to isolate collective effects from particle motion. \cite{tsuji-morikawa-terashima-1982} and \cite{tsuji-etal-2003} investigated pairs aligned with or transverse to the flow, while \cite{kim-elghobashi-sirignano-1993} and \cite{schouveiler-etal-2004} focused on side-by-side arrangements. These works consistently showed that the presence of a neighbouring sphere modifies the wake structure and its stability, affecting drag and lift forces depending on the relative positions. Additionally, flow past random arrays of fixed spheres over a broad Reynolds number range has been studied \citep[e.g.,][]{hill-koch-ladd-2001,beetstra-vanDerHoef-kuipers-2007,tenneti-garg-subramaniam-2011}, resulting in drag correlations dependent on suspension volume fraction and Reynolds number.

In recent years, several studies have investigated the sedimentation of particles in a quiescent fluid, using both experiments and direct numerical simulations (DNS) coupled with the immersed boundary method (IBM). Particular attention has been paid to how the microstructure of the suspension evolves across parameter space, and how these changes influence the average settling velocity.
One key contribution is that of \citet{uhlmann-doychev-2014}, who provided a detailed description of particle motion in dilute suspensions with a volume fraction of $\Phi_V = 0.005$ and a solid-to-fluid density ratio of $\rho_p/\rho = 1.5$. They considered two Galileo numbers, $Ga = 121$ and $Ga = 178$, to explore the influence of spatial particle distributions on collective dynamics. For the higher $Ga$, strong clustering was observed in the form of vertically elongated columnar structures. As a result of this clustering, the mean particle settling velocity increased by approximately 12\% relative to the terminal velocity of an isolated sphere. This enhancement was attributed to particles preferentially sampling regions of downward fluid motion, which are generated by the clusters themselves.
\citet{zaidi-etal-2014} investigated the effects of particle Reynolds number and volume fraction on the settling behaviour of spherical particles. They varied the Reynolds number, based on the terminal velocity of an isolated particle, from $Re = 1$ to $Re = 300$, and the volume fraction from $\Phi_V = 0.005$ to $\Phi_V = 0.05$. At high $Re$, particles exhibited strong clustering, primarily becoming entrained in regions of high fluid shear. At intermediate $Re$, particle pairs were observed to form due to the presence of relatively weak wakes, which promote the draft–kissing–tumbling mechanism \citep{fortes-etal-1987,wu-manasseh-1998}.
\citet{huisman-etal-2016} experimentally investigated the effect of the Galileo number ($110 \leq Ga \leq 310$) on the settling of spherical particles in a quiescent fluid, at a fixed solid-to-fluid density ratio of $\rho_p/\rho = 2.5$ and a volume fraction of $\Phi_V = \mathcal{O}(10^{-3})$. Consistent with the findings of \citet{uhlmann-doychev-2014}, they used Vorono\"{\i} tessellation to demonstrate that clustered particles preferentially align vertically and settle faster than isolated particles.
In a recent study, \citet{jiang-etal-2025} investigated the settling of spheroidal particles with a density ratio of $\rho_p/\rho = 2$ at a relatively low Galileo number, $Ga = 80$. They found that spherical particles do not cluster and settle with hindered velocity, whereas oblate and prolate particles form columnar clusters that enhance the mean settling speed. Additionally, they reported a $k^{-3}$ scaling in the fluid-phase energy spectrum \citep{lance-bataille-1991}, providing insight into the distinct mechanisms sustaining fluid velocity fluctuations.

The turbulence intensity of the surrounding fluid affects the settling velocity of particles. This effect was first identified by \cite{maxey-1987}, who studied the settling of a single point particle in a homogeneous Gaussian random flow. He observed that inertial particles settle faster in turbulent flows than in quiescent fluids. The classical explanation attributes this enhancement to the preferential concentration of particles outside regions of high vorticity, causing them to sample predominantly downward-moving sides of vortices.
More recently, \cite{fornari-etal-2016} and \cite{fornari-etal-2016b} investigated finite-size spherical particles settling in homogeneous isotropic turbulence at a Taylor microscale Reynolds number of $Re_\lambda = u' \lambda/\nu = 90$, where $u'$ is the root-mean-square velocity fluctuation. They considered particles slightly heavier than the fluid, with density ratios $\rho_p/\rho \gtrapprox 1$, volume fractions in the range $\Phi_V \in [0.005,0.01]$, and Galileo numbers between $20 \leq Ga \leq 200$. Their results showed that the average settling velocity decreases compared to quiescent conditions, with the reduction becoming more pronounced as the ratio of the particle's isolated terminal velocity to the characteristic turbulent velocity fluctuation decreases. They attributed this reduction to non-stationary vortex shedding effects, which hinder the mean settling velocity in turbulent environments.
 
\subsection{Non-spherical particles}

The settling dynamics of non-spherical particles introduce additional complexities due to their anisotropic shapes and orientation-dependent interactions with the surrounding flow \citep{voth-soldati-2017,chiarini-rosti-mazzino-2024}. In the present study, we focus on fibre-like (or rod-like) rigid bodies, characterised by one dimension significantly larger than the others. Throughout, the subscript ``$f$'' denotes fibre-related quantities.

Unlike spheres, the orientation of fibre-like particles strongly influences their settling behaviour in a quiescent fluid. In the absence of hydrodynamic interactions, the sedimentation velocity of a fibre with length $\ell_f$, diameter $d_f$, and density $\rho_f$ can be predicted by slender body theory as $\bm{u}_f = \Delta \rho d_f^2 \left[ A(\AR) \bm{g} + B(\AR) (\bm{g} \cdot \bm{p}) \bm{p} \right]$, where $A(\AR)$ and $B(\AR)$ are coefficients that depend on the fibre aspect ratio $\AR = \ell_f / d_f$, $\boldsymbol{p}$ is the unit vector denoting the fibre orientation, and $\Delta \rho = \rho_f - \rho$ \citep{batchelor-1970,mackaplow-shapfeh-1998}. 
It follows that the velocity components parallel, $\bm{u}_f \cdot \bm{g} / g$, and perpendicular, $\bm{u}_f - (\bm{u}_f \cdot \bm{g}) \bm{g} / g^2$, to gravity depend explicitly on the fibre orientation. In suspensions of settling fibres, this implies that both the mean settling velocity and the structure of fluid-phase velocity fluctuations are strongly influenced by the orientations of the falling fibres.

The settling behaviour of fibres depends strongly on the effective particle volume fraction, commonly defined as $n \ell_f^3$, where $n = N / V_t$ is the fibre number density (with $N$ the total number of fibres). This definition is motivated by the observation that fibres disturb the surrounding fluid over length scales on the order of their length $\ell_f$. Several studies, primarily experimental, have investigated how concentration influences the settling dynamics of rigid fibre suspensions.
For rigid fibres settling in Stokes flow, \cite{herzhaft-guazzelli-1999} observed that under dilute conditions, fibres tend to align with gravity and aggregate into vertically oriented clusters known as streamers, consistent with earlier findings \citep[e.g.,][]{koch-shaqfeh-1989,herzhaft-etal-1996,mackaplow-shapfeh-1998}. Notable contributions to the characterisation of streamers in fibre suspensions include those by \cite{metzger-guazzelli-butler-2005} and \cite{saintillan-shaqfeh-darve-2006}. In this regime, the mean vertical sedimentation velocity is not hindered by the upward volume flux and exceeds the Stokes settling velocity of an isolated vertical fibre.
However, \cite{herzhaft-guazzelli-1999} also experimentally reported that at higher concentrations, although fibres still tend to orient vertically, the formation of streamers weakens and the average settling velocity decreases. This observation was later corroborated by simulations from \cite{butler-shaqfeh-2002}, who showed that an initially homogeneous settling suspension may develop streamer structures surrounded by clarified fluid. They conjectured that this instability arises purely from hydrodynamic interactions coupling particle orientation and sedimentation rate within clusters. Furthermore, they confirmed that, depending on particle concentration and aspect ratio, streamers can enhance the sedimentation rate of the suspension beyond the maximum settling speed of an isolated fibre.
\cite{salmela-martinez-kataja-2005} experimentally measured the motion of individual fibres in settling suspensions across volume fractions ranging from $\Phi_V = 3.7 \times 10^{-4}$ to approximately $0.01-0.4$. Their study provided insights into how suspension concentration affects the average settling velocity. They observed that changes in settling velocity are accompanied by shifts in average fibre orientation: fibres preferentially orient horizontally in the dilute limit, while increasing concentration leads to alignment closer to the vertical.
More recently, \cite{banaei-etal-2020} conducted simulations of inertial settling for both rigid and flexible fibres. Fixing the Galileo number at $Ga = 160$ and varying the suspension concentration, they explored dilute to semi-dilute regimes. Their results showed that streamers form at high concentrations ($n \ell_f^3 > 10$), but in agreement with \citet{herzhaft-guazzelli-1999}, the streamers weaken as concentration increases due to reduced fibre clustering and mobility. Furthermore, they investigated the correlation between individual fibre velocities and the local suspension concentration at fibre locations, finding that the fastest settling fibres tend to reside in regions of higher concentration, whereas fibres exhibiting the fastest upward motion are predominantly found in areas with local concentrations below the average distribution value.

\subsection{The present study}

As discussed, much of the existing literature on fibre sedimentation in quiescent fluids has focused on the effects of $Ga$ and $n \ell_f^3$ on the suspension microstructure and average settling velocity. However, the modulation of fluid-phase velocity fluctuations by settling fibres, and their dependence on these parameters, remains relatively unexplored. This study addresses this gap through DNS coupled with the IBM.
We consider a suspension of heavy rigid fibres with density ratio $\rho_f/\rho = \mathcal{O}(100)$ settling under gravity in a quiescent fluid. The parameter space spans concentrations in the range $ n \ell_f^3 \in [0.36,23.15]$, covering dilute to semi-dilute regimes, and Galileo numbers in the range $Ga \in [180,900]$. The lower bound of $Ga$ aligns with recent work by \citet{banaei-etal-2020}, while the upper bound extends beyond commonly studied values. The elevated $Ga$ values emphasise buoyancy-driven effects and enhance fluid velocity fluctuations induced by the settling fibres.
The specific objectives of this study are: 
(i) to investigate how $Ga$ and $n \ell_f^3$ influence the organisation of fluid-phase velocity fluctuations induced by settling fibres;
(ii) to elucidate the mechanisms responsible for the generation and sustainment of these fluctuations under varying conditions;
(iii) to provide insight into the local structure of the fluid-phase velocity field; and
(iv) as a secondary aim, characterise the single-fibre statistics and collective spatial arrangement of settling fibres within the considered parameter range.

The remainder of the paper is organised as follows. \S\ref{sec:method} details the numerical methods and computational setup. The results are presented in \S\ref{sec:collective}, \S\ref{sec:fluid-phase} and \S\ref{sec:QR}: \S\ref{sec:collective} explores the collective fibre dynamics, while \S\ref{sec:fluid-phase} and \S\ref{sec:QR} analyse the sustaining mechanisms and local structure of the fluid-phase velocity fluctuations.  Finally, concluding remarks and an outlook are provided in \S\ref{sec:conclusion}.

 \section{Methods}
\label{sec:method}

We study the motion of $N$ slender, rigid fibres settling under gravity in a Newtonian, incompressible fluid that is otherwise at rest. The fibres are modelled as one-dimensional slender objects, i.e., their length $\ell_f$ is much larger than their diameter $d_f$, such that $\ell_f \gg d_f$. Besides $N$ and $\ell_f$, the system is characterised by the fluid density $\rho$, the kinematic fluid viscosity $\nu$, and the magnitude of the gravitational acceleration $g = |\bm{g}|$. The flow physics is entirely determined by the concentration of the suspension $n \ell_f^3$ ($n = N/V_t$ is the fibre number density with $V_t$ the volume of the entire suspension), the density ratio $\rho_f/\rho$, and the Galileo number $Ga$. As an alternative to using $n \ell_f^3$, it is possible to look at the volume fraction of the suspension $\Phi_V = V_f/V_t$, where $V_f$ denotes the volume occupied by the solid phase. The Galileo number can be written in the form of a Reynolds number, i.e. $Ga = u_g \ell_f/\nu$, where $u_g$ is the gravitational velocity scale given by $u_g = (|\rho_f/\rho-1| \ell_f g)^{1/2}$.

The fluid phase is governed by the incompressible Navier--Stokes equations for a Newtonian fluid,
\begin{equation}
  \frac{\partial u_i}{\partial t} + \frac{ \partial u_i u_j}{ \partial x_j } = 
  - \frac{\partial p}{\partial x_i} + \frac{1}{Ga} \frac{ \partial^2 u_i }{\partial x_j \partial x_j} + g_i + f_{fs,i}; \;\;\;\;\; \frac{\partial u_i}{\partial x_i} = 0,
\end{equation}
where $\bm{u} = (u,v,w)$ is the velocity field and $p$ is the reduced pressure. In the momentum equation, the term $\bm{f}_{fs}$ represents an additional volumetric forcing that models the presence of the solid phase. 

The fibre dynamics is governed by an extended Euler-Bernoulli beam equation and by the inextensibility constraint,
\begin{equation}
  \Delta \tilde{\rho} \frac{\partial^2 X_i}{\partial t^2} = 
  \frac{\partial}{\partial s} \left( T \frac{\partial X_i}{\partial s} \right) -
  \gamma_f \frac{\partial^4 X_i}{\partial s^4} + \Delta \tilde{\rho} g_i - F_i; \;\;\;\;\; \frac{\partial X_i}{\partial s}\frac{\partial X_i}{\partial s} = 1,
  \label{eq:fib}
\end{equation}
where $\bm{X}(s,t)$ denotes the (Lagrangian) position of a fibre point parameterised by the curvilinear coordinate $s$, $\Delta \tilde{\rho} = \tilde{\rho}_f - \tilde{\rho}$ is the difference between the linear densities (mass per unit length, based on the fibre cross-sectional area $A_f$) of the fibre and the fluid, $T$ represents the tension enforcing the inextensibility constraint, and $\gamma_f$ is the fibre bending rigidity. The bending rigidity $\gamma_f$ is chosen sufficiently large so that the fibres behave effectively as rigid bodies, exhibiting negligible deviation of their end-to-end distance from their nominal length \citep{brizzolara-etal-2021}. The resulting Cauchy number, defined as $Ca = (\rho_f d_f \ell_f^3 u_g^2)/(2 \gamma_f)$ is below $\mathcal{O}(10)$ for all cases. Specifically, the maximum relative variation in end-to-end distance remains below $0.4\%$ across all cases, with average variations under $0.03\%$, both smaller than those reported in previous studies \citep[e.g.,][]{rosti-etal-2020c}.
Additionally, $\bm{F}$ denotes the forcing accounting for the fluid's influence on the fibres. Since the fibres are freely moving within the carrier fluid, the governing equations are subject to the following boundary conditions at both ends of the fibre, i.e. $\partial^2 X_i/\partial s^2|_{s=0,\ell_f} = \partial^3 X_i/\partial s^3|_{s=0,\ell_f} = T|_{s=0,\ell_f} = 0$. 

The governing equations are numerically integrated in time using the in-house solver \href{https://www.oist.jp/research/research-units/cffu/fujin}{\textit{Fujin}}, based on second-order finite differences. The same numerical method has been adopted and validated in previous numerical studies; see for example \cite{olivieri-etal-2020-2, olivieri-mazzino-rosti-2022} and \cite{aswathy-rosti-2024} for suspensions of fibres in Newtonian and viscoelastic turbulent flows. The momentum equation is advanced in time by a fractional step method using a second-order Adams-Bashforth scheme, and the Poisson equation for the pressure is solved using an approach based on a fast Poisson solver. The dynamics of the flow and the fibres are coupled by enforcing the no-slip condition, i.e., $\partial X_i/\partial t = u_i(\bm{X}(s,t),t)$, with an immersed boundary method proposed by \cite{huang-etal-2007} and then modified by \cite{banaei-rosti-brandt-2020}, which is shortly described here. At each time step, the value of the fluid velocity is evaluated at the discretised Lagrangian points of the fibres $U_i = u_i(\bm{X}(s,t),t)$ by interpolating the values of the Eulerian velocity field $u_i$ at the grid points surrounding the Lagrangian point, i.e.,
\begin{equation}
  U_i(\bm{X}(s,t),t) = \int_{\Omega} u_i(\bm{x},t) \delta (\bm{x}-\bm{X}(s,t)) \text{d}V,
\end{equation} 
where $\delta$ is the Dirac delta function and $\Omega$ denotes the computational domain. Then, the interpolated velocity $\bm{U}$ is used to compute the fluid structure forcing $\bm{F}$ as
\begin{equation}
  F_i(s,t) = \beta \left( U_i - \frac{\partial X_i}{\partial t} \right),
\end{equation}
with $\beta$ being a very large negative constant \citep{huang-etal-2007, rosti-etal-2018, banaei-rosti-brandt-2020}. Eventually, the volume forcing acting on the fluid is evaluated by spreading $\bm{F}$ to the surrounding Eulerian grid points as
\begin{equation}
  f_{fs,i}(\bm{x},t) = \frac{1}{\tilde{\rho}} \int_S F_i(s,t) \delta (\bm{x}- \bm{X}(s,t)) \text{d} s.
\end{equation}
The Dirac operator is numerically treated with the regularised function proposed by \cite{roma-peskin-berger-1999}; in doing this, the diameter of the fibres is effectively around $d_f = 3 \Delta x$, where $\Delta x$ is the grid spacing. Equation \eqref{eq:fib} is solved following the same approach of \cite{huang-etal-2007}, the only difference being that the bending term is dealt implicitly \citep{banaei-etal-2020}. To prevent interpenetration between fibres, we use a fibre-to-fibre collision model, and we adopt the minimal collision model introduced by \cite{snook-guazzelli-butler-2012} by applying a constant forcing $\bm{F}_{col} = \bm{F}_0 \hat{\bm{d}}$ when the distance $\bm{d}$ between Lagrangian points of two distinct fibres is $|\bm{d}|<\Delta_{col}$; following \cite{olivieri-mazzino-rosti-2022}, we set $\bm{F}_0=1.0$ and $\Delta_{col} = 3 \Delta x$. A fixed-radius near-neighbors algorithm is used for the interaction between fibres to avoid an otherwise prohibitive computational cost \citep{monti-etal-2021}.

The fibres are released in a domain with periodic boundary conditions in all directions. Gravity acts in the vertical direction, aligned with the $z$-axis, such that $\bm{g} = -g \hat{\text{e}}_z$. Due to the presence of gravity, the system does not reach a statistically steady state on its own. To compensate for the resulting net flow, a uniform body force is applied in the vertical direction to maintain zero mean flow in the periodic domain, following the approach of \citet{uhlmann-doychev-2014}.

The computational domain has dimensions $(L_x, L_y, L_z) = (3\pi, 3\pi, 6\pi)$, which is larger than that used in \citet{banaei-etal-2020} under similar conditions. The number of fibres is fixed at $N = 10^4$, and the fibre length is varied within the set $\ell_f \in \{0.125\pi,\, 0.25\pi,\, 0.5\pi\}$ to explore different concentrations. The corresponding values of the concentration $n\ell_f^3$ are $0.36$, $2.89$, and $23.15$. Since $n$ is kept constant across cases, longer fibres correspond to higher concentrations, while shorter fibres correspond to lower concentrations. For the intermediate concentration ($n \ell_f^3 = 2.89$), we systematically vary the Galileo number within the set $Ga \in \{180,225,450,675,900\}$. For the lower concentration ($n \ell_f^3 = 0.36$), we consider $Ga \in \{180,450\}$, while for the higher concentration ($n \ell_f^3 = 23.15$), we explore $Ga \in \{180,450,900\}$. To focus on the generation of fluid-phase fluctuations, we consider heavy fibres with a density ratio $\rho_f / \rho \approx 234$, i.e. $\mathcal{O}(10^2)$.

The computational domain is discretised using $(N_x, N_y, N_z) = (384, 384, 768)$ grid points in the three directions. This resolution ensures that $\Delta x / \langle \eta \rangle = \mathcal{O}(1)$ in all cases, where $\langle \eta \rangle = (\nu^3/\varepsilon)^{1/4}$ is the average Kolmogorov length scale and $ \langle \varepsilon \rangle$ is the average energy dissipation rate; $\aver{\cdot}$ represents an average over realisations, time and along the three homogeneous directions. Specifically, the condition $\Delta x / \langle \eta \rangle < 1$ is satisfied for all simulations, except for the cases $(n \ell_f^3 = 0.36,\ Ga = 450)$ and $(n \ell_f^3 = 2.89,\ Ga = 900)$, where $\Delta x / \langle \eta \rangle \approx 1.4$.
This choice results in a fixed fibre diameter of $d_f = 3\Delta x $ across all cases.
 Consequently, the fibre aspect ratio $\AR = \ell_f / d_f$ varies with fibre length, taking values $\AR = 5.3$, $10.7$, and $21.3$ for $n \ell_f^3 = 0.36$, $2.89$, and $23.15$, respectively. Each fibre is discretised using $N_L$ Lagrangian points, selected such that the inter-point spacing $\Delta s = \ell_f / (N_L - 1)$ approximately matches the Eulerian grid spacing $\Delta x$. Accordingly, the number of Lagrangian points varies with fibre length: $N_L = 17$, $33$, and $65$ for $n \ell_f^3 = 0.36$, $2.89$, and $23.15$, respectively. All simulations are advanced for a duration of $\mathcal{O}(1000)\, \ell_f/u_g$ time units using a constant time step of $\Delta t \approx 0.001\, \ell_f/u_g$. The full set of simulation parameters is summarised in Table~\ref{tab:sim}.
Data are collected after discarding the initial transient, once the system reaches a statistically steady state. After this point, flow statistics fluctuate around their mean values. All reported quantities and visualisations are robust and representative of the long-time dynamics. Data are sampled every $50 \ell_f/u_g$.

\begin{table}
\centering
\begin{tabular}{ccccccccccccccccccc}
$n\ell_f^3$  & &  $\Phi_V$  & & $\Phi_M$ & &  $\AR$  & & $\rho_f/\rho$ & & $Ga$  & & $|\aver{ w_f }_f|/u_g$ & & $\aver{\eta}/\ell_f$ & & $\aver{\varepsilon}\ell_f/u_g^3$ & & $\aver{E}/u_g^2$ \vspace{0.2cm} \\
$0.36$     & &  $0.0032$      & & $0.43$ & &  $5.3$  & &  $234$         & & $180$  & & $0.66$             & & $ 0.076$     & & $0.0054$    & & $0.089$ \\
$0.36$     & &  $0.0032$      & & $0.43$ & &  $5.3$  & &  $234$         & & $450$  & & $0.87$             & & $ 0.037$     & & $0.0063$    & & $0.14$ 
\vspace{0.2cm} \\
$2.89$     & &  $0.0064$      & & $0.60$ & &  $10.7$ & &  $234$         & & $180$  & & $0.26$                 & & $ 0.078$     & & $0.0043$    & & $0.016$ \\
$2.89$     & &  $0.0064$      & & $0.60$ & &  $10.7$  & & $234$         & & $225$  & & $0.29$                 & & $ 0.065$     & & $0.0049$    & & $0.035$ \\
$2.89$     & &  $0.0064$      & & $0.60$ & &  $10.7$  & & $234$         & & $450$  & & $0.52$                 & & $ 0.035$     & & $0.0079$    & & $0.093$ \\
$2.89$     & &  $0.0064$      & & $0.60$ & &  $10.7$  & & $234$         & & $675$  & & $0.55$                 & & $ 0.025$     & & $0.0087$    & & $0.096$ \\
$2.89$     & &  $0.0064$      & & $0.60$ & &  $10.7$  & & $234$         & & $900$  & & $0.67$                 & & $ 0.019$     & & $0.010$    & & $0.13$ 
\vspace{0.2cm} \\
$23.15$    & &  $0.013$      & & $0.75$ & &  $21.3$ & &  $234$         & & $180$  & & $0.12$                 & & $ 0.081$     & & $0.0041$    & & $0.0064$ \\
$23.15$    & &  $0.013$      & & $0.75$ & &  $21.3$  & & $234$         & & $225$  & & $0.18$                 & & $ 0.037$     & & $0.0062$    & & $0.0098$ \\
$23.15$    & &  $0.013$      & & $0.75$ & &  $21.3$  & & $234$         & & $450$  & & $0.23$                 & & $ 0.021$     & & $0.0075$    & & $0.0086$ \\
\end{tabular}
\caption{
Summary of the numerical simulations conducted in this study. Here, $n \ell_f^3$ denotes the dimensionless fibre concentration, $\Phi_V = V_f / V_t$ is the volume fraction, and $\Phi_M = (\rho_f V_f)/((\rho_f - \rho) V_f + \rho V_t)$ is the mass fraction. The aspect ratio is defined as $\AR = \ell_f / d_f$, where $\ell_f$ and $d_f$ are the fibre length and diameter, respectively. The Galileo number is $Ga = u_g \ell_f / \nu$, with the buoyancy velocity $u_g = \sqrt{|\rho_f / \rho - 1| \, \ell_f g} $.
The quantity $\langle w_f \rangle_f$ denotes the mean settling velocity of the fibre midpoint in the vertical ($z$) direction; here, $\langle \cdot \rangle_f$ indicates an ensemble and time average over all fibres. The Kolmogorov length scale is defined as $\eta = (\nu^3 / \varepsilon)^{1/4}$, where the dissipation rate is given by $\varepsilon = 2 \nu s_{ij} s_{ij}$ with the strain-rate tensor $s_{ij} = ( \partial u_i/\partial x_j + \partial u_j/\partial x_i)/2$. $E$ denotes the kinetic energy of the fluid phase, and $\langle \cdot \rangle$ represents an average over realisations, time, and the three homogeneous spatial directions.
}
\label{tab:sim}
\end{table}

 \section{Collective dynamics of the fibres}
\label{sec:collective}

\begin{figure}
  \centering
  \includegraphics[width=1\textwidth]{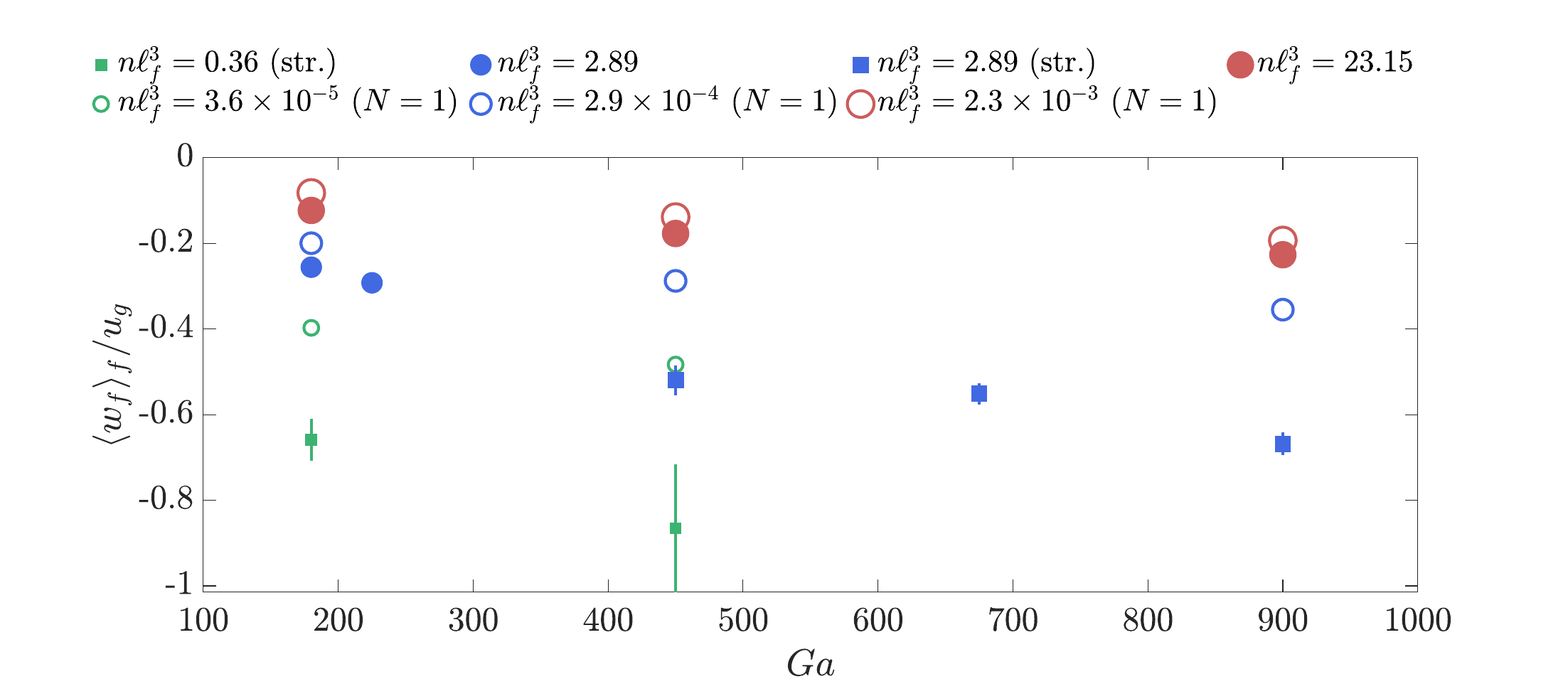}
  \caption{Settling velocity of the fibres. The filled symbols refer to the suspension of fibres with $N=10^4$. The empty symbols refer to the terminal velocity of single fibres settling in the same conditions. These velocities are obtained by running additional simulations with the same grid resolution and same domain size, but with $N=1$. Here and in the following, circles refer to cases where streamers are not formed or are rather weak, while squares refer to the cases where the streamers are evident and strong}
  \label{fig:settling_velocity}
\end{figure}

Before discussing the organisation of the fluid-phase velocity fluctuations, we examine the formation of streamers and/or clusters at the considered parameters. Streamers are regions where fibres concentrate, correlating with high local settling velocities. Outside the streamers the local concentration of the suspension is low, and the fibres may also move in the direction opposite to gravity.

Figure \ref{fig:settling_velocity} shows the average settling velocity of the fibres, $\langle w_f \rangle_f$, where the operator $\langle \cdot \rangle_f$ denotes an ensemble average over the $N$ fibres and in time. For comparison, the terminal velocity of a single fibre, $w_{f,t}$, is also shown (empty symbols), obtained from additional simulations with $N = 1$ under the same conditions. The error bars indicate the variance of the fibre settling velocity.
As shown in figure \ref{fig:settling_velocity}, fibres settle faster at lower concentrations, and increasing $Ga$ enhances the magnitude of $\langle w_f \rangle_f$. In all cases considered, the average settling velocity exceeds the terminal velocity of an isolated fibre, i.e., $|\langle w_f \rangle_f| > |w_{f,t}|$, consistent with the tendency of fibres to cluster and form regions of intensified downward motion.
A closer inspection reveals that the difference between $\langle w_f \rangle_f$ and $w_{f,t}$ is small at high concentrations and/or low $Ga$ ($n\ell_f^3 > 2.89$ or $Ga \le 450$), but becomes more pronounced at lower concentrations and/or higher $Ga$ ($n\ell_f^3 \le 2.89$ and/or $Ga \ge 450$). This behaviour is linked to the formation of columnar clusters or streamers by the settling fibres; see also \cite{herzhaft-guazzelli-1999,butler-shaqfeh-2002,metzger-guazzelli-butler-2005}.

\begin{figure}
  \centering
  \includegraphics[trim={0 260 0 0},clip,width=0.6\textwidth]{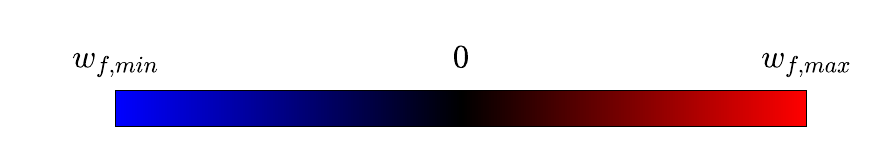}  
  \begin{tikzpicture}
    \node at (-1,0) {\includegraphics[width=0.32\textwidth]{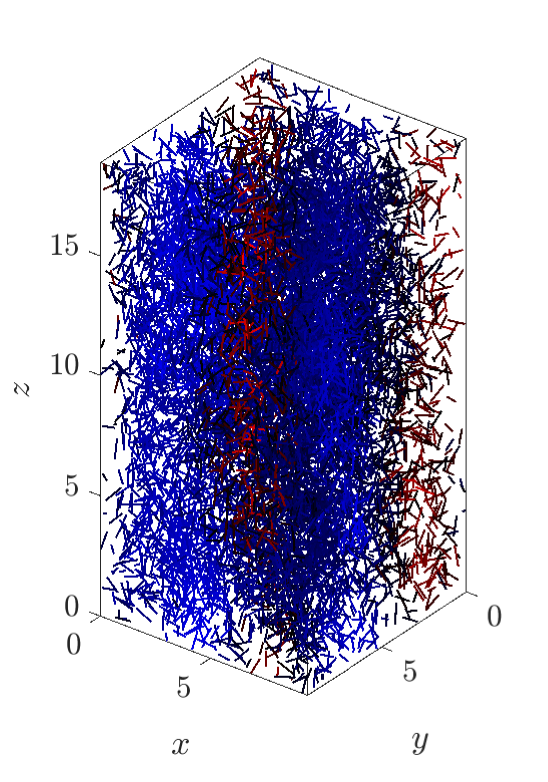}};
    \node at (3,0){\includegraphics[width=0.32\textwidth]{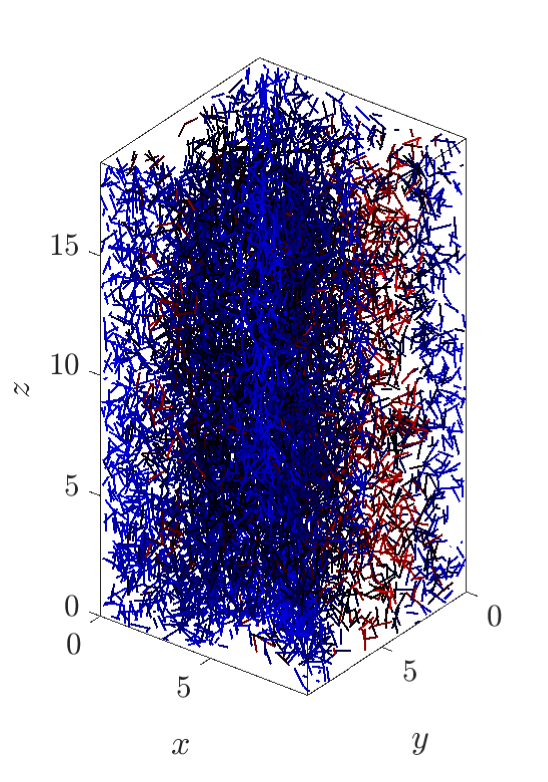}};
    \node at (7,0)  {\includegraphics[width=0.32\textwidth]{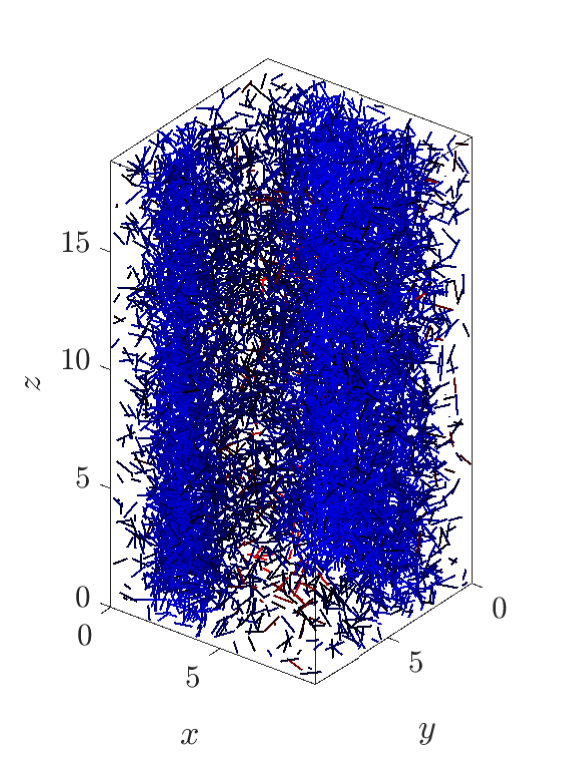}};
    \node at (-1,-5.25) {\includegraphics[trim={0 30 0 60},clip,width=0.31\textwidth]{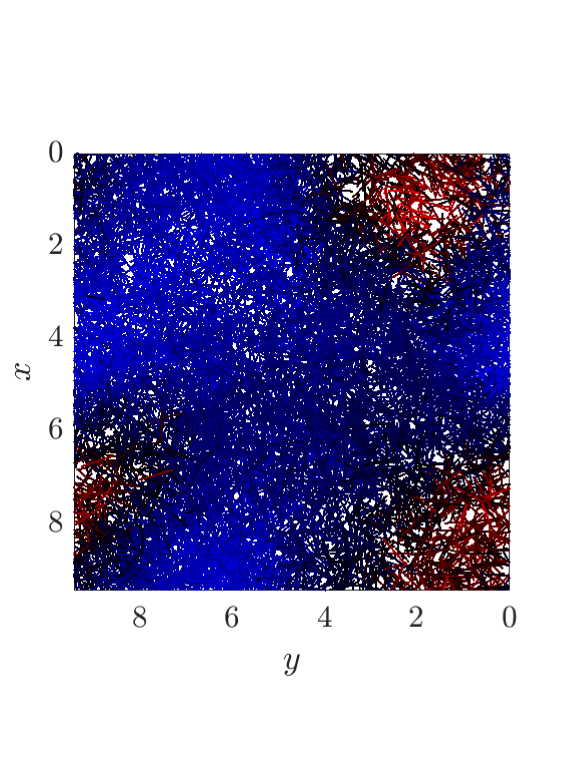}};
    \node at (3,-5.25){\includegraphics[trim={0 30 0 60},clip,width=0.31\textwidth]{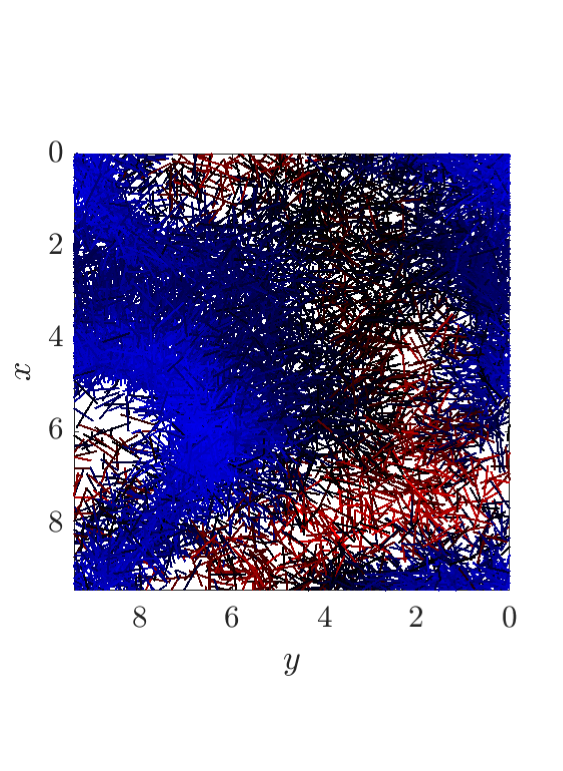}};
    \node at (7,-5.25)  {\includegraphics[trim={0 38 0 60},clip,width=0.31\textwidth]{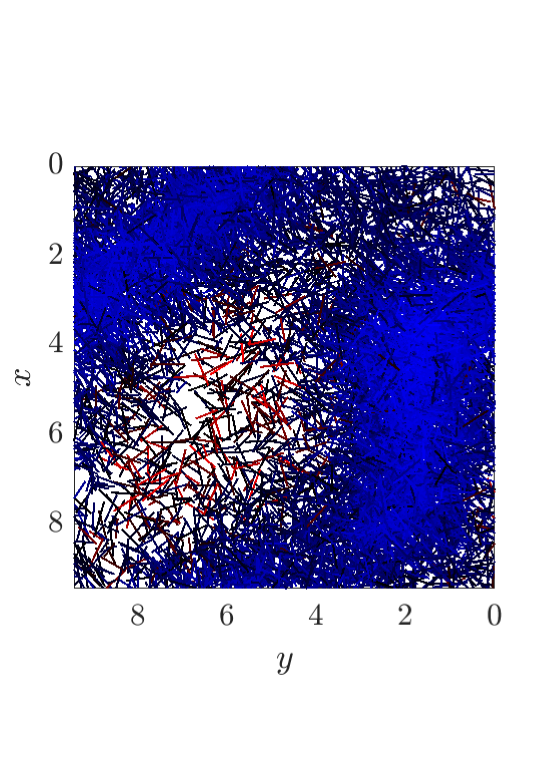}}; 
    \node at (-2.5,2.2){$(a)$};
    \node at ( 1.5,2.2){$(b)$};
    \node at ( 5.5,2.2){$(c)$};
    \node at (-2.75,-3){$(d)$};
    \node at ( 1.25,-3){$(e)$};
    \node at ( 5.25,-3){$(f)$};     
    \node at ( -1,-7.5){$Ga=180$};
    \node at (  3,-7.5){$Ga=450$};
    \node at (7.2,-7.5){$Ga=900$};
   \end{tikzpicture}
  \caption{Effect of the Galileo number on streamer formation for $n\ell_f^3=2.89$. Fibres are coloured according to the vertical velocity of their midpoints: blue indicates downward motion (negative velocity), red indicates upward motion (positive velocity), and black corresponds to zero vertical velocity. From left to right: $Ga = 180$ ($w_{f,\mathrm{min}} = -0.6335\,u_g$, $w_{f,\mathrm{max}} = 0.3156\,u_g$), $Ga = 450$ ($w_{f,\mathrm{min}} = -1.4282\,u_g$, $w_{f,\mathrm{max}} = 0.4251\,u_g$), and $Ga = 900$ ($w_{f,\mathrm{min}} = -1.2117\,u_g$, $w_{f,\mathrm{max}} = 0.5019\,u_g$).
}
  \label{fig:snap_c025}
\end{figure}
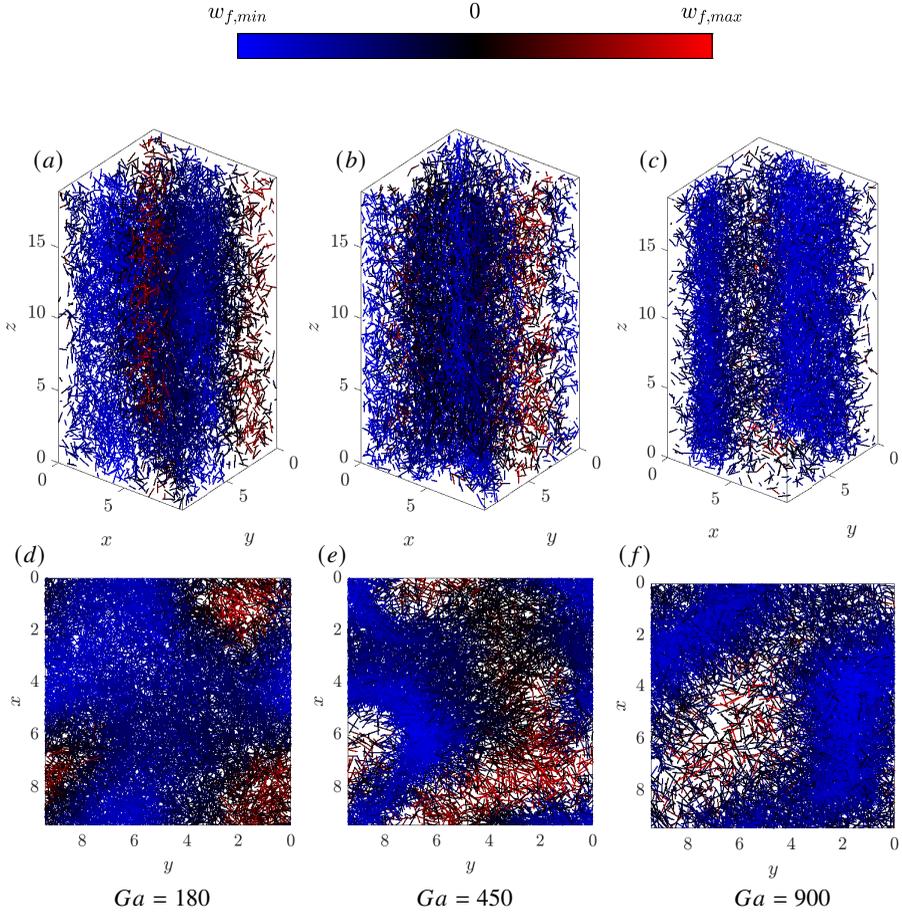

\begin{figure}
  \centering
  \includegraphics[trim={0 260 0 0},clip,width=0.6\textwidth]{./fig/legend_w_fib-eps-converted-to.pdf}
  \begin{tikzpicture}
    \node at (-1,0) {\includegraphics[trim={0 30 0 60},clip,width=0.31\textwidth]{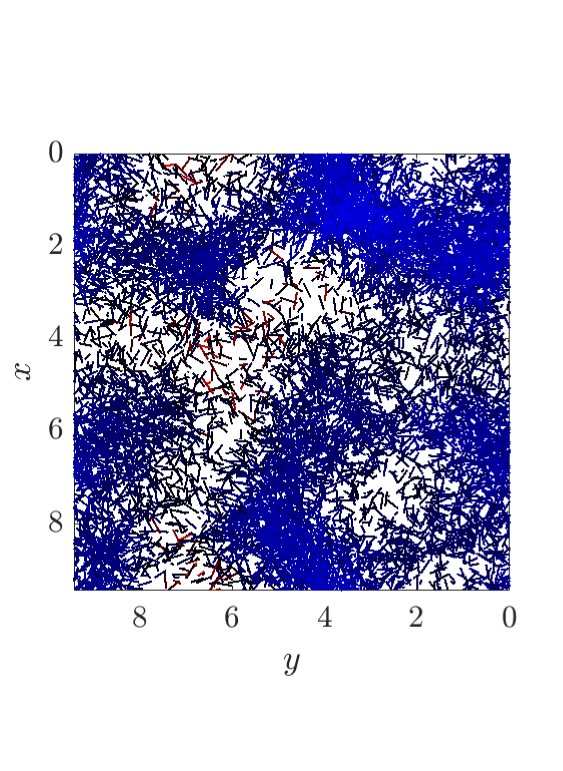}};
    \node at (3,0){\includegraphics[trim={0 30 0 60},clip,width=0.31\textwidth]{./fig/FibVel_c025pi_Ga180_top-eps-converted-to.pdf}};
    \node at (7,0)  {\includegraphics[trim={0 30 0 60},clip,width=0.31\textwidth]{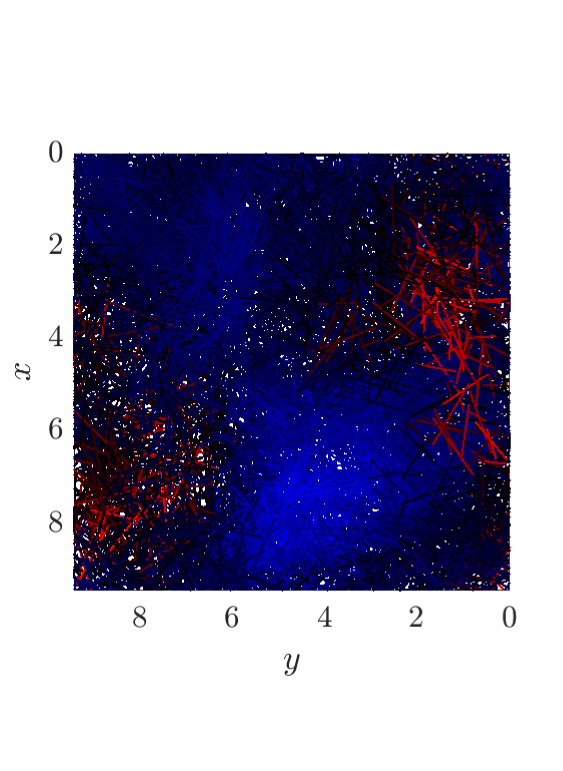}};
    \node at (-1,-4.25) {\includegraphics[trim={0 30 0 60},clip,width=0.31\textwidth]{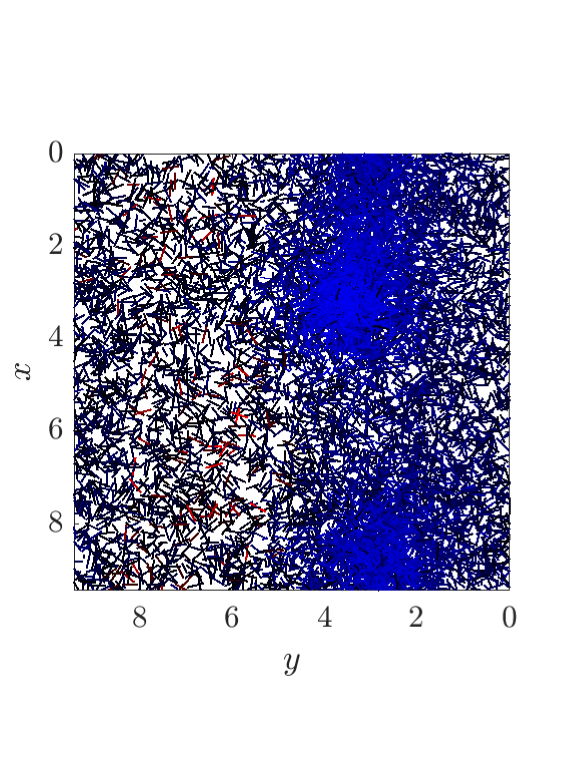}};
    \node at ( 3,-4.25){\includegraphics[trim={0 30 0 60},clip,width=0.31\textwidth]{./fig/FibVel_c025pi_Ga450_top-eps-converted-to.pdf}};
    \node at ( 7,-4.25)  {\includegraphics[trim={0 30 0 60},clip,width=0.31\textwidth]{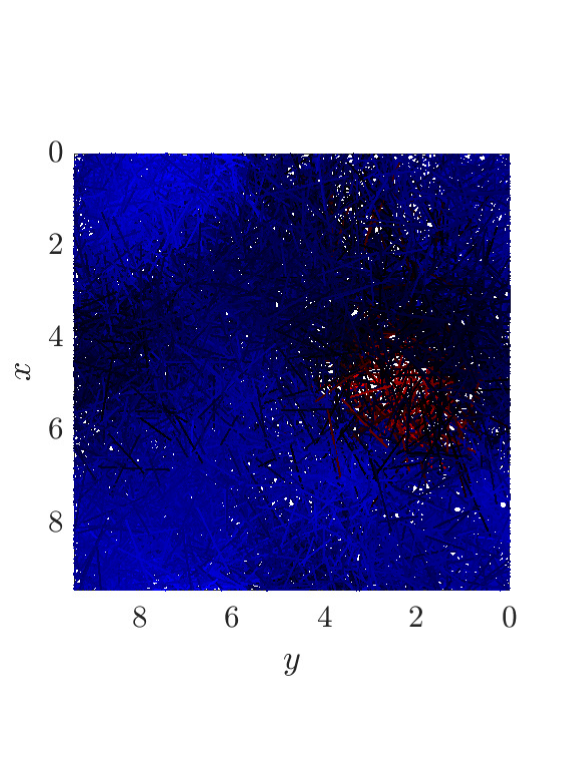}}; 
    \node at (-2.75, 2.35){$(a)$};
    \node at ( 1.25, 2.35){$(b)$};
    \node at ( 5.25, 2.35){$(c)$};
    \node at (-2.75,-1.9){$(d)$};
    \node at ( 1.25,-1.9){$(e)$};
    \node at ( 5.25,-1.9){$(f)$};  
    \node at ( -1,-6.5){$n \ell_f^3=0.36$};
    \node at (  3,-6.5){$n \ell_f^3=2.89$};
    \node at (7.2,-6.5){$n \ell_f^3=23.15$};    
    \node[rotate=90] at (9.3,0.5){$Ga=180$};
    \node[rotate=90] at (9.3,-3.75){$Ga=450$};   
  \end{tikzpicture}
  \caption{Effect of concentration $n \ell_f^3$ on streamer formation for (top) $Ga = 180$ and (bottom) $Ga = 450$. Fibres are coloured according to the vertical velocity of their midpoint Lagrangian point: blue indicates negative velocity, red indicates positive velocity, and black corresponds to near-zero velocity. From left to right, the concentration is $n \ell_f^3 = 0.36$, $n \ell_f^3=2.89$ and $n \ell_f^3 = 23.15$. For $n \ell_f^3=0.36$, $w_{f,\mathrm{min}} = -1.3549\,u_g$, $w_{f,\mathrm{max}} = 0.2906\,u_g$ at $Ga=180$ and $w_{f,\mathrm{min}} = -2.0396\,u_g$, $w_{f,\mathrm{max}} = 0.1505\,u_g$ at $Ga=450$. For $n \ell_f^3 = 2.89$, $w_{f,\mathrm{min}} = -0.6335\,u_g$, $w_{f,\mathrm{max}} = 0.3156\,u_g$ at $Ga=180$ and $w_{f,\mathrm{min}} = -1.4282\,u_g$, $w_{f,\mathrm{max}} = 0.4251\,u_g$ at $Ga=450$. For $n \ell_f^3 = 23.15$, $w_{f,\mathrm{min}} = -0.4025\,u_g$, $w_{f,\mathrm{max}} = 0.1302\,u_g$ at $Ga=180$ and $w_{f,\mathrm{min}} = -0.4527\,u_g$, $w_{f,\mathrm{max}} = 0.1767\,u_g$ at $Ga=450$.
}
  \label{fig:snap_Ga450}
\end{figure}

Figures \ref{fig:snap_c025} and \ref{fig:snap_Ga450} show instantaneous visualisations of fibre positions, coloured by their instantaneous settling velocity $w_f$. 
Figure \ref{fig:snap_c025} corresponds to $n \ell_f^3 = 2.89$ and illustrates the dependence of streamer formation on $Ga$. A clear columnar accumulation of fibres is observed only for $Ga \ge 450$.
At higher $Ga$, the top-view visualisation reveals alternating regions: some where fibres are highly concentrated and settle with large negative velocities, and others that are nearly devoid of fibres. In the latter regions, the few fibres present occasionally exhibit upward motion.
Figure \ref{fig:snap_Ga450} illustrates the influence of $n \ell_f^3$ on streamer formation. In these visualisations, the fibre concentration varies while the Galileo number is fixed at $Ga = 180$ (top panels) and $Ga = 450$ (bottom panels). At both values of $Ga$, fibres form vertical columns for $n \ell_f^3 = 0.36$ and $n \ell_f^3 = 2.89$, but not for $n \ell_f^3 = 23.15$.
More specifically, at $Ga = 180$, streamers are observed only at the lowest concentration ($n \ell_f^3 = 0.36$), whereas at $Ga = 450$, they also emerge at the intermediate concentration ($n \ell_f^3 = 2.89$). The absence of streamers at the highest concentration ($n \ell_f^3 = 23.15$) is consistent with previous findings by \cite{herzhaft-guazzelli-1999} and \cite{banaei-etal-2020}, and is attributed to the reduced and hindered mobility of the fibres under these conditions.

Clustering can be quantified using various metrics. One such metric is the radial distribution function (RDF) \citep{salazar-etal-2008,saw-etal-2008}, defined as
\begin{equation}
  g(r) = \frac{N_s(r)/\Delta V_i(r)}{N_p/V_t},
\end{equation}
where $N_s(r)$ is the number of fibre pairs separated by a distance $d$ such that $r - \Delta r/2 \le d \le r + \Delta r/2 $, $\Delta V_i $ is the volume of the spherical shell located at radius $r$ and $N_p = N(N-1)/2$ is the total number of fibre pairs. The RDF describes the probability of finding a pair of fibres separated by a distance $r$. For a perfectly uniform distribution (allowing for fibre overlap), $g(r) = 1$ for all $r$. In this analysis, each fibre is reduced to its midpoint Lagrangian marker.
As an alternative metric, we consider the local concentration $C_i$ associated with the $i_{th}$ fibre, defined as
\begin{equation}
  C_i = \frac{1}{(N-1)N_L^2} \sum_{j=1}^{N_L} \sum_{\substack{l=1\\ l \ne i}}^{N} \sum_{m=1}^{N_L} 
  \left( \frac{L_x}{d^{il}_{jm}} \right)^2,
\end{equation}
where $d^{il}_{jm}$ is the distance between the $j_{th}$ Lagrangian point of the $i_{th}$ fibre and the $m_{th}$ Lagrangian point of the $l_{th}$ fibre. A higher value of $C_i$ indicates that the fibre is located in a region of elevated local concentration and is likely part of a cluster. A similar approach was adopted by \cite{banaei-etal-2020}.

\begin{figure}
  \centering
  \begin{tikzpicture}
    \node at (3.5,1.7) {\includegraphics[trim={0 30 0 20},clip,width=0.8\textwidth]{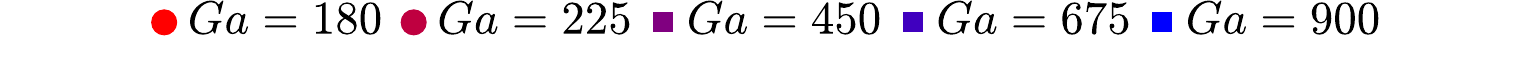}};
    \node at (0,0) {\includegraphics[width=0.49\textwidth]{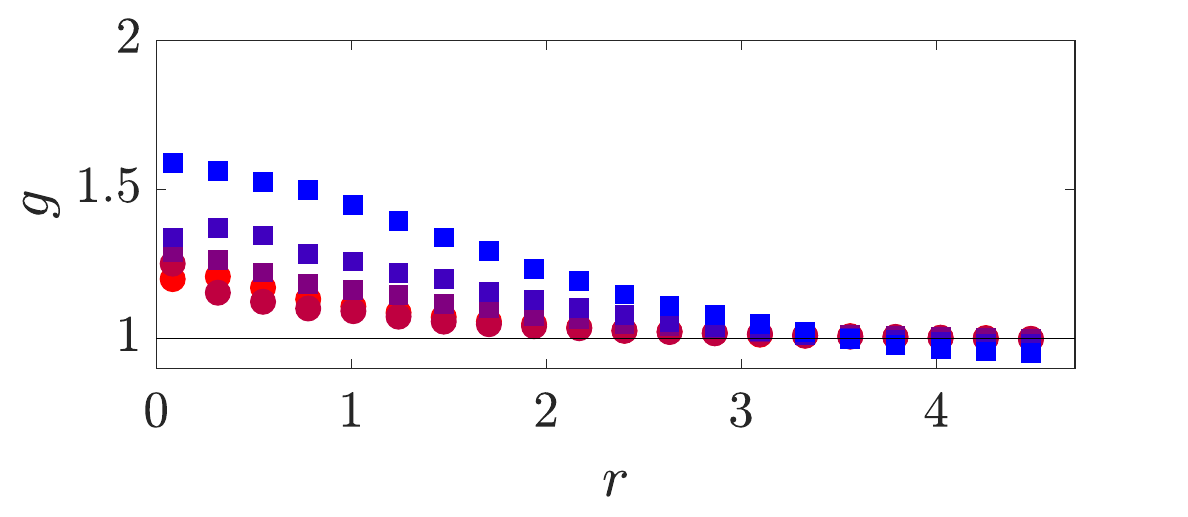}};
    \node at (7,0) {\includegraphics[width=0.49\textwidth]{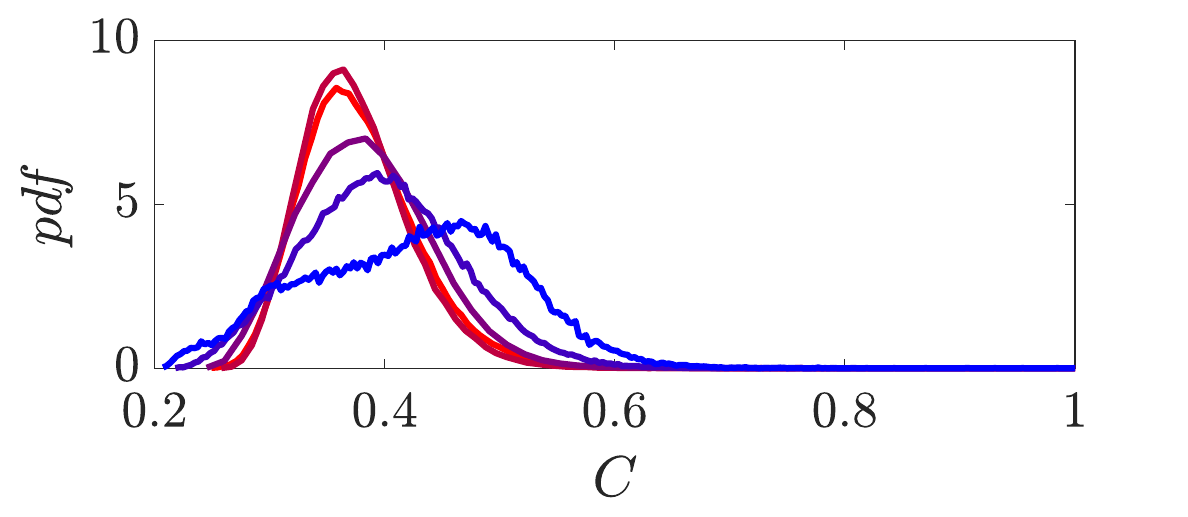}};
    \node at (3.5,-1.9) {\includegraphics[trim={0 30 0 5},clip,width=0.65\textwidth]{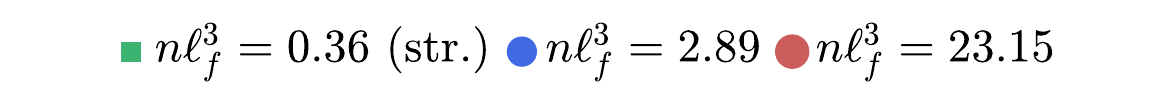}};   
    \node at (0,-3.6) {\includegraphics[width=0.49\textwidth]{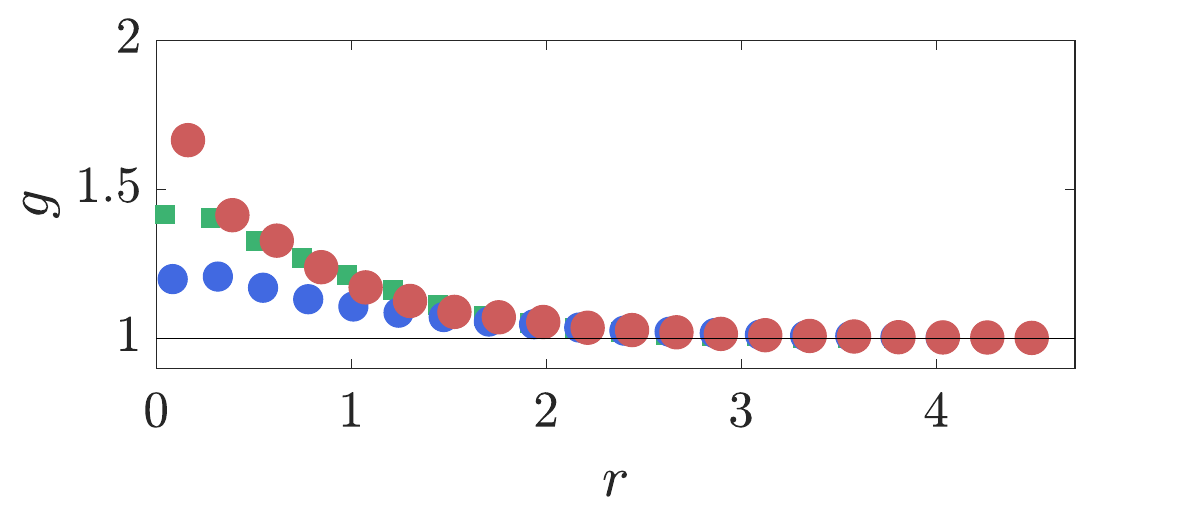}};
    \node at (7,-3.6) {\includegraphics[width=0.49\textwidth]{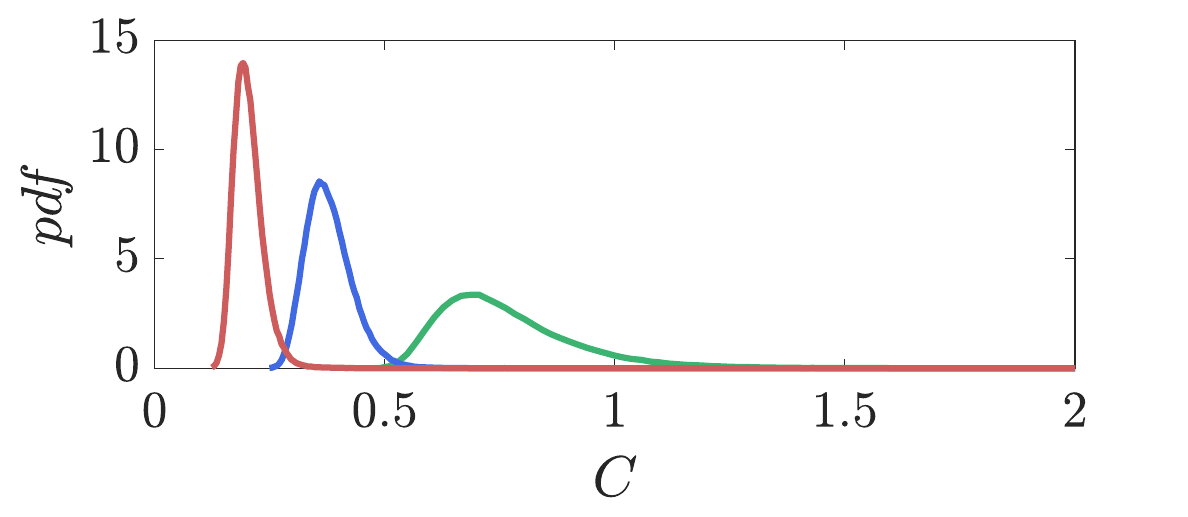}};  
    \node at (0,-6) {\includegraphics[width=0.49\textwidth]{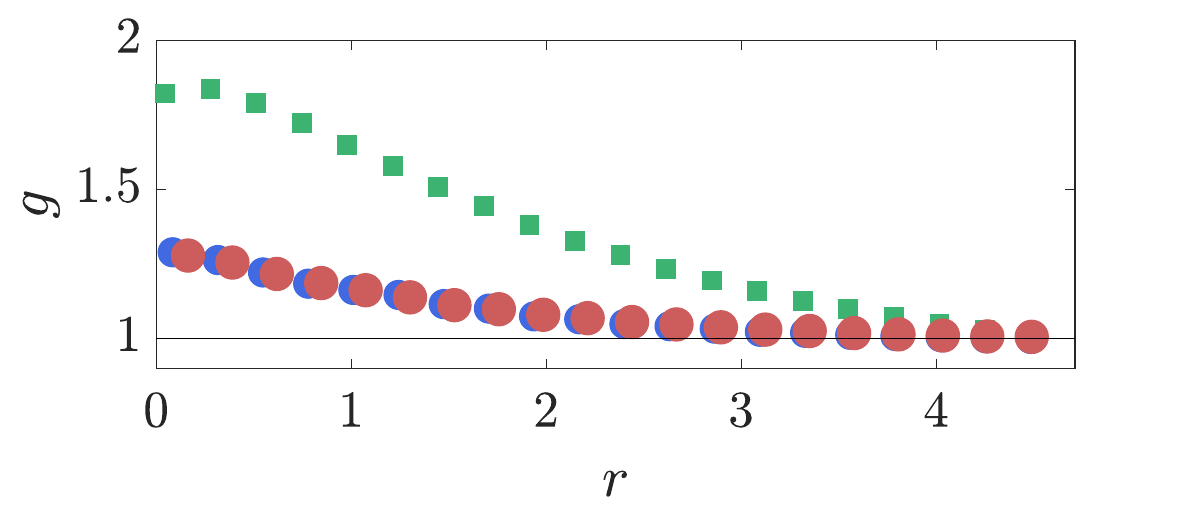}};
    \node at (7,-6) {\includegraphics[width=0.49\textwidth]{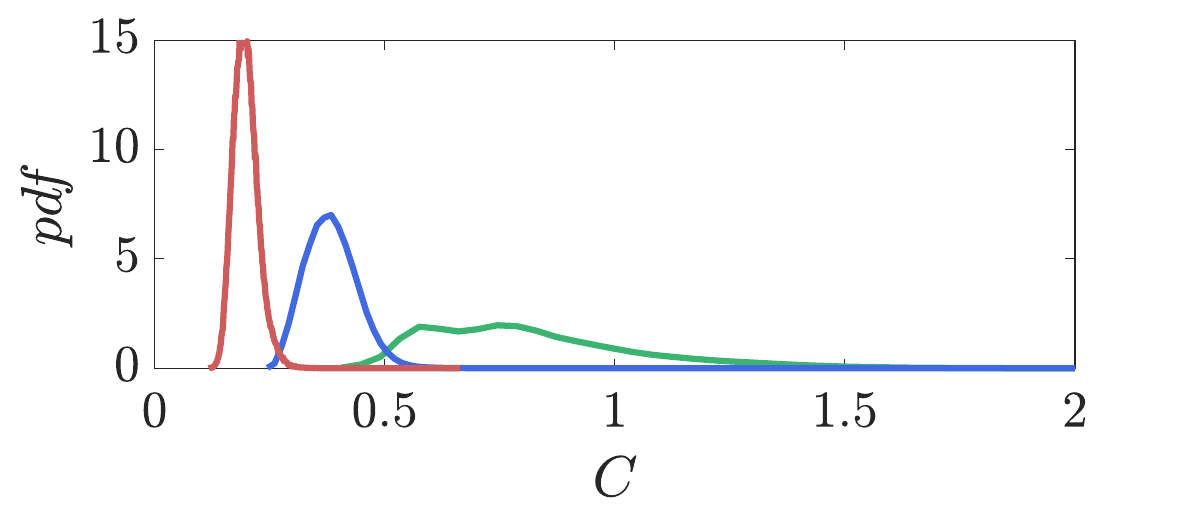}};     
    \node at (-3,1.7){$(a)$};
    \node at (-3,-1.9){$(b)$};
  \end{tikzpicture}
  \caption{Effect of $Ga$ and $n \ell_f^3$ on the (left) radial distribution function and (right) local concentration. Panel (a): variation of $Ga$ at fixed $n \ell_f^3 = 2.89$. Panel (b): variation of $n \ell_f^3$ at fixed $Ga = 180$ (top) and $Ga = 450$ (bottom). Symbols follow the convention in figure~\ref{fig:settling_velocity}, where squares indicate cases with streamer formation, and circles indicate cases without.}
  \label{fig:rdf_Gaeffect}
\end{figure}
The influence of $Ga$ and $n \ell_f^3$ on fibre clustering is illustrated in figure~\ref{fig:rdf_Gaeffect}. 
Figure \ref{fig:rdf_Gaeffect}$(a)$ shows that, once streamers form ($Ga > 225$), increasing the Galileo number enhances clustering.
Specifically, the probability of finding a pair of fibres at a distance $r\lessapprox 3$ increases with $Ga$, while it decreases at larger separations. This trend reflects the development of void regions between streamers, as captured by the RDF, which satisfies $g(r) < 1$ for $r \gtrapprox 3$.
Correspondingly, the left and right tails of the distribution of the local concentration metric $C$ extend as $Ga$ increases, with the smallest and largest values of $C$ identifying void regions and densely packed regions, respectively.
 Notably, for $Ga \leq 225$, i.e. in the absence of streamers, increasing $Ga$ leads to a reduction in clustering. This non-monotonic behaviour is robust, as confirmed by the standard deviation of clustering metrics across different snapshots (not shown for brevity).
Such trends are consistent with previous observations in turbulent flows, where particle clustering is maximised at intermediate values of particle inertia or Stokes number, i.e., when inertia is neither negligible nor dominant \citep[see, e.g.][]{brandt-coletti-2022,chiarini-rosti-2024}.
The effect of $n \ell_f^3$ on clustering is more subtle. On one hand, increasing $n \ell_f^3$ raises the volume fraction of the suspension, promoting greater clustering \citep[see, e.g.,][]{olivieri-mazzino-rosti-2022}. On the other hand, as visualised in figure \ref{fig:snap_Ga450}, lower $n \ell_f^3$ values favour streamer formation, leading to locally higher concentrations. These competing effects result in the non-monotonic dependence of $g(r)$ on $n \ell_f^3$ shown in figure~\ref{fig:rdf_Gaeffect}$(b)$. Nevertheless, the distributions of $g(r)$ and $C$ for $n \ell_f^3 = 0.36$ consistently correspond to streamer formation at both $Ga = 180$ and $Ga = 450$.

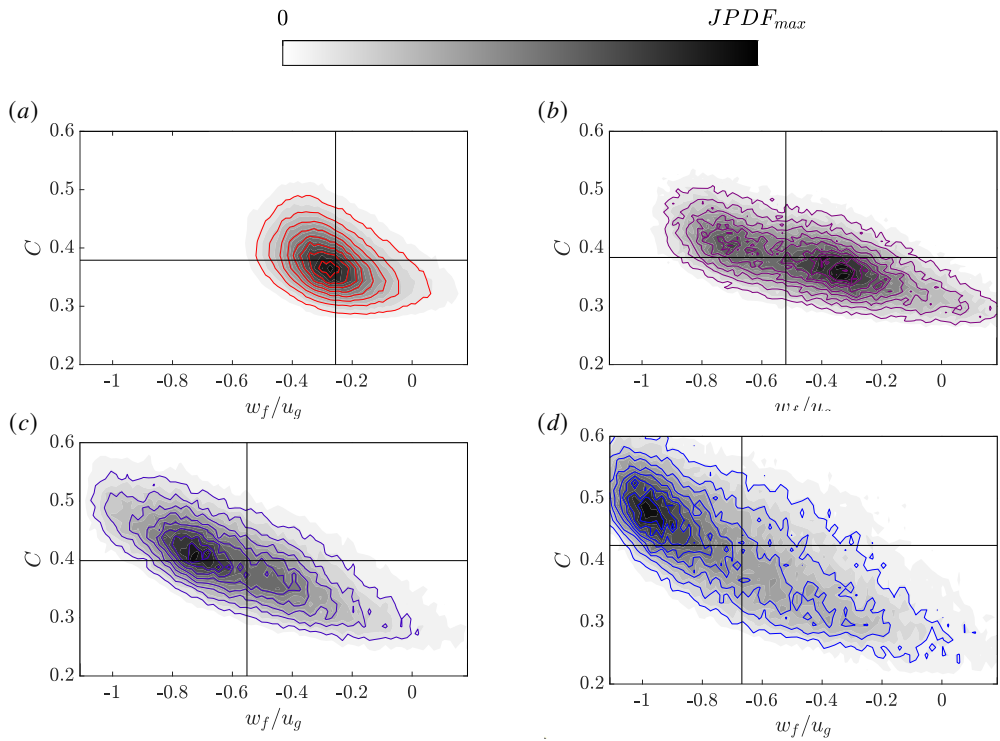
\begin{figure}
  \includegraphics[trim={0 260 0 0},clip,width=0.6\textwidth]{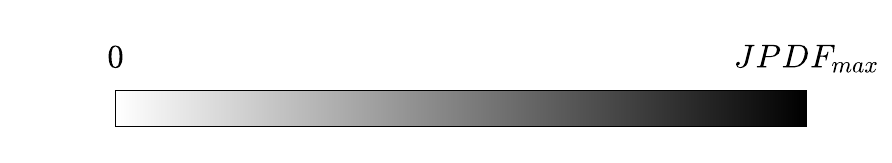}
  \centering
    \begin{tikzpicture}
    \node at (0,0) {\includegraphics[width=0.49\textwidth]{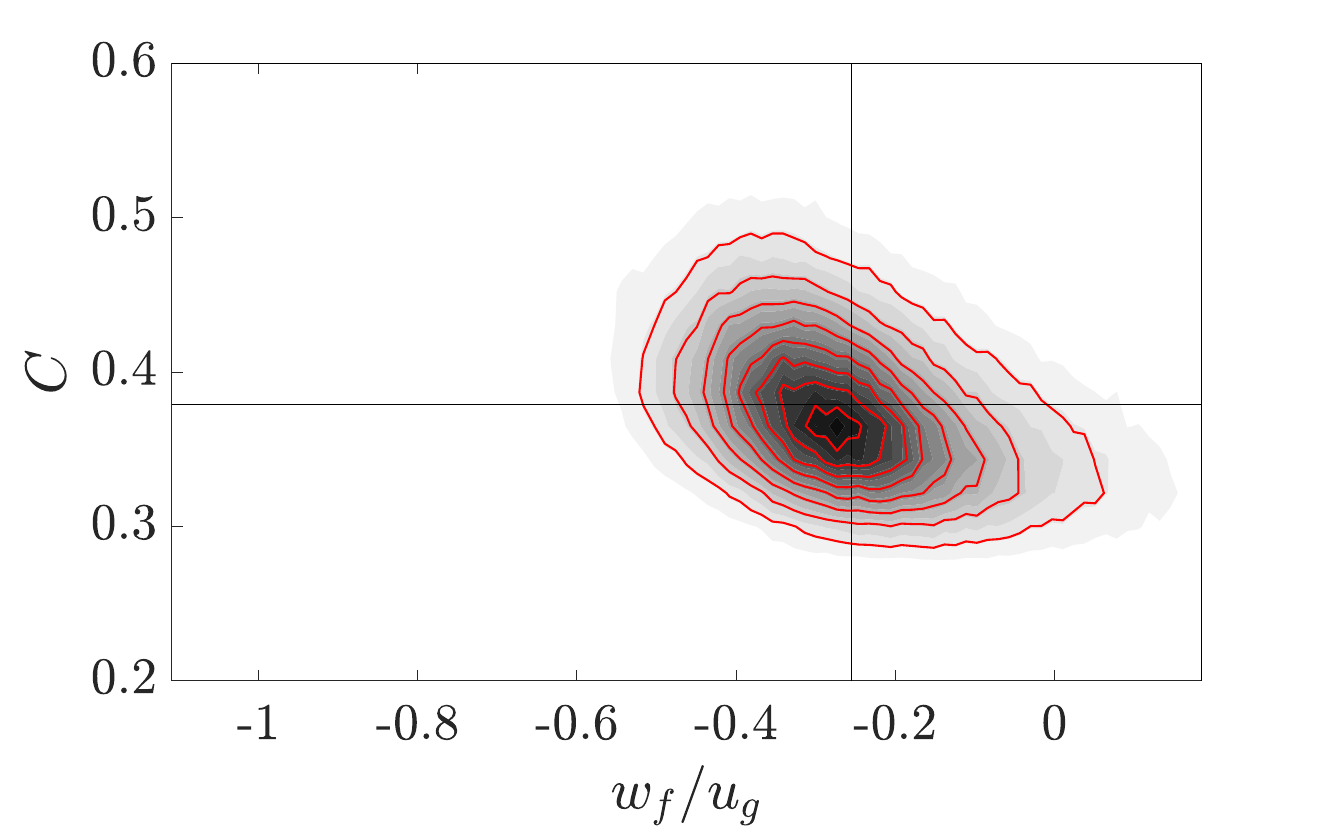}};
    \node at (7,0) {\includegraphics[width=0.49\textwidth]{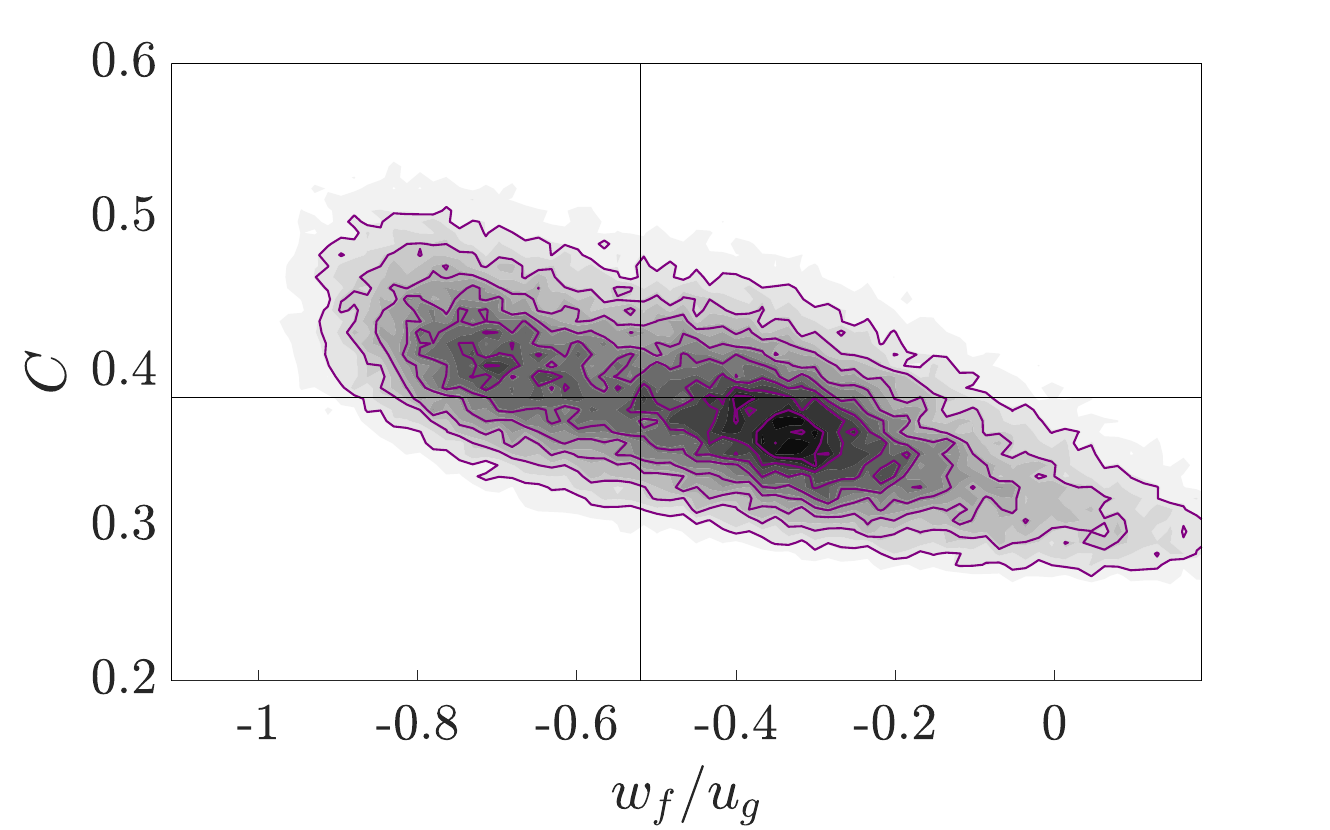}};  
    \node at (0,-4.1) {\includegraphics[width=0.49\textwidth]{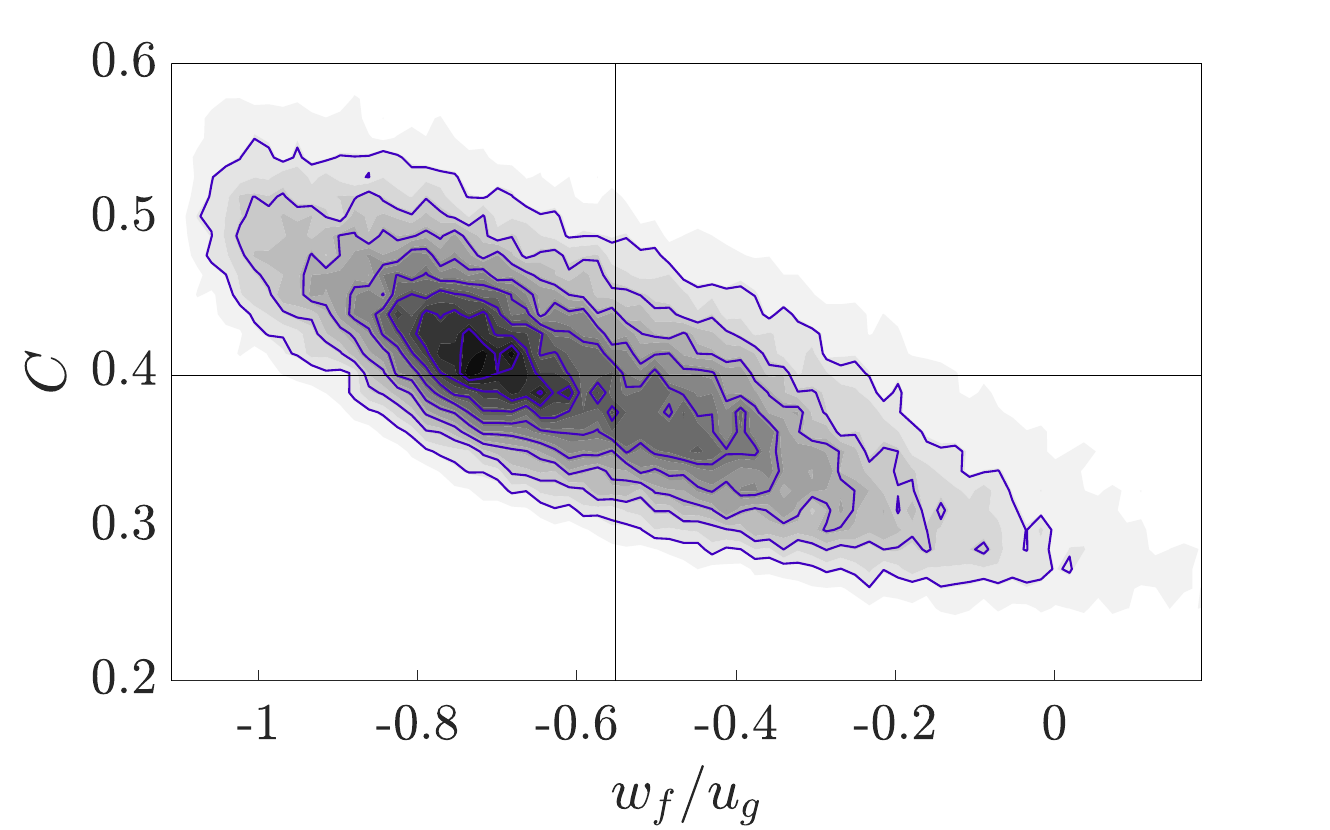}};
    \node at (7,-4.1) {\includegraphics[width=0.49\textwidth]{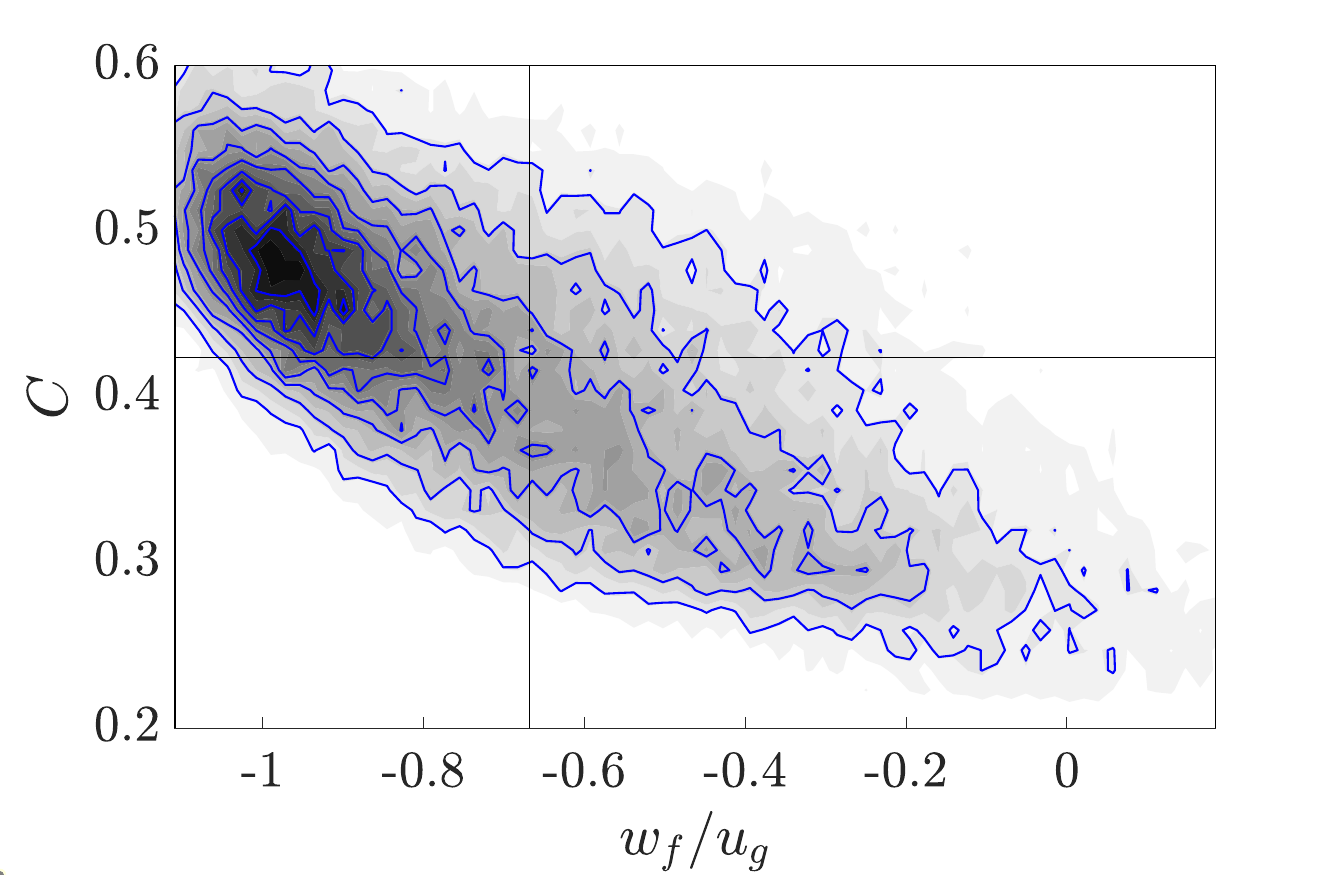}};   
    \node at (-3.2,2){$(a)$};
    \node at ( 3.8,2){$(b)$};
    \node at (-3.2,-2.1){$(c)$};
    \node at ( 3.8,-2.1){$(d)$};  
  \end{tikzpicture}        
  \caption{JPDF of fibre concentration $C$ and fibre velocity $w_f$ for $n \ell_f^3 = 2.89$, illustrating the effect of Galileo number. Panels (a–d) correspond to $Ga = 180$, 450, 675, and 900, respectively. Maximum JPDF values are $JPDF_{\mathrm{max}} = 30.59$, 16.94, 16.48, and 12.91 for $Ga = 180$, 450, 675, and 900.}
  \label{fig:Cw_Gaeffect}
\end{figure}
To further characterise the streamers formation, figure \ref{fig:Cw_Gaeffect} considers the correlation between the settling velocity $w_f$ of the fibres and their local concentration $C$. 
For the sake of brevity, we only focus on $n \ell_f^3 = 2.89$, but the discussion naturally extends to the other concentrations. 
We divide the $w_f-C$ planes into four quadrants by using the averages of the settling velocity and of the local fibres concentration, i.e. $\langle w_f \rangle_f$ and $\langle C \rangle_f$. 
The second quadrant with high fibre settling velocity and high local concentration ($w_f<\langle w_f \rangle_f$ and $C>\langle C \rangle_f$) contains the streamers, while the fourth quadrant with positive vertical relative velocities and low local fibre concentration ($w_f > \langle w_f \rangle_f$ and $C<\langle C \rangle_f$) denotes the void regions where fibres either settle slowly or move in the direction opposite to the gravity. 
As expected, the joint probability density function (JPDF) of $C$ and $w_f$ exhibits a strong dependence on $Ga$. For low values of $Ga$ ($Ga \le 225$), i.e., in the absence of streamers, the JPDF is approximately symmetric with respect to inversion of both axes. A slight bias is nevertheless observed toward events where $(w_f - \langle w_f \rangle_f)(C - \langle C \rangle_f) < 0$.
For $Ga \ge 450$, this bias becomes more pronounced, with the distribution skewed toward the second and fourth quadrants, consistent with the formation of streamers. Interestingly, at $Ga = 450$, the distribution peak lies in the fourth quadrant, indicating a significant number of fibres settling more slowly than the average despite the presence of streamers; see the central panels in figure \ref{fig:snap_c025}.
At even higher $Ga$ ($Ga \ge 675$), the peak shifts to the second quadrant. In this regime, the probability of events with $w_f \ge \langle w_f \rangle_f$ and $C < \langle C \rangle_f$ diminishes, suggesting that most fibres are tightly packed within the streamers and few remain in low-concentration regions.

\begin{figure}
  \centering
  \begin{tikzpicture}
    \node at (0,0) {\includegraphics[width=0.49\textwidth]{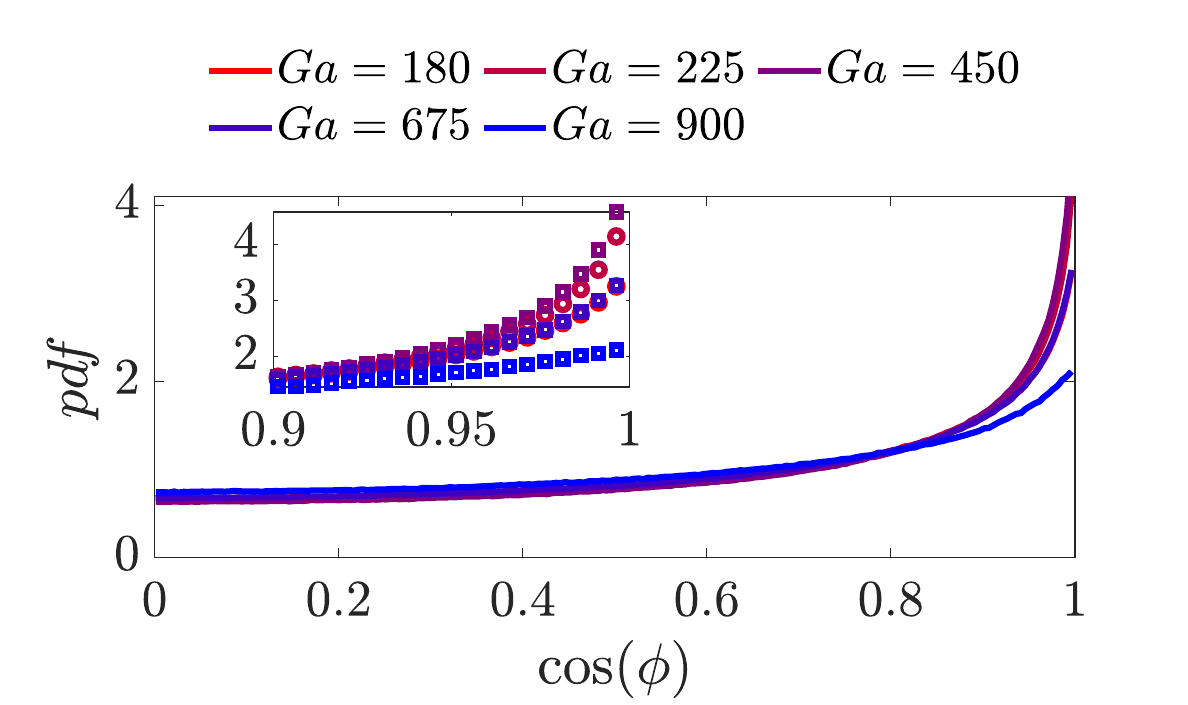}};
    \node at (7,0) {\includegraphics[width=0.49\textwidth]{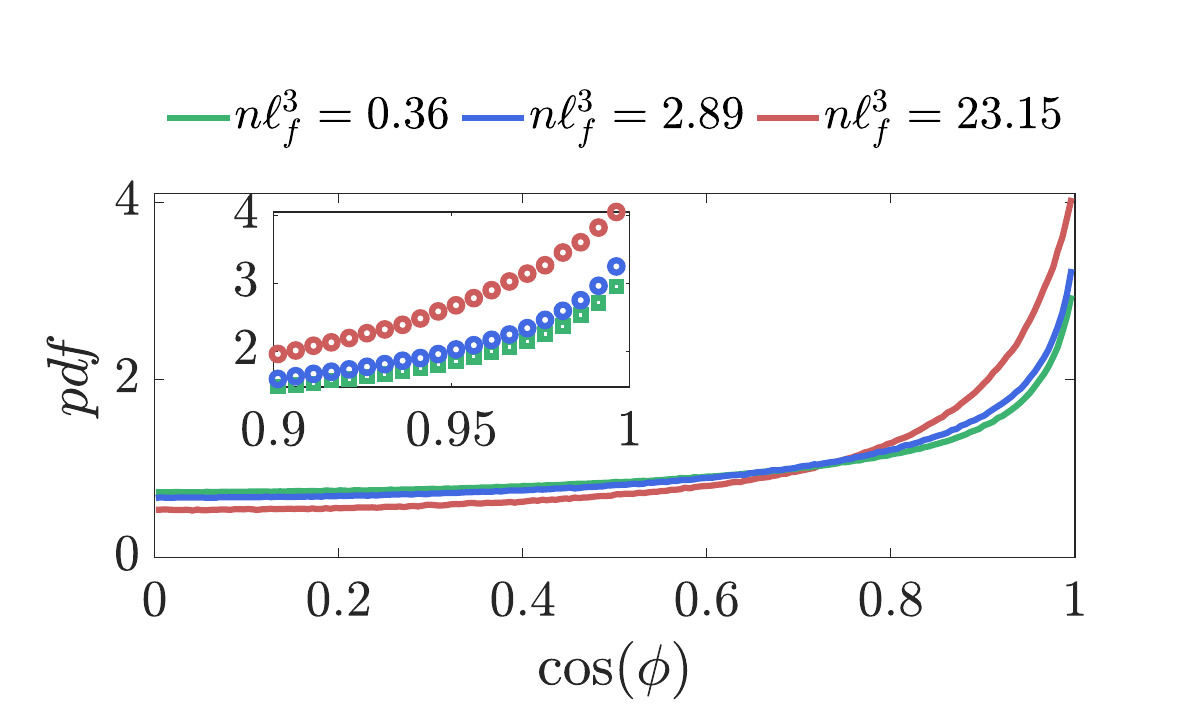}};
    \node at (-3.2,1.4){$(a)$};
    \node at ( 3.8,1.4){$(b)$};
  \end{tikzpicture}  
  \caption{Distribution of the alignment of the fibres with the $z$ direction. When fibres  align with the $z$ direction $\cos(\phi)=1$. When fibres are perpendicular to the $z$ direction $\cos(\phi)=0$. $(a)$: influence of the Galileo number on the distribution of $\cos(\phi)$ at $n \ell_f^3 = 2.89$. $(b)$: influence of the concentration $n \ell_f^3$ on $\cos(\phi)$ at $Ga=180$.}
  \label{fig:cth}
\end{figure}
This section concludes with an analysis of the fibres settling angle. We define the settling angle $\phi$ such that $\cos(\phi) = \hat{\bm{e}}_z \cdot \bm{\ell}_f/\ell_f$, where $\bm{\ell}_f$ is the vector that connects the free ends of the fibres; $\cos(\phi)=1$ ($\cos(\phi)=0$) indicates fibres that settle aligned with (perpendicular to) the gravity. Figure \ref{fig:cth} shows that at the present parameters, fibres tend to align with the direction of the gravity, in agreement with the experimental results of previous authors \citep[see for example][]{herzhaft-guazzelli-1999}. Figure \ref{fig:cth}$(a)$ details the influence of $Ga$ on the distribution of $\cos(\phi)$ by fixing $n \ell_f^3 = 2.89$, while figure \ref{fig:cth}$(b)$ characterises the influence of $n \ell_f^3$ by fixing $Ga=180$. The influence of $Ga$ on $\cos(\phi)$ is not monotonous. When increasing the Galileo number from $Ga=180$ to $Ga=450$ the probability of events with $\cos(\phi)=1$ increases and then decreases for larger $Ga$, being minimum for $Ga=900$. Arguably, this is because once the streamers form and $Ga$ is large enough, the backreaction of the chaotic fluid phase on the fibre motion becomes relevant (see \S\ref{sec:fluid-phase}), tending to restore a more uniform distribution. Turning to figure \ref{fig:cth}$(b)$, the data show that, for fixed $Ga$, fibres tend to align more strongly with the direction of gravity as the concentration increases.

 \section{The fluid-phase fluctuations}
\label{sec:fluid-phase}

We now shift our focus to the fluid phase. The objectives of this section are: (i) to examine the influence of $Ga$ and $n\ell_f^3$ on the scale-by-scale distribution of kinetic energy, and (ii) to elucidate the mechanisms responsible for the generation of fluid-phase velocity fluctuations. We first analyse integral flow quantities in \S\ref{sec:flow-mean}, then investigate the energy distribution across scales in \S\ref{sec:scale-by-scale}, and finally discuss the mechanisms sustaining velocity fluctuations in \S\ref{sec:bud}.

\subsection{Flow snapshots and bulk quantities}
\label{sec:flow-mean}

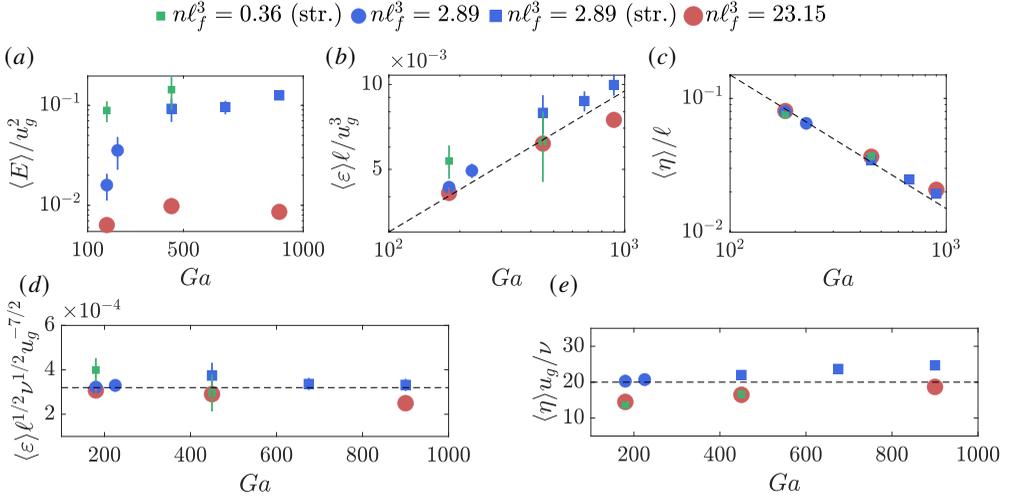
\begin{figure}
  \includegraphics[trim={0 30 0 5},clip,width=0.8\textwidth]{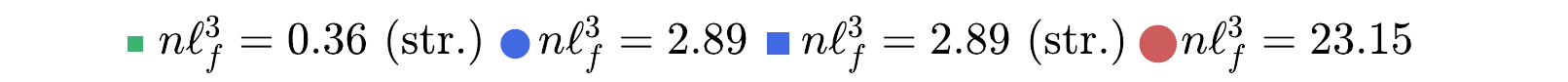}  
  \centering
  \begin{tikzpicture}
\node at (-1,0) {\includegraphics[width=0.32\textwidth]{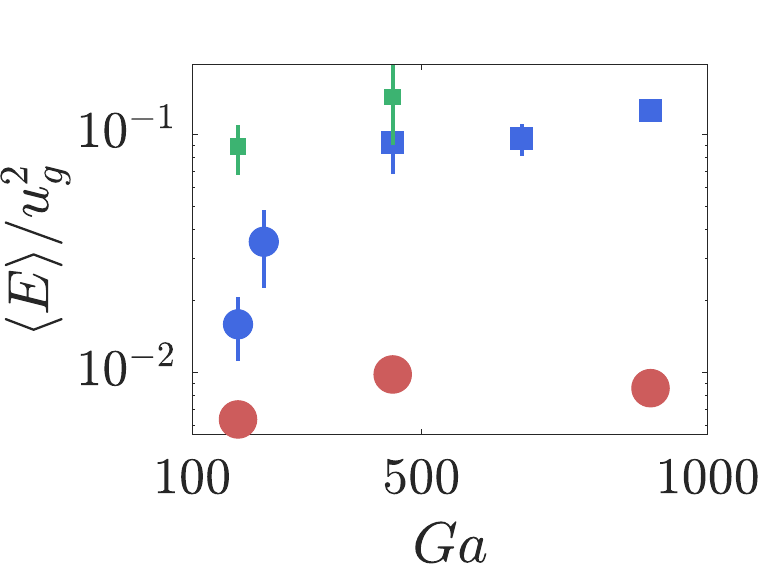}};
    \node at (3.25,0) {\includegraphics[width=0.32\textwidth]{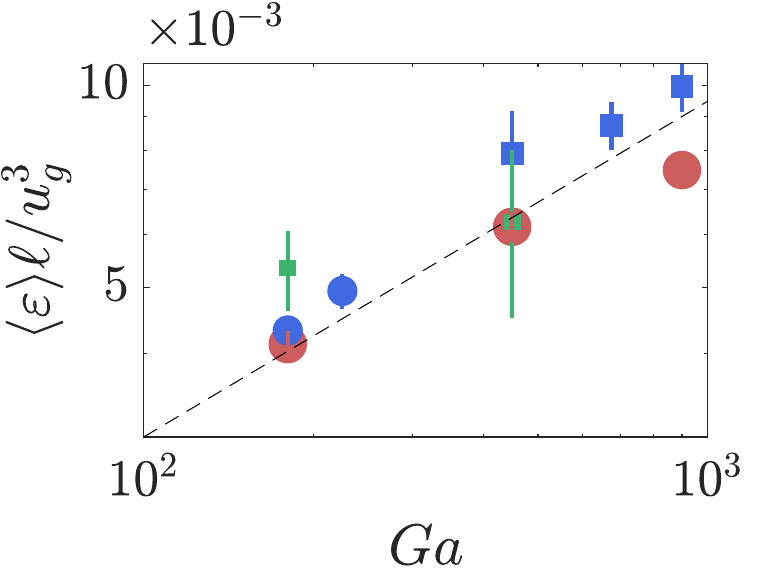}};
    \node at (7.5,0) {\includegraphics[width=0.32\textwidth]{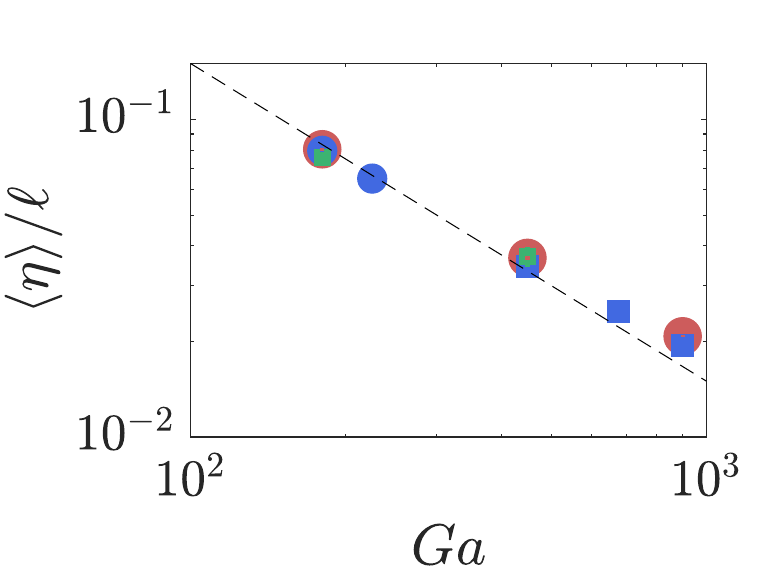}};
    \node at (0,-3) {\includegraphics[width=0.49\textwidth]{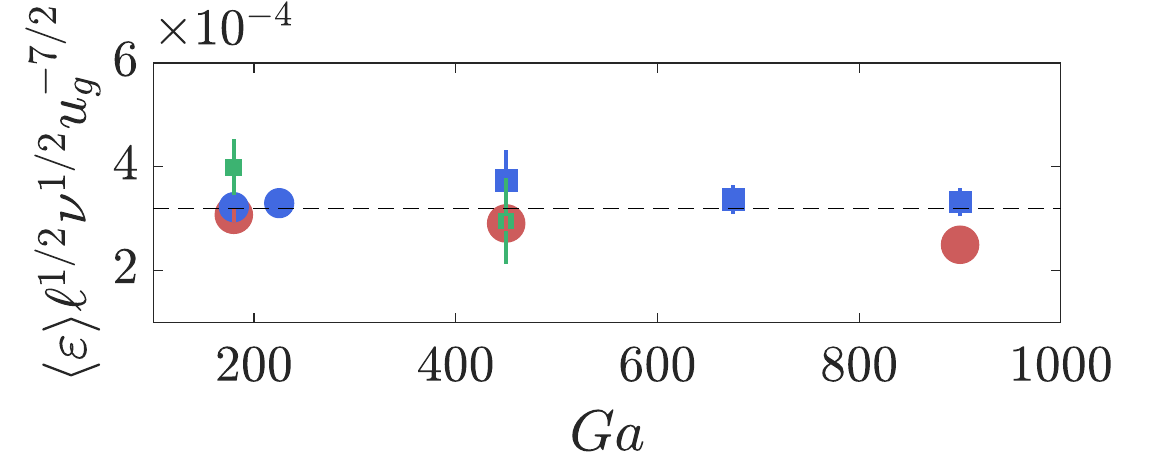}};
    \node at (7,-3) {\includegraphics[width=0.49\textwidth]{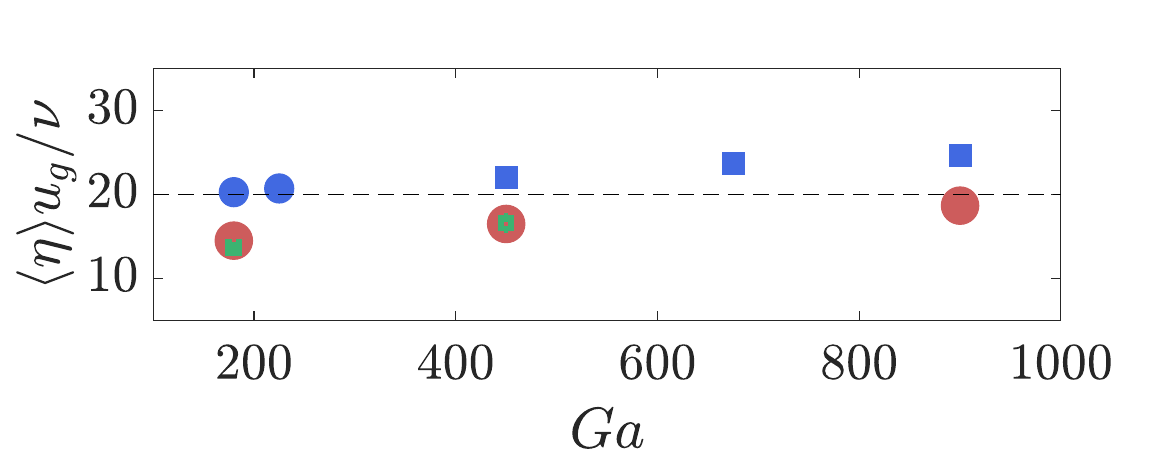}};
    \node at (-3,1.5){$(a)$};
    \node at ( 1.25,1.5){$(b)$};
    \node at ( 5.5,1.5){$(c)$};
    \node at (-2.8,-1.5){$(d)$};
    \node at ( 4.2,-1.5){$(e)$};
  \end{tikzpicture}  
  \caption{Dependence of mean flow quantities on $Ga$ and $n \ell_f^3$. Blue symbols are for $n \ell_f= 0.36$, red symbols for $n \ell_f^3 = 2.89$, and green ones for $n \ell_f^3 = 23.15$. Circles/squares indicate the absence/presence of the streamers. Top left: energy of the fluctuations $E$. Top centre: viscous dissipation $\varepsilon$. Top right: Kolmogorov length scale $\eta$. In the central panel, the dashed line indicates a $Ga^{1/2}$ power law scaling. In the right panel, the dashed line indicates a $Ga^{-1}$ power law scaling. The vertical bars refer to the root mean square values of the time evolution of the three quantities. Bottom: Dependence of (left) $\langle \varepsilon \rangle$ and (right) $\langle \eta \rangle$ on $Ga$ and $n \ell_f^3$. Symbols follow the convention in figure~\ref{fig:settling_velocity}, where squares indicate cases with streamer formation, and circles indicate cases without.}
  \label{fig:bulk}
\end{figure}

Figure \ref{fig:bulk}$(a)$ shows the dependence of the fluid-phase kinetic energy, $\langle E \rangle$, on $Ga$ and $n \ell_f^3$. The fluid-phase kinetic energy increases with decreasing $n \ell_f^3$ and increasing $Ga$, and exhibits a plateau for $Ga \gtrapprox 450$.
At lower $n \ell_f^3$ and higher $Ga$, fibres settle with larger vertical velocities (see figure \ref{fig:settling_velocity}), generating stronger momentum deficits in the fluid phase. This enhances the production of fluid-phase velocity fluctuations (see \S\ref{sec:scale-by-scale}).
Figures \ref{fig:bulk}$(b)$ and \ref{fig:bulk}$(c)$ show the average dissipation rate, $\langle \varepsilon \rangle$, and Kolmogorov length scale, $\langle \eta \rangle$, both of which are key quantities in turbulence theory. 
The data corresponding to different values of $n \ell_f^3$ collapse reasonably well when $u_g$ and $\ell_f$ are used as characteristic velocity and length scales.
Specifically, we observe that the normalised dissipation rate, $\langle \varepsilon \rangle \ell_f / u_g^3$, follows a power-law increase with the Galileo number, i.e., $\langle \varepsilon \rangle \ell_f/u_g^3 \sim Ga^{\alpha_{\varepsilon}}$, where $\alpha_{\varepsilon} \approx 1/2$. The increased production of velocity fluctuations at higher $Ga$ leads to enhanced viscous dissipation. This scaling implies $\langle \varepsilon \rangle \sim u_g^{7/2} \ell_f^{-1/2} \nu^{-1/2}$, as shown in figure~\ref{fig:bulk}$(d)$.
In contrast, the average Kolmogorov length scale decreases with increasing $Ga$, exhibiting a power-law behaviour $\langle \eta \rangle/ \ell_f \sim Ga^{-\alpha_{\eta}}$, with $\alpha_{\eta} \approx 1$. This $Ga^{-1}$ scaling implies $\langle \eta \rangle/\ell_f \sim Ga^{-1} \sim \left( u_g \ell_f/\nu \right)^{-1} \sim \nu/(u_g \ell_f)$, and thus $\langle \eta \rangle \sim \nu/u_g$, as illustrated in figure~\ref{fig:bulk}$(e)$. 
This suggests that at the present parameters the smallest flow scales are independent of the fibre length $\ell_f$, and instead depend on the fibre diameter $d_f$, which is held constant across all cases (see \S\ref{sec:method}). This finding is consistent with the observation that fibres tend to align with gravity (see figure \ref{fig:cth}), implying that the characteristic scale of the velocity fluctuations around the fibres is set by their effective frontal area ($\sim d_f^2$). 
In contrast, no such collapse is observed for $\langle E \rangle$, which is associated with the larger scales of the flow, inherited from the collective motion of the fibres.

\begin{figure}
  \centering
\centering
  \begin{tikzpicture}
    \node at (-1,0) {\includegraphics[width=0.32\textwidth]{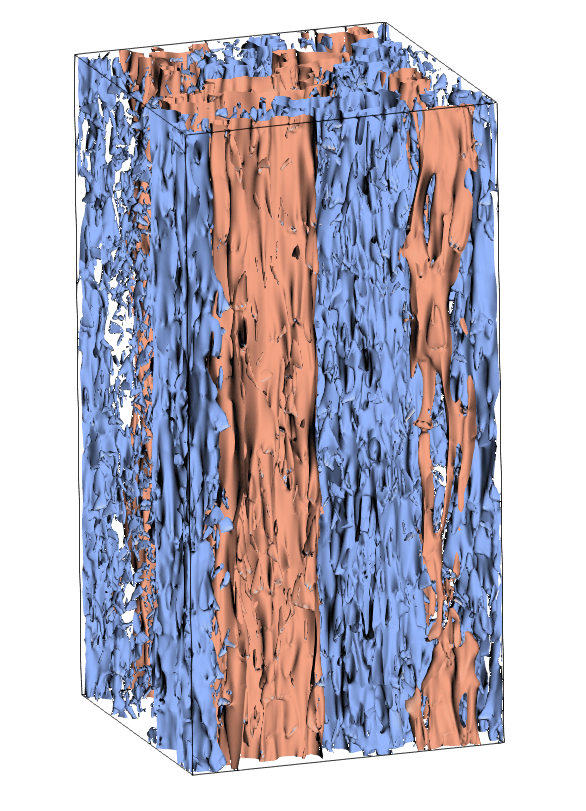}};
    \node at (3,0) {\includegraphics[width=0.32\textwidth]{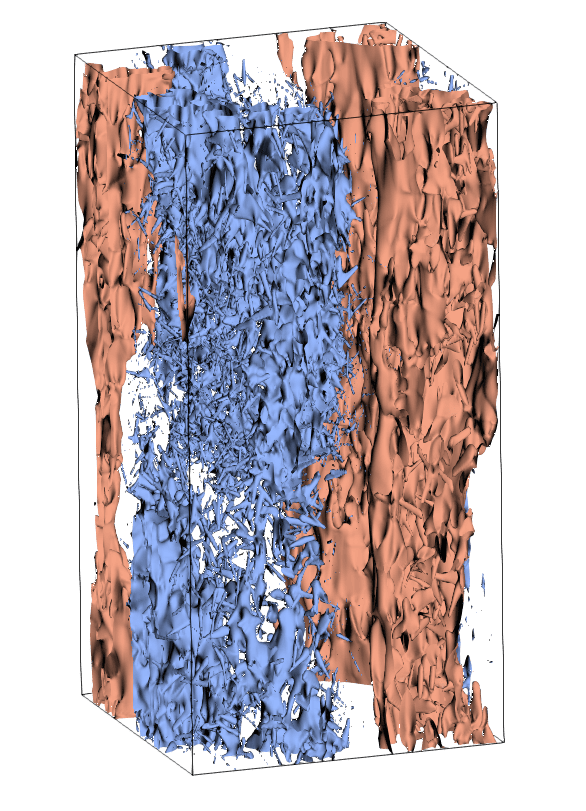}};
    \node at (7,0) {\includegraphics[width=0.32\textwidth]{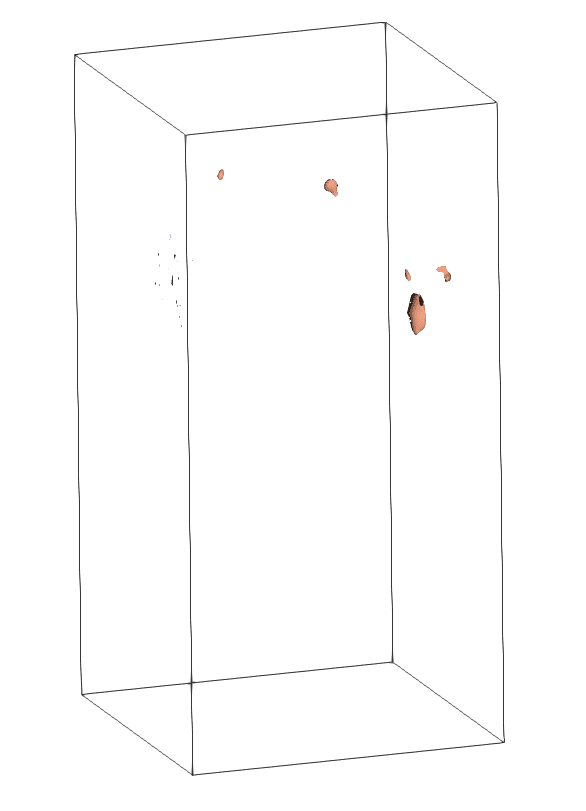}};  
    \node at (-1,-5.7) {\includegraphics[width=0.32\textwidth]{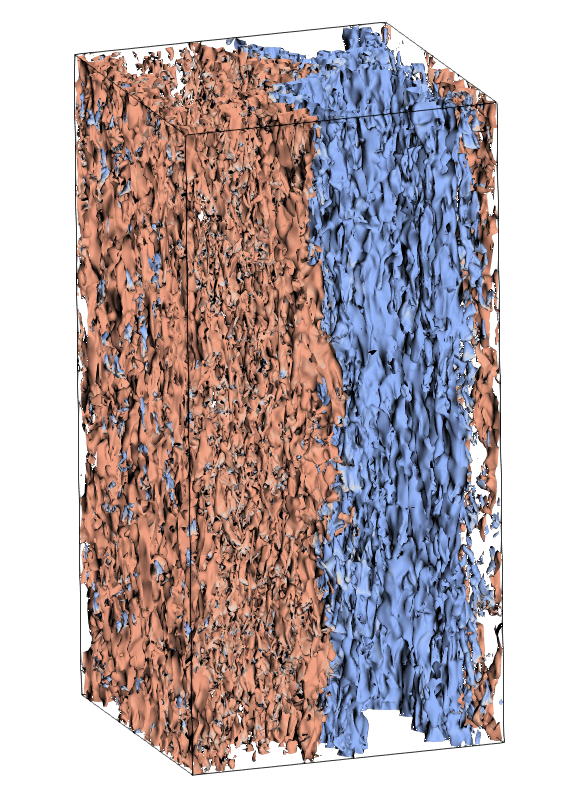}};
    \node at (3,-5.7) {\includegraphics[width=0.32\textwidth]{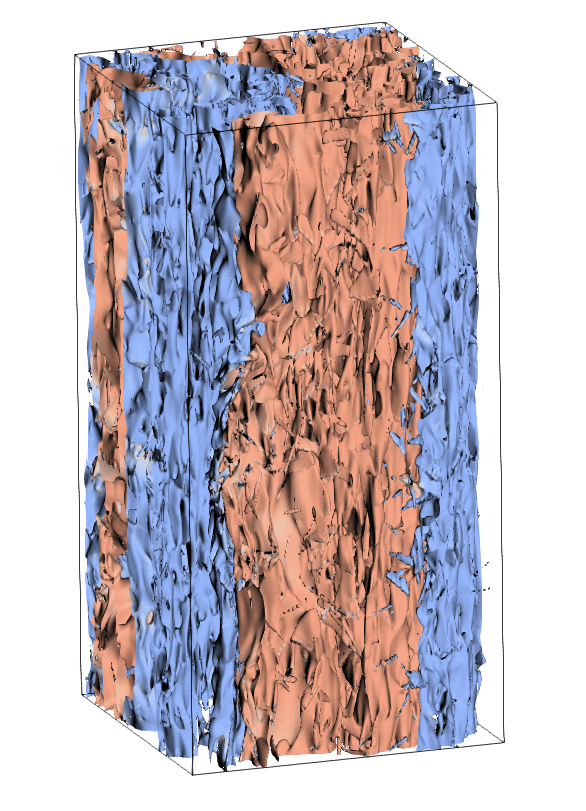}};
    \node at (7,-5.7) {\includegraphics[width=0.32\textwidth]{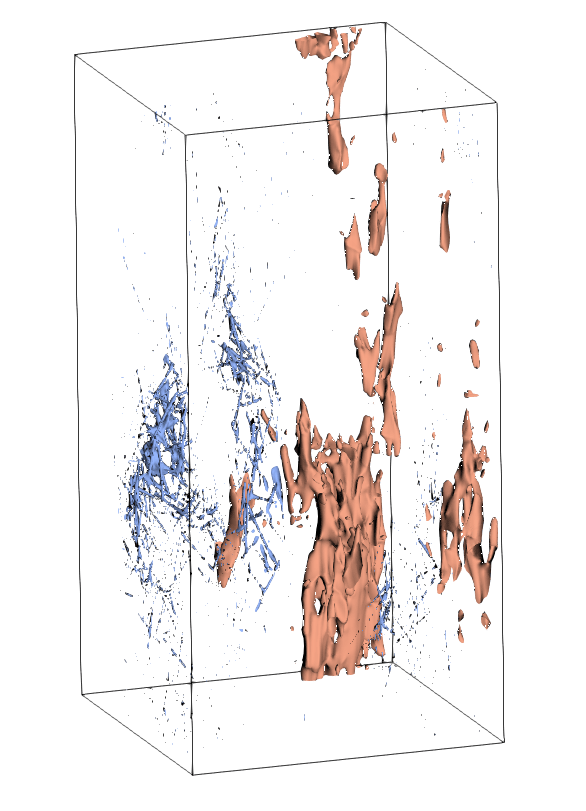}};
    \node at (-3,2.8){$(a)$};
    \node at ( 1,2.8){$(b)$};
    \node at ( 5,2.8){$(c)$};
    \node at (-3,-2.9){$(d)$};
    \node at ( 1,-2.9){$(e)$};
    \node at ( 5,-2.9){$(f)$};
  \end{tikzpicture}
  \caption{Visualisation of the fluctuating velocity field for (left to right) $n \ell_f^3 = 0.36$, 2.89, and 23.15, at (top) $Ga = 180$ and (bottom) $Ga = 450$. Shown are isosurfaces of kinetic energy $E = 0.05 u_g^2$, coloured by the vertical velocity component $w$: red indicates positive $w$, and blue indicates negative $w$}\label{fig:snapEner}
\end{figure}
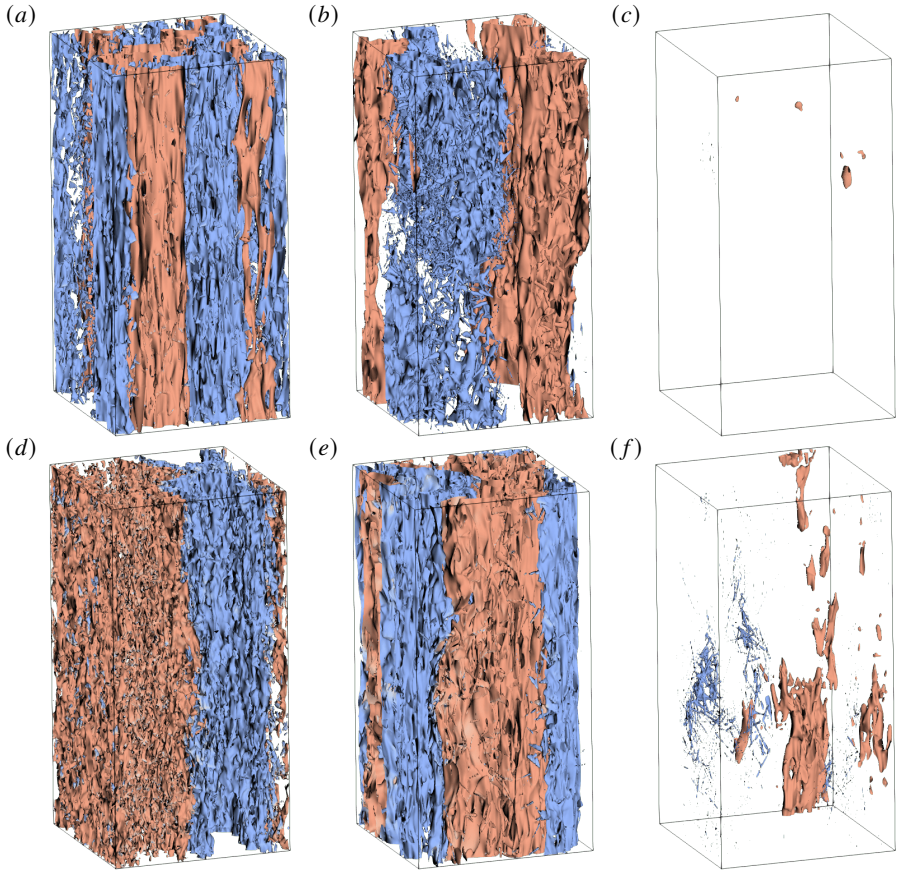

The influence of $Ga$ and $n \ell_f^3$ on the velocity fluctuations is also illustrated in figure \ref{fig:snapEner}, where isosurfaces of the instantaneous fluid kinetic energy are coloured by the vertical velocity $w$. Consistent with the previous discussion, when fixing $Ga$ fibres at lower concentrations generate more intense fluctuations and energise a broader range of scales. This is particularly evident by comparing figures \ref{fig:snapEner}$(d)$ and \ref{fig:snapEner}$(e)$, where smaller-scale structures are more pronounced in the former. 
Conversely, the weaker velocity fluctuations observed at lower $Ga$ manifest as larger void regions in the top panels. Importantly, due to the formation of streamers and the zero mean vertical velocity $\langle w \rangle = 0$, the flow exhibits an alternating pattern of ascending and descending regions.

\subsection{Scale-by-scale energy spectrum}
\label{sec:scale-by-scale}

We now investigate the scale-by-scale energy content of the fluid phase, i.e., the fluid energy spectrum. To properly account for the flow anisotropy, we consider scales/wavenumbers in the $z$ directions, $r_\perp$ and $\kappa_\perp$, and scales/wavenumbers in the $x-y$ plane, $r_\parallel$ and $\kappa_\parallel$, separately. Following \cite{pope-2000}, we define the velocity-spectrum tensor $\Phi_{ij}(\bm{\kappa})$ as
\begin{equation}
  \Phi_{ij}(\bm{\kappa}) =  \aver{ \hat{u}_i(\bm{\kappa},t)^* \hat{u}_j(\bm{\kappa},t)  },
\end{equation}
and introduce 
\begin{equation}
  \mathcal{E}_{ij}(\kappa_\parallel,\kappa_\perp) = \iint\limits_{-\infty}^{+\infty} \Phi_{ij}(\bm{\kappa}) \delta ( |\bm{\kappa}_{2D}|- \kappa_\parallel ) \text{d} \bm{\kappa}_{2D}.
\end{equation}
Here, $\bm{\kappa}_{2D} \equiv (\kappa_x, \kappa_y, 0)$, $\kappa_\parallel \equiv \sqrt{\kappa_x^2 + \kappa_y^2}$, and $\kappa_\perp \equiv \kappa_z$; the operator $\hat{\cdot}$ denotes the Fourier transform of a physical quantity, and the superscript ``$*$'' indicates complex conjugation.
The in-plane ($\mathcal{E}^{\parallel}_{ij}$) and out-of-plane ($\mathcal{E}^{\perp}_{ij}$) velocity-spectrum tensors are thus defined as:
\begin{equation}
  \mathcal{E}^{\parallel}_{ij}(\kappa_\parallel) = \int_{-\infty}^{\infty} \mathcal{E}_{ij}(\kappa_\parallel,\kappa_\perp) \text{d} \kappa_\perp,
\end{equation}
and
\begin{equation}
  \mathcal{E}^{\perp}_{ij}(\kappa_\perp) = \int_{-\infty}^{\infty} \mathcal{E}_{ij}(\kappa_\parallel,\kappa_\perp) \text{d} \kappa_\parallel.
\end{equation}
For notational simplicity, the wavenumber dependence in the definitions of $\mathcal{E}^{\parallel}_{ij}$ and $\mathcal{E}^{\perp}_{ij}$ is omitted hereafter. 
To properly account for flow anisotropy, we define $\mathcal{E}_{1c} \equiv \mathcal{E}_{zz}$ and $\mathcal{E}_{2c} \equiv \mathcal{E}_{xx} + \mathcal{E}_{yy}$. The total energy spectrum is then given by $\mathcal{E}_{3c} = \mathcal{E}_{1c} + \mathcal{E}_{2c}$.

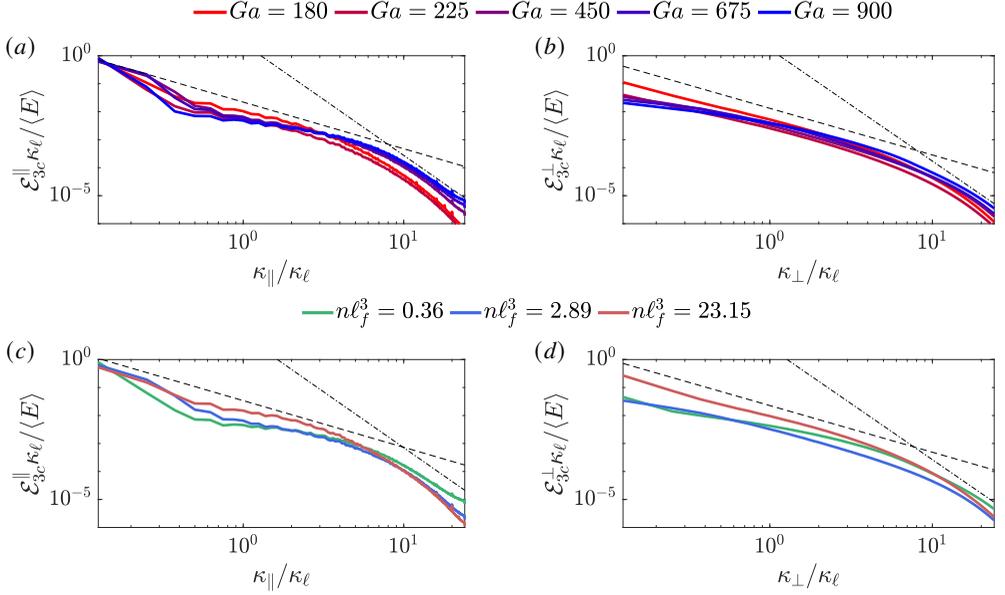
\begin{figure}
  \centering
  \begin{tikzpicture}
    \node at (3.5,2) {\includegraphics[trim={0 14 0 5},clip,width=0.9\textwidth]{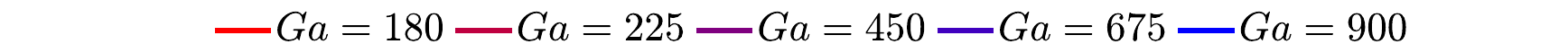}};    
    \node at (0,0) {\includegraphics[width=0.49\textwidth]{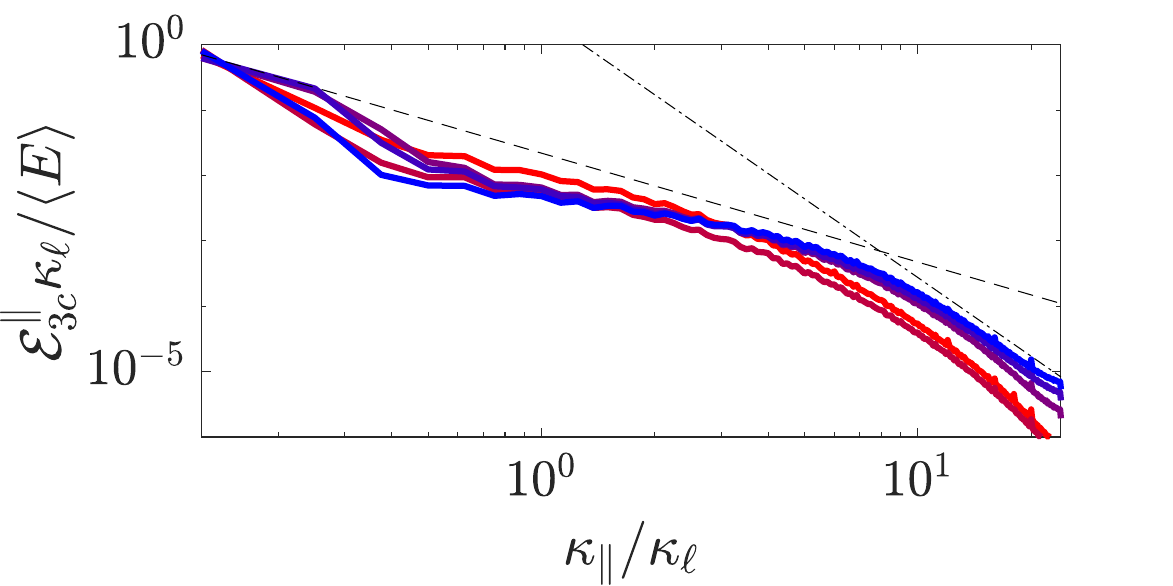}};
    \node at (7,0) {\includegraphics[width=0.49\textwidth]{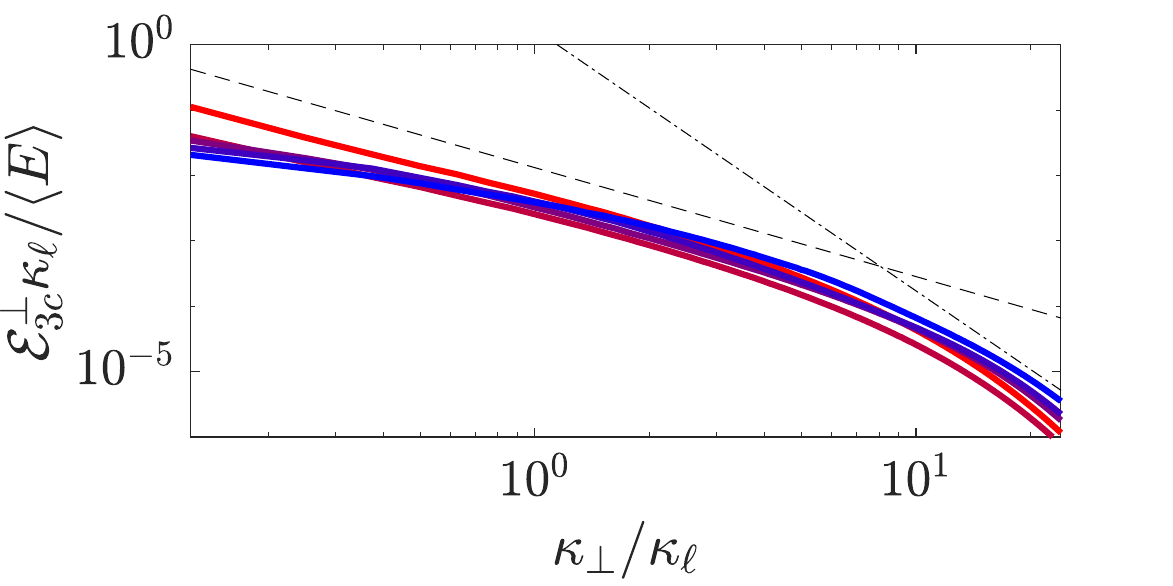}};
 \node at (3.5,-2) {\includegraphics[trim={0 30 0 10},clip,width=0.55\textwidth]{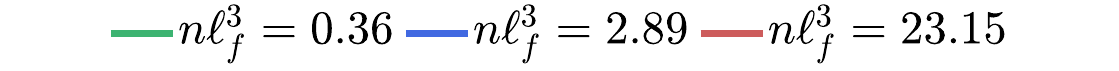}};     
    \node at (0,-4) {\includegraphics[width=0.49\textwidth]{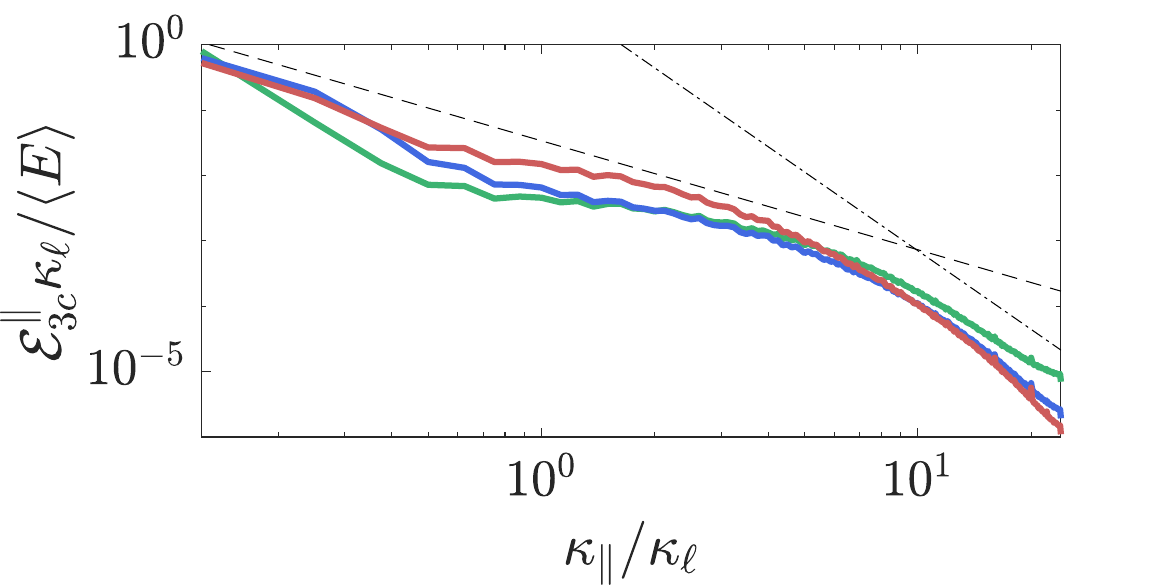}};
    \node at (7,-4) {\includegraphics[width=0.49\textwidth]{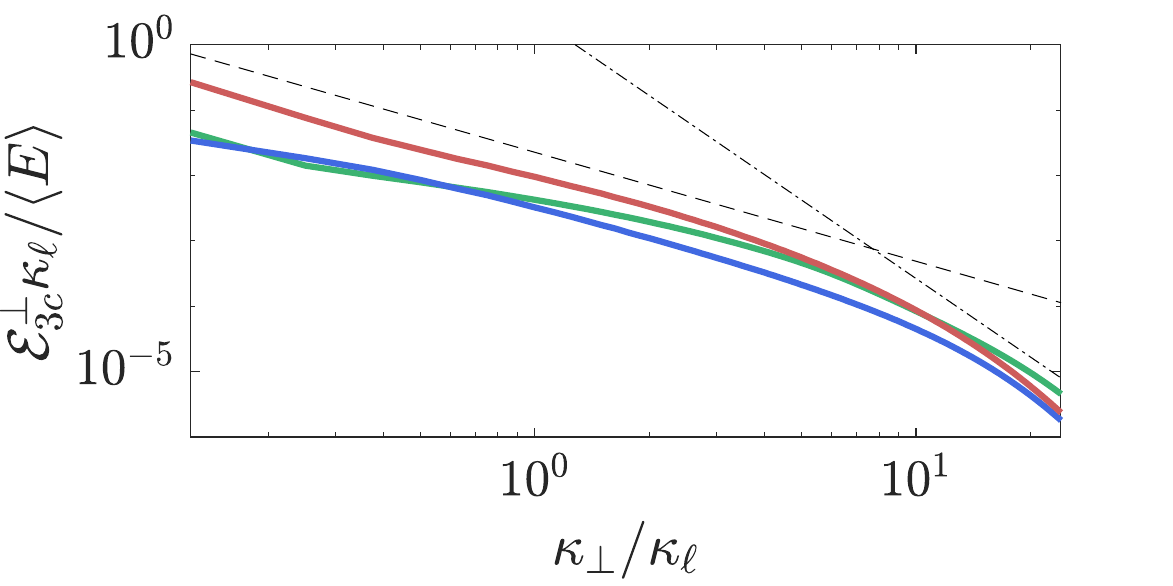}};
    \node at (-3.2,1.5){$(a)$};
    \node at ( 3.8,1.5){$(b)$};
    \node at (-3.2,-2.5){$(c)$};
    \node at ( 3.8,-2.5){$(d)$};    
  \end{tikzpicture} 
  \caption{Top panels: energy spectra (a) $\mathcal{E}_{3c}^\parallel$ and (b) $\mathcal{E}_{3c}^\perp$ as functions of $Ga$ for $n \ell_f^3 = 2.89$. Bottom panels: (c) $\mathcal{E}_{3c}^\parallel$ and (d) $\mathcal{E}_{3c}^\perp$ as functions of $n \ell_f^3$ for $Ga = 450$. The black dashed and dash-dotted lines indicate $\kappa^{-5/3}$ and $\kappa^{-4}$ scaling, respectively. All spectra are normalised such that $\int_0^\infty \mathcal{E}_{3c}^{\parallel,\perp} / \langle E \rangle \mathrm{d} \kappa_{\parallel,\perp} = 1$. The $\kappa_{\parallel,\perp} = 0$ mode is excluded due to the logarithmic horizontal axis; $\mathcal{E}_{3c}^\perp(0) $ is larger in cases exhibiting streamer formation, owing to their strong vertical coherence.} \label{fig:spec3c_Ga}
\end{figure}  
Figure \ref{fig:spec3c_Ga} presents the dependence of the total energy spectra, $\mathcal{E}_{3c}^\parallel$ (a,c) and $\mathcal{E}_{3c}^\perp$ (b,d), on $Ga$ (a,b) and fibre concentration $n \ell_f^3$ (c,d). The spectra are normalised by the total kinetic energy, such that $\int_0^\infty \mathcal{E}_{3c}^{\parallel,\perp}/\langle E \rangle \text{d}\kappa_{\parallel,\perp} = 1$; this allows us to examine the relative scale-by-scale distribution of energy across cases.
All cases exhibit broadband spectra, indicating that fibre settling induces chaotic motion and excites a wide range of spatial scales. Notably, the spectra do not collapse when the wavenumbers are rescaled using the Kolmogorov length scale $\langle \eta \rangle$.

Two main effects are observed as $Ga$ increases:
(i) a stronger separation of scales, reflected in a shift of energy toward higher wavenumbers, and
(ii) a redistribution of energy, with enhanced small-scale content and depleted large-scale content.
These trends are especially apparent in $\mathcal{E}_{3c}^\parallel$ (a), where the suppression of energy for $\kappa \lessapprox \kappa_\ell$ and the enhancement for $\kappa \gtrapprox \kappa_\ell$ is more pronounced; here, $\kappa_\ell = 2\pi/\ell_f$ denotes the wavenumber associated with the fibre length.
This energy redistribution is consistent with the spectral shortcut mechanism \citep{brandt-coletti-2022}, previously observed for spherical particles \citep{tenCate-etal-2004} and rigid/flexible fibres in turbulence \citep{olivieri-mazzino-rosti-2022,cannon-olivieri-rosti-2024}. The fibres affect the fluid in two main ways: (i) their settling induces a local momentum deficit, generating velocity fluctuations across a broad range of scales, and (ii) they disrupt large-scale structures ($\kappa \lessapprox \kappa_\ell$), transferring energy to smaller scales ($\kappa \gtrapprox \kappa_\ell$). Both effects intensify with increasing $Ga$, in line with the rise in total energy (figure \ref{fig:bulk}) and the enhanced small-scale energy content.
At larger scales, the spectra approach a Kolmogorov-like $\kappa^{-5/3}$ scaling, particularly in $\mathcal{E}_{3c}^\perp$ and at lower $Ga$, where large-scale energy depletion is less pronounced. At higher wavenumbers, the spectra steepen, indicating a dissipative range.
Varying the concentration $n \ell_f^3$ produces qualitatively similar effects: decreasing $n \ell_f^3$ leads to stronger scale separation, reduced energy at large scales, and enhanced energy at small scales. These observations are consistent with earlier results in figures \ref{fig:bulk} and \ref{fig:snapEner}.

\begin{figure}
  \centering
  \begin{tikzpicture}
    \node at (0,0) {\includegraphics[width=0.49\textwidth]{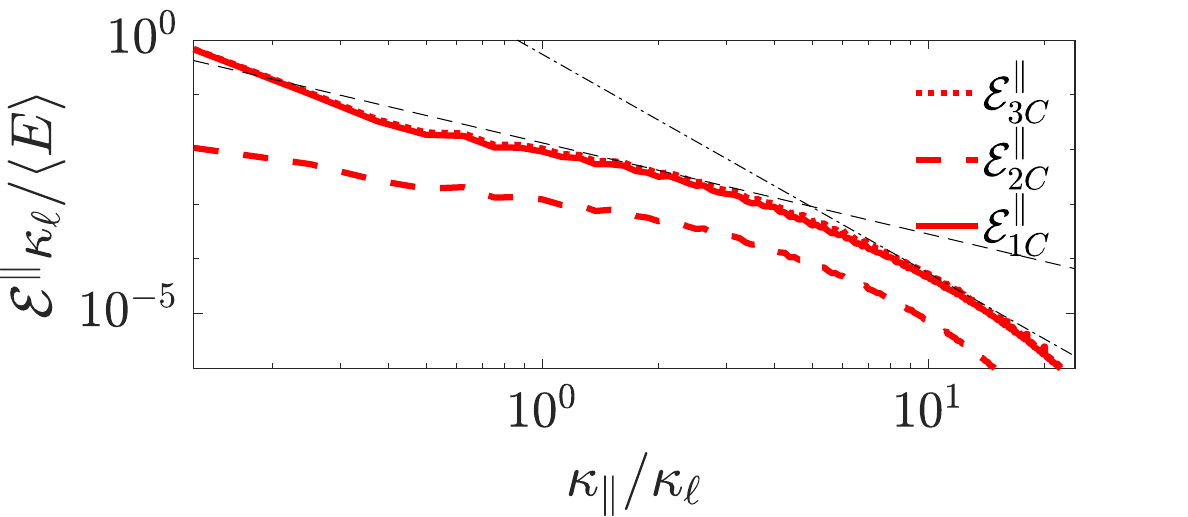}};
    \node at (7,0) {\includegraphics[width=0.49\textwidth]{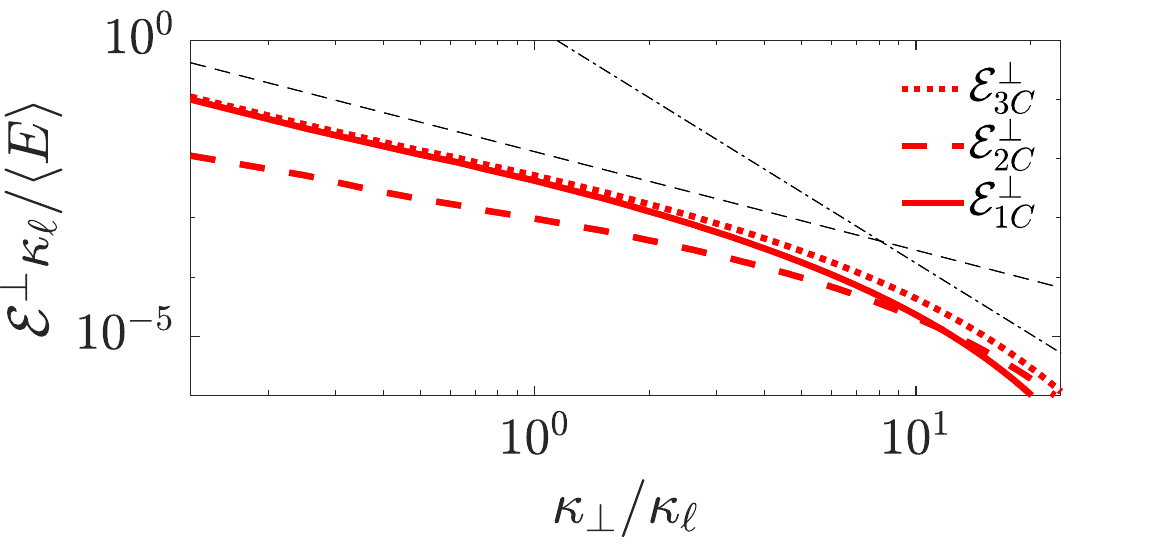}};
    \node at (0,-3.2) {\includegraphics[width=0.49\textwidth]{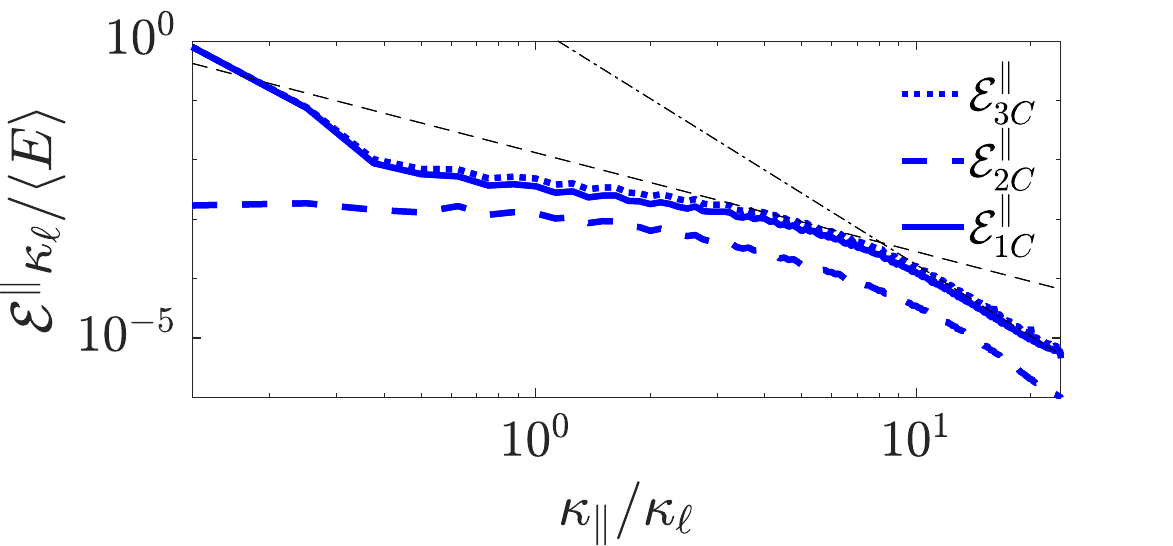}};
    \node at (7,-3.2) {\includegraphics[width=0.49\textwidth]{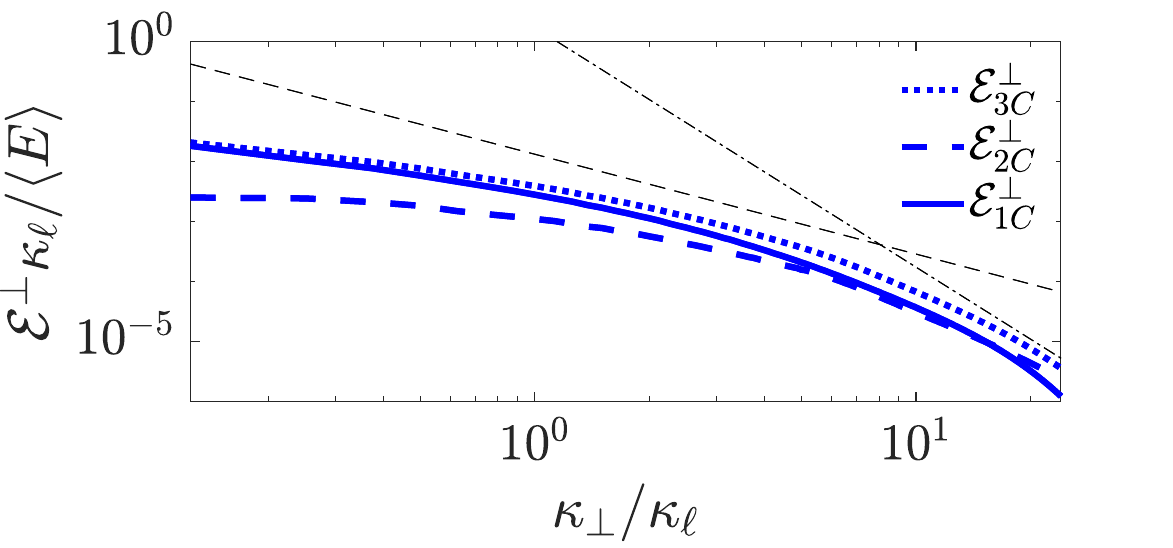}};
    \node at (-3.2,1.5){$(a)$};
    \node at ( 3.8,1.5){$(b)$};
    \node at (-3.2,-1.7){$(c)$};
    \node at ( 3.8,-1.7){$(d)$};    
  \end{tikzpicture}   
  \caption{Anisotropy of the velocity fluctuations in the $\kappa_\parallel$ ($a$ and $c$) and $\kappa_\perp$ ($b$ and $d$) wavenumber space for $n \ell_f^3 = 2.89$ and $Ga=180$ ($a$ and $b$) and $Ga=900$ ($c$ and $d$). The coloured solid lines are for $\mathcal{E}_{1C}$, the coloured dashed lines for $\mathcal{E}_{2C}$ and the coloured dotted lines for $\mathcal{E}_{3C}$. The black dashed line indicates $\kappa^{-5/3}$, while the black dashed-dotted one indicates $\kappa^{-4}$.}
  \label{fig:spec1c2c}
\end{figure}
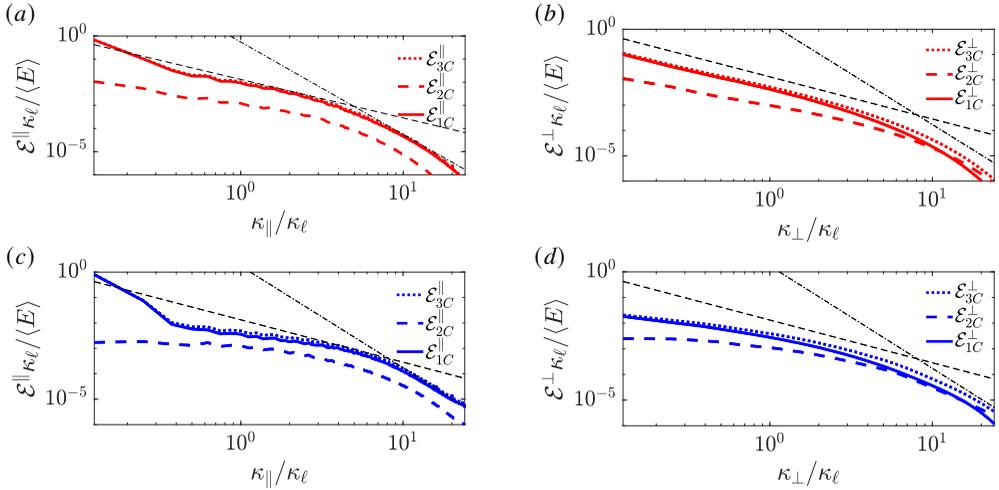  

Motivated by the observed anisotropy, we analyse the separate contributions of $\mathcal{E}_{1c}$ and $\mathcal{E}_{2c}$ to the energy spectrum in figure \ref{fig:spec1c2c}. The spectra are normalised by the mean total energy, $\langle E \rangle$, and shown for $n \ell_f^3 = 2.89$ at $Ga = \{180, 900\}$; similar trends are observed across other concentrations.
Figure \ref{fig:spec1c2c} confirms that the total energy spectrum $\mathcal{E}_{3c}$ is predominantly governed by $\mathcal{E}_{1c}$, both exhibit similar spectral decay in $\kappa_\parallel$ and $\kappa_\perp$. This reflects the stronger vertical velocity fluctuations compared to the in-plane components.
The degree of anisotropy varies across scales. At low wavenumbers, $\mathcal{E}_{2c}^\parallel$ flattens, suggesting weak correlations in large-scale horizontal motions. This behaviour is consistent with the formation of vertically aligned fibre clusters, which primarily enhance vertical $w$ fluctuations at low $\kappa_\parallel$ (see figure \ref{fig:Cw_Gaeffect}). The vertical fluctuations remain dominant at all horizontal scales, as $\mathcal{E}_{1c}^\parallel \gg \mathcal{E}_{2c}^\parallel$ across all $\kappa_\parallel$. However, at high $\kappa_\perp$, isotropy is progressively recovered, with $\mathcal{E}_{1c}^\perp \lessapprox \mathcal{E}_{2c}^\perp$ at the smallest vertical scales. 
Overall, $\mathcal{E}_{1c}^{\parallel,\perp}$ exceeds $\mathcal{E}_{2c}^{\parallel,\perp}$ at all scales, except at large $\kappa_\perp$, where $\mathcal{E}_{1c}^\perp$ becomes comparable to, or slightly smaller than, $\mathcal{E}_{2c}^\perp$. The relative contribution of $\mathcal{E}_{2c}^{\parallel,\perp}$ increases with $Ga$, reflecting the emergence of a more chaotic, three-dimensional flow (see figure \ref{fig:snapEner}). 
To quantify this anisotropy, table \ref{tab:spec} reports the ratios $\mathcal{E}_{1c}^\parallel / \mathcal{E}_{2c}^\parallel$ and $\mathcal{E}_{1c}^\perp / \mathcal{E}_{2c}^\perp$ at representative values of $\kappa_\parallel$ and $\kappa_\perp$.

\begin{table}
\centering
\begin{tabular}{cccccccccccc}
                                        & & & \multicolumn{3}{c}{$Ga=180$} & & &      \multicolumn{3}{c}{$Ga=900$} \\                     
  $\kappa_{\parallel,\perp}/\kappa_\ell$&  & & $\mathcal{E}_{1c}^\parallel/\mathcal{E}_{2c}^\parallel$ & & $\mathcal{E}_{1c}^\perp/\mathcal{E}_{2c}^\perp$ & & & $\mathcal{E}_{1c}^\parallel/\mathcal{E}_{2c}^\parallel$ & & $\mathcal{E}_{1c}^\perp/\mathcal{E}_{2c}^\perp$  \vspace{0.2cm}\\
  $0.25$                                & & & $19.01$ & & $6.25$      & & & $40.01$ & &  $4.33$ \\
  $3.625$                               & & & $ 5.64$ & & $2.20$      & & & $3.86 $ & & $1.57$ \\
  $22.375$                              & & & $14.89$ & &  $0.44$     & & & $5.31 $ & &  $0.60$ \\
\end{tabular}
\caption{Ratios of $\mathcal{E}_{1c}^\parallel / \mathcal{E}_{2c}^\parallel$  and $\mathcal{E}_{1c}^\perp / \mathcal{E}_{2c}^\perp $ at selected wavenumbers $\kappa_\parallel$ and $\kappa_\perp$ for $n \ell_f^3 = 2.89$, comparing $Ga = 180$ (left) and $ Ga = 900$ (right).
}
\label{tab:spec}
\end{table}

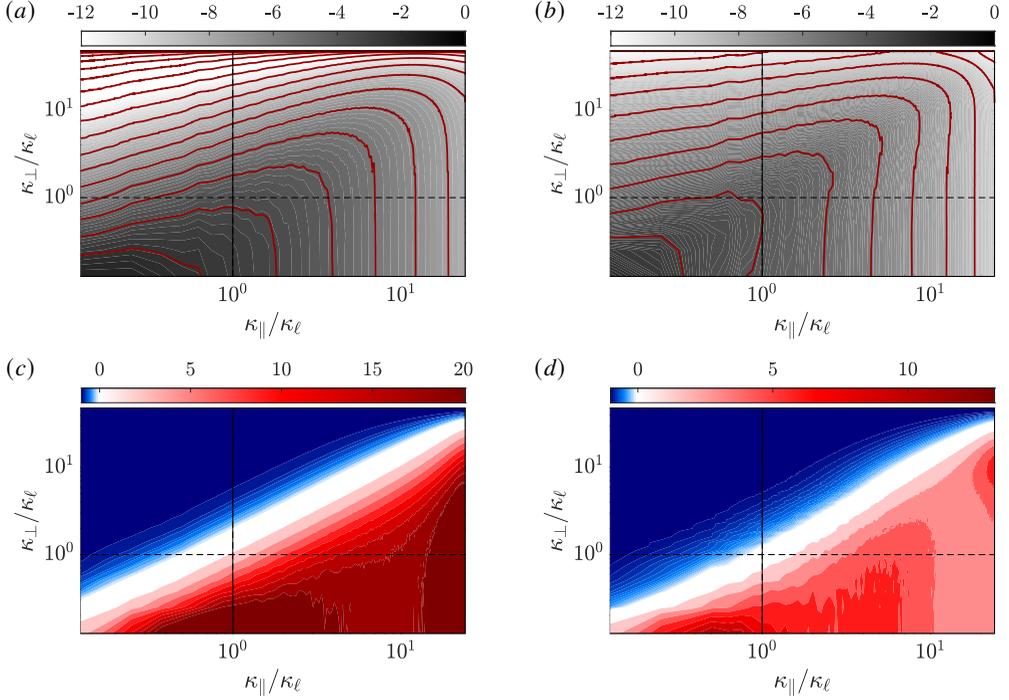
\begin{figure}
  \centering
  \begin{tikzpicture}
    \node at (0,0) {\includegraphics[width=0.49\textwidth]{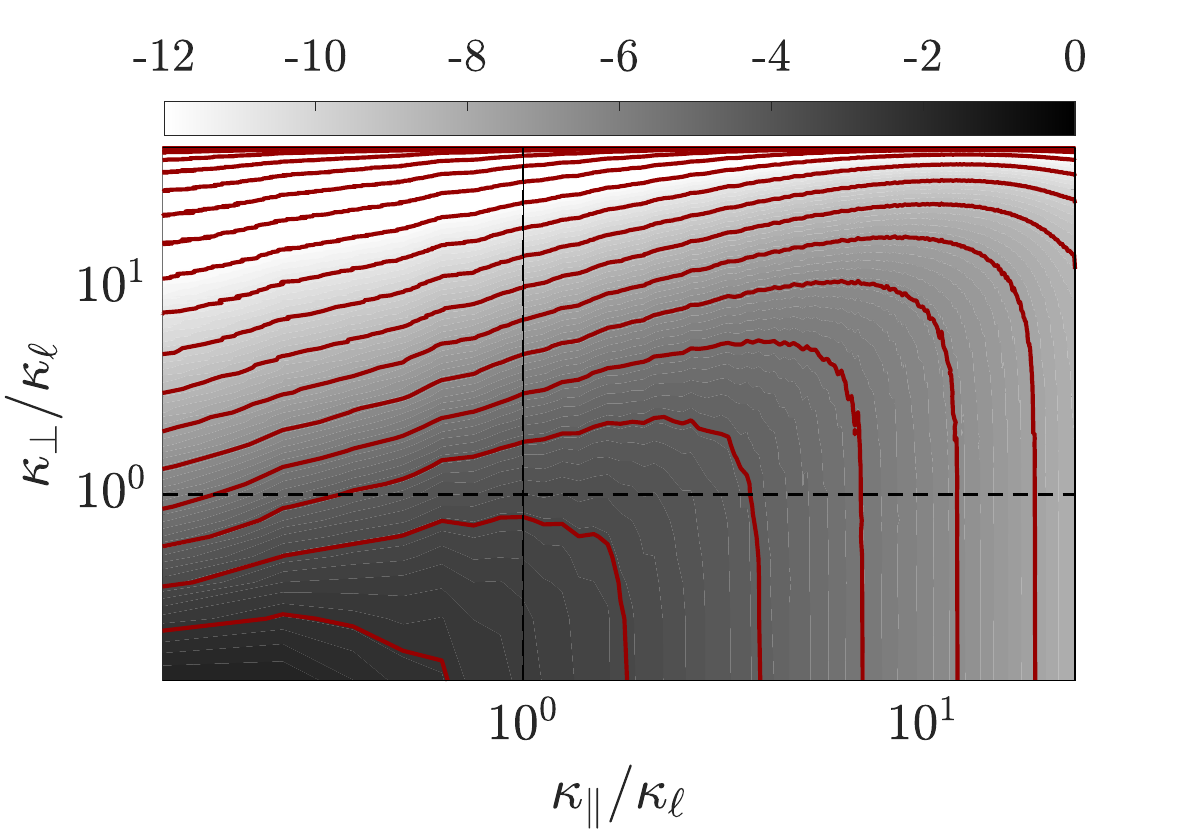}};
    \node at (7,0) {\includegraphics[width=0.49\textwidth]{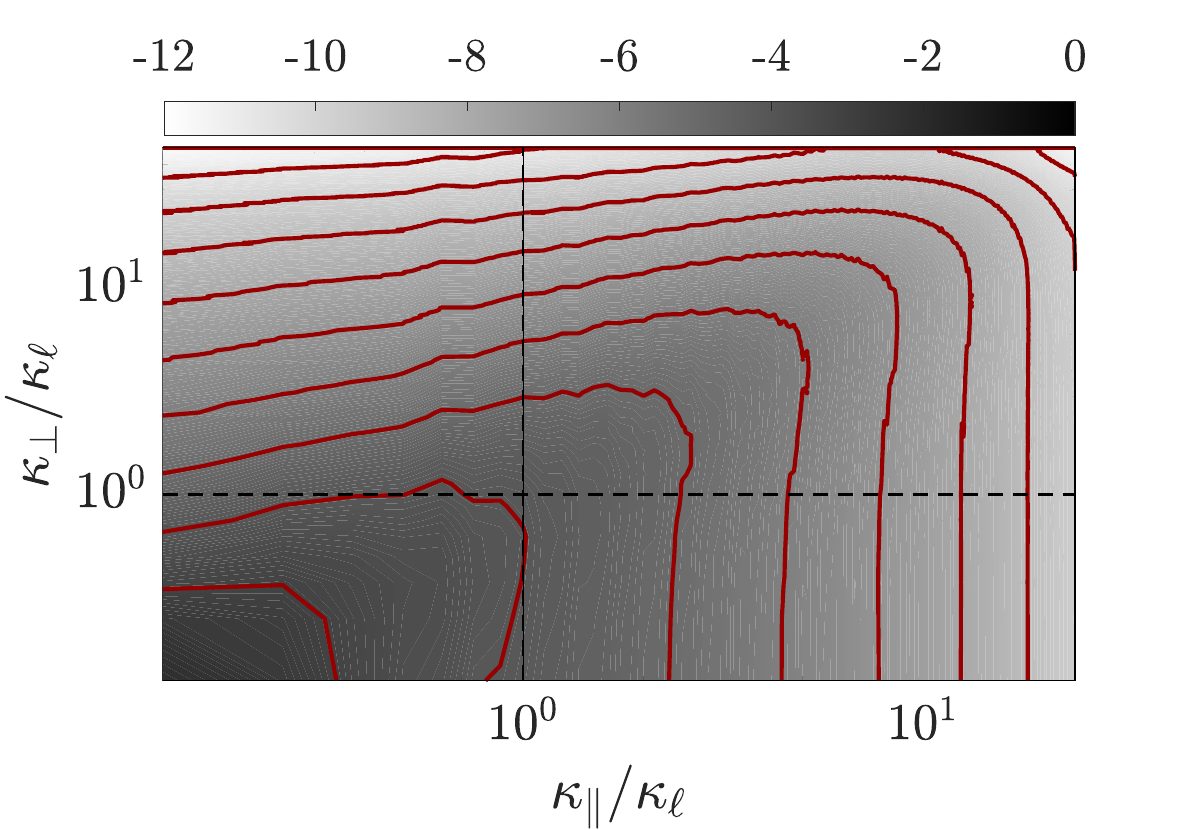}};
    \node at (0,-4.7) {\includegraphics[width=0.49\textwidth]{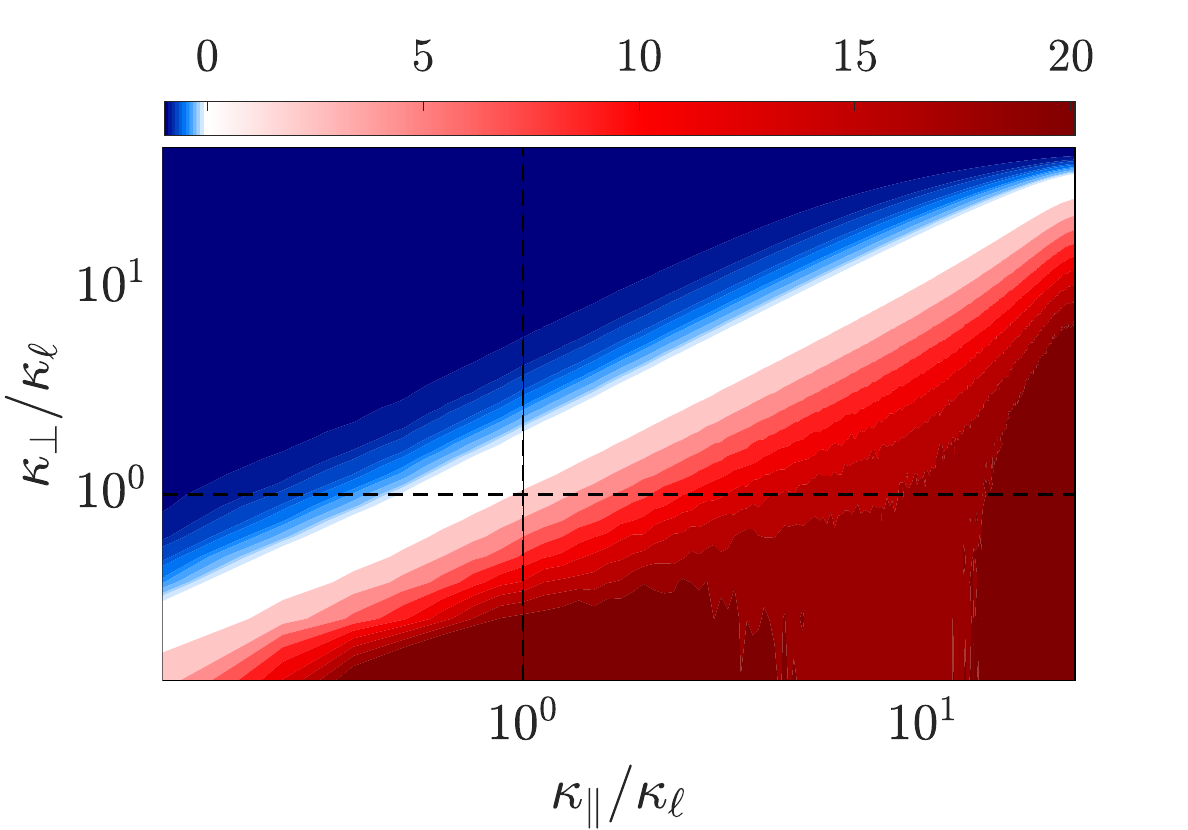}};
    \node at (7,-4.7) {\includegraphics[width=0.49\textwidth]{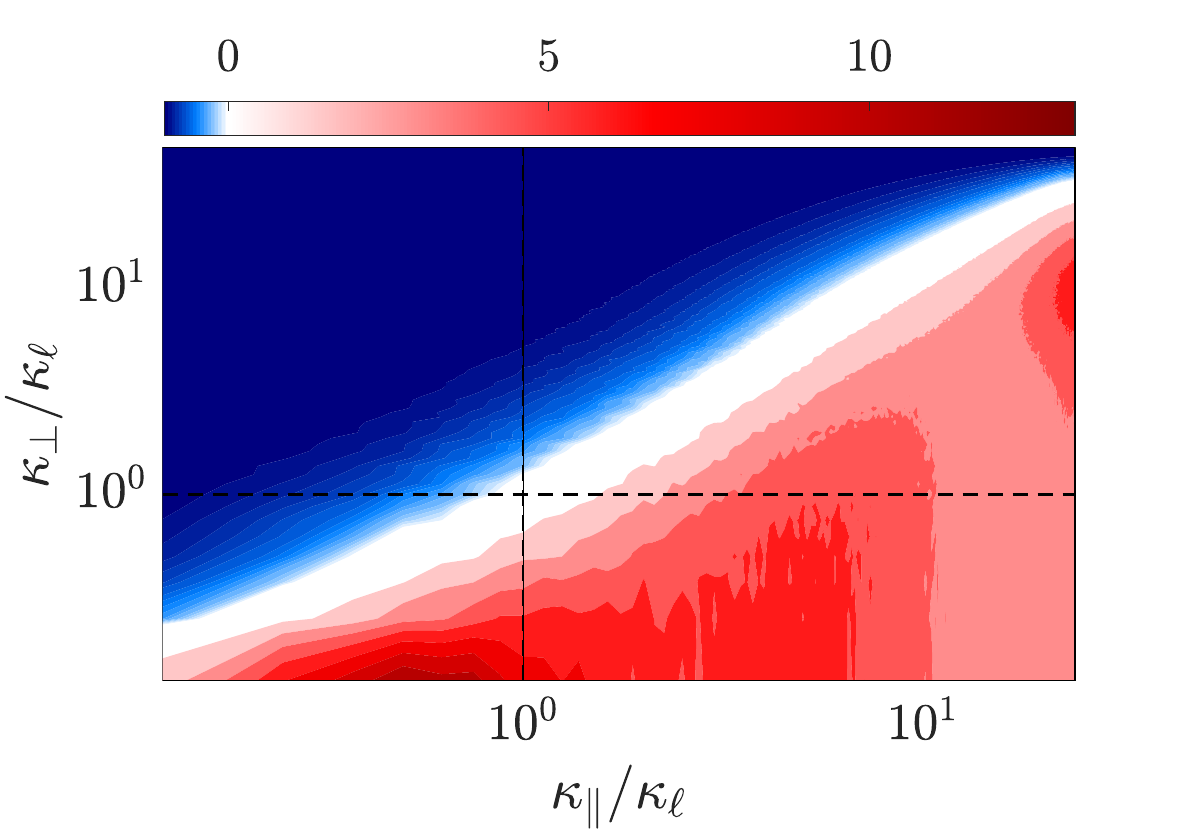}};  
    \node at (-3.2,2){$(a)$};
    \node at ( 3.8,2){$(b)$};
    \node at (-3.2,-2.7){$(c)$};
    \node at ( 3.8,-2.7){$(d)$};    
  \end{tikzpicture}  
  \caption{Panels (a) and (b): Two-dimensional spectral maps of (a) $\mathcal{E}_{1c}(\kappa_\parallel,\kappa_\perp)$ and (b) $\mathcal{E}_{2c}(\kappa_\parallel,\kappa_\perp)$ for $Ga = 180$. The white-to-black colormap represents $\log(\mathcal{E}{\cdot c}/E)$. Panels (c) and (d): Relative difference maps of $\mathcal{E}_{1c}/\mathcal{E}_{2c} - 1$ in $(\kappa_\parallel,\kappa_\perp)$ space, for $n \ell_f^3 = 2.89$, at (c) $Ga = 180$ and (d) $Ga = 900$. The dashed black lines indicate the fibre wavenumber $\kappa_\ell = 2\pi/\ell_f$.}
  \label{fig:Spec2D}
\end{figure}  

Figure \ref{fig:Spec2D} presents the two-dimensional spectra $\mathcal{E}_{1c}$ and $\mathcal{E}_{2c}$ in the $(\kappa_\parallel, \kappa_\perp)$ wavenumber space, for $n \ell_f^3 = 2.89$ and $Ga = \{180, 900\}$. Consistent with previous observations, figure \ref{fig:Spec2D}(a,b) show that most of the energy is concentrated at small $|\bm{\kappa}|$. Due to the anisotropic nature of the flow, the spectral decay is markedly steeper along the $\kappa_\perp$ direction than along $\kappa_\parallel$, reflecting the elongated and vertically coherent structure of the streamers (see figure \ref{fig:snapEner}). As a result, energetic motions are primarily aligned along the vertical direction and span a broader range of horizontal scales.
At higher wavenumbers, the contours of both $\mathcal{E}_{1c}$ and $\mathcal{E}_{2c}$ tend to align along $\kappa_\perp \approx \kappa_\parallel$, suggesting a progressive recovery of isotropy at small scales. 
The flow anisotropy is further illustrated in figure \ref{fig:Spec2D}(c,d), which shows maps of the energy ratio $\mathcal{E}_{1c}/\mathcal{E}_{2c}$. Red regions indicate $\mathcal{E}_{1c} > \mathcal{E}_{2c}$, while blue indicates the opposite. The $(\kappa_\parallel, \kappa_\perp)$ plane is effectively partitioned: for $\kappa_\parallel \gtrapprox \kappa_\perp$, vertical fluctuations dominate ($\mathcal{E}_{1c} > \mathcal{E}_{2c}$), whereas for $\kappa_\perp \gtrapprox \kappa_\parallel$, in-plane fluctuations prevail. This implies that flow structures elongated in the vertical direction are associated with stronger $w$ fluctuations, while structures broader in the horizontal plane are dominated by $u$ and $v$ components.
Finally, as $Ga$ increases, the magnitude of the $\mathcal{E}_{1c}/\mathcal{E}_{2c}$ ratio decreases, indicating a reduction in anisotropy as the system becomes more chaotic and fluctuation intensity increases.

 \subsection{Scale-by-scale energy budget}
\label{sec:bud}

To uncover the mechanism responsible for the organisation of the fluctuations, we examine the scale-by-scale energy budget. To account for the flow anisotropy, we again separate the in-plane and out-of-plane contributions, and formulate distinct budget equations for $\mathcal{E}_{1c}$ and $\mathcal{E}_{2c}$. The budget for the total energy $\mathcal{E}_{3c}$ is then obtained by summing the two contributions.
 
The scale-by-scale budget equations for $\mathcal{E}_{1c}$ and $\mathcal{E}_{2c}$ are obtained by manipulating the Fourier-transformed form of the Navier--Stokes equations and, in a compact form, they read
\begin{equation}
\underbrace{\mathcal{N}_{ic,\parallel}(\kappa_\parallel,\kappa_\perp) +
            \mathcal{N}_{ic,\perp}(\kappa_\parallel,\kappa_\perp) }_{\mathcal{N}_{ic}} =
\mathcal{D}_{ic}(\kappa_\parallel,\kappa_\perp) + \mathcal{P}_{ic}(\kappa_\parallel,\kappa_\perp) +
 \mathcal{F}_{ic}(\kappa_\parallel,\kappa_\perp) + \mathcal{G}_{ic}(\kappa_\parallel,\kappa_\perp),
 \label{eq:budEi}
\end{equation}
where $i=1,2$. For a detailed derivation of the energy equation in the case of a homogeneous and isotropic flow, we refer the reader to \cite{pope-2000}. At the left-hand side of equation \eqref{eq:budEi}, the terms
\begin{equation}
  \mathcal{N}_{1c,\parallel} = \left \langle \widehat{\frac{\partial uw}{\partial x}}\hat{w}^* + \widehat{\frac{\partial vw}{\partial y}}\hat{w}^* + \widehat{\frac{\partial uw}{\partial x}}^*\hat{w} + \widehat{\frac{\partial vw}{\partial y}}^*\hat{w}  \right \rangle
\end{equation}
and
\begin{equation}
  \mathcal{N}_{2c,\parallel} = \left \langle \widehat{\frac{\partial uu}{\partial x}}\hat{u}^* + \widehat{\frac{\partial uv}{\partial y}}\hat{u}^* + \widehat{\frac{\partial uv}{\partial x}}\hat{v}^* + \widehat{\frac{\partial vv}{\partial y}}\hat{v}^* + \widehat{\frac{\partial uu}{\partial x}}^*\hat{u} + \widehat{\frac{\partial uv}{\partial y}}^*\hat{u} + \widehat{\frac{\partial uv}{\partial x}}^*\hat{v}^* + \widehat{\frac{\partial vv}{\partial y}}^*\hat{v} \right \rangle,
\end{equation}
denote the in-plane ($x-y$) scale-by-scale transfers of $\mathcal{E}_{1c}$ and $\mathcal{E}_{2c}$, while 
\begin{equation}
  \mathcal{N}_{1c,\perp} = \left \langle \widehat{\frac{\partial ww}{\partial z}}\hat{w}^* + \widehat{\frac{\partial ww}{\partial z}}^* \hat{w} \right \rangle,
\end{equation}
and
\begin{equation}
  \mathcal{N}_{2c,\perp} = \left \langle \widehat{\frac{\partial uw}{\partial z}}\hat{u}^* + \widehat{\frac{\partial vw}{\partial z}}\hat{v}^* + \widehat{\frac{\partial uw}{\partial z}}^*\hat{u} + \widehat{\frac{\partial vw}{\partial z}}^*\hat{v} \right \rangle,
\end{equation}
are the out-of-plane ($z$) scale-by-scale transfers. At the right-hand side,
\begin{equation}
  \mathcal{D}_{1c} = -\frac{1}{Ga} (\kappa_\perp^2 + \kappa_\parallel^2 ) \mathcal{E}_{1c} \ \text{and} \ 
  \mathcal{D}_{2c} = -\frac{1}{Ga} (\kappa_\perp^2 + \kappa_\parallel^2 ) \mathcal{E}_{2c}
\end{equation}
denote the scale-by-scale viscous dissipation, while 
\begin{equation}
  \mathcal{F}_{1c} =  \left \langle \hat{f}_{fs,z} \hat{w}^* + \hat{f}_{fs,z}^* \hat{w} \right \rangle \ \text{and} \
  \mathcal{F}_{2c} = \left \langle \hat{f}_{fs,x} \hat{u}^* + \hat{f}_{fs,y} \hat{v}^* + \hat{f}_{fs,x}^* \hat{u} + \hat{f}_{fs,y}^* \hat{v} \right \rangle
\end{equation}
are the scale-by-scale fluid-solid coupling terms, and
\begin{equation}
  \mathcal{G}_{1c} = \left \langle \hat{g}_z \hat{w}^* + \hat{g}_z^* \hat{w} \right \rangle \ \text{and} \
  \mathcal{G}_{2c} = \left \langle \hat{g}_x \hat{u}^* + \hat{g}_y \hat{v}^* + \hat{g}_x^* \hat{u} + \hat{g}_y^* \hat{v} \right \rangle
\end{equation}
are the gravity contributions to the budgets. When looking separately at the equations for $\mathcal{E}_{1c}$ and $\mathcal{E}_{2c}$, an additional term arises at the right-hand side, i.e., the pressure strain term $\mathcal{P}_{ic}$, which is not present in the budget for $\mathcal{E}_{3c}$ because of the incompressibility constraint. The pressure strain details how energy is redistributed among the different (in-plane and out-of-plane) contributions at each ($\kappa_\parallel,\kappa_\perp$) scale and reads
\begin{equation}
  \mathcal{P}_{1c} = - \left \langle \widehat{ \frac{\partial p}{\partial z} } \hat{w}^* + \widehat{ \frac{\partial p}{\partial z} }^* \hat{w} \right \rangle \ \text{and} \ \mathcal{P}_{2c} = - \left \langle \widehat{ \frac{\partial p}{\partial x} } \hat{u}^* + \widehat{\frac{\partial p}{\partial y} } \hat{v}^* +\widehat{ \frac{\partial p}{\partial x} }^* \hat{u} + \widehat{\frac{\partial p}{\partial y} }^* \hat{v} \right \rangle.
\end{equation}
Equation \eqref{eq:budEi} describes the balance at each $(\kappa_\parallel,\kappa_\perp)$ scale between the different mechanisms at play: the energy source/sink due to the fluid-solid interaction $\mathcal{F}$, the viscous dissipation $\mathcal{D}$, the transport $\mathcal{N}$, and the redistribution $\mathcal{P}$. Similarly to what has been done with the energy spectra, we first look at the budgets integrated along the $\kappa_\perp$ and $\kappa_\parallel$ wavenumbers (whose terms are indicated using the $``\perp"$ and $``\parallel"$ superscripts),
and then look at the complete two-dimensional $(\kappa_\parallel,\kappa_\perp)$ space. 

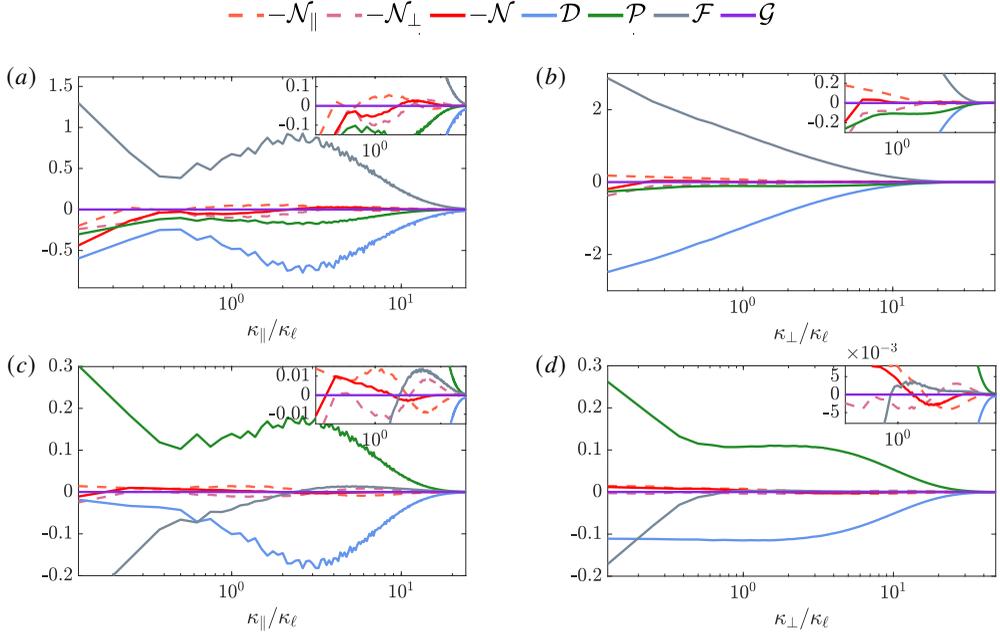
\begin{figure}
  \centering  
  \begin{tikzpicture}
    \node at (3,2.5) {\includegraphics[trim={0 7 0 0},clip,width=0.7\textwidth]{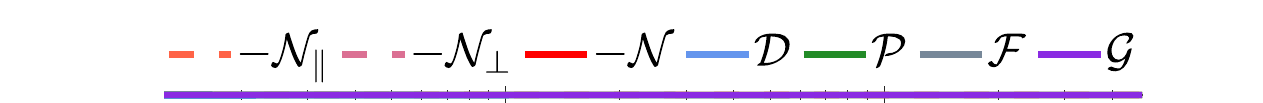}};
    \node at (0,0) {\includegraphics[width=0.49\textwidth]{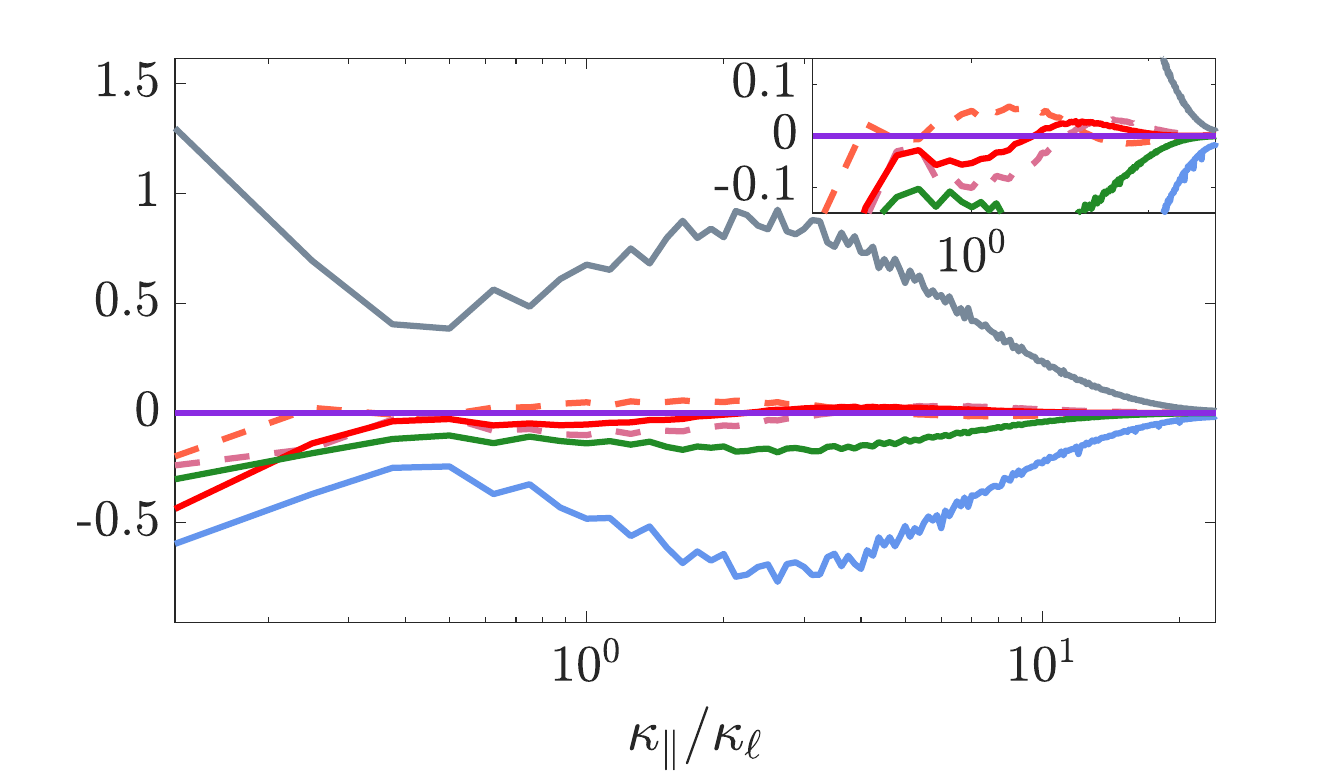}};
    \node at (7,0) {\includegraphics[width=0.49\textwidth]{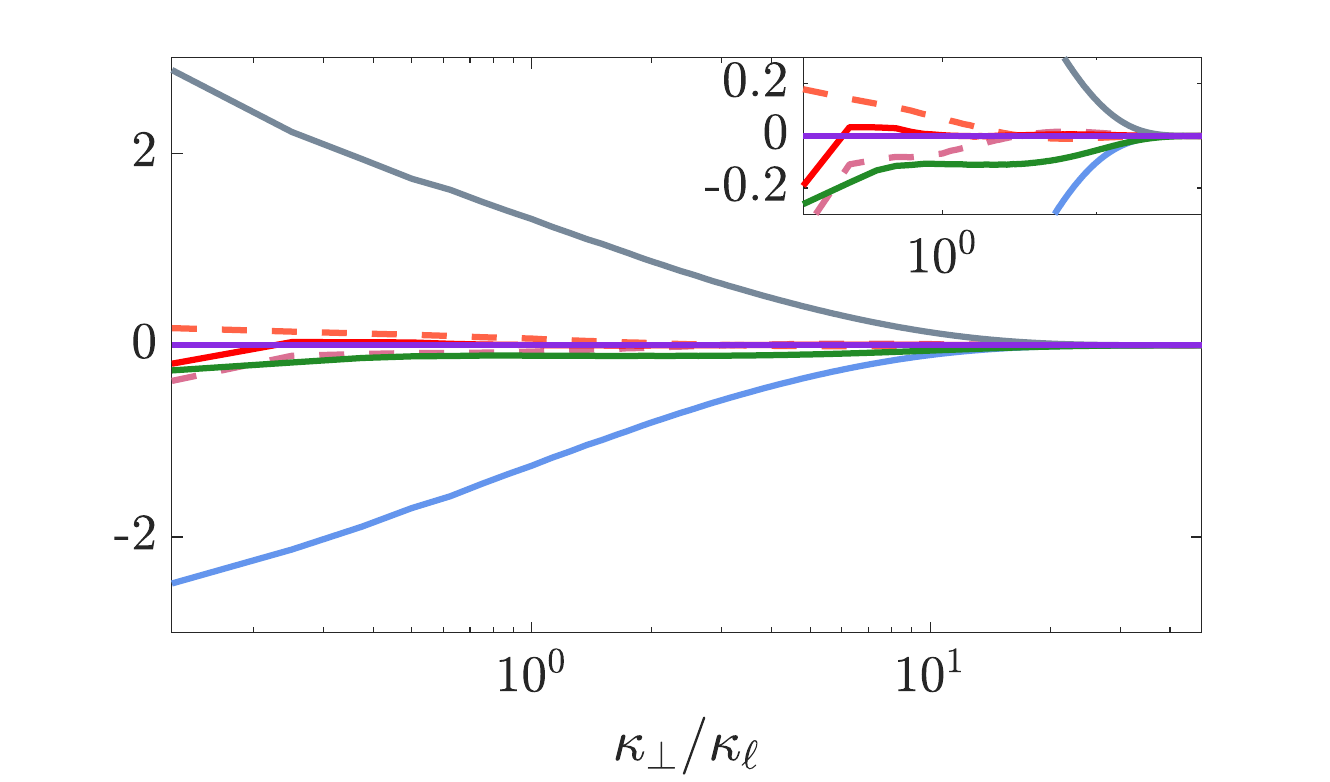}};
    \node at (0,-3.8) {\includegraphics[width=0.49\textwidth]{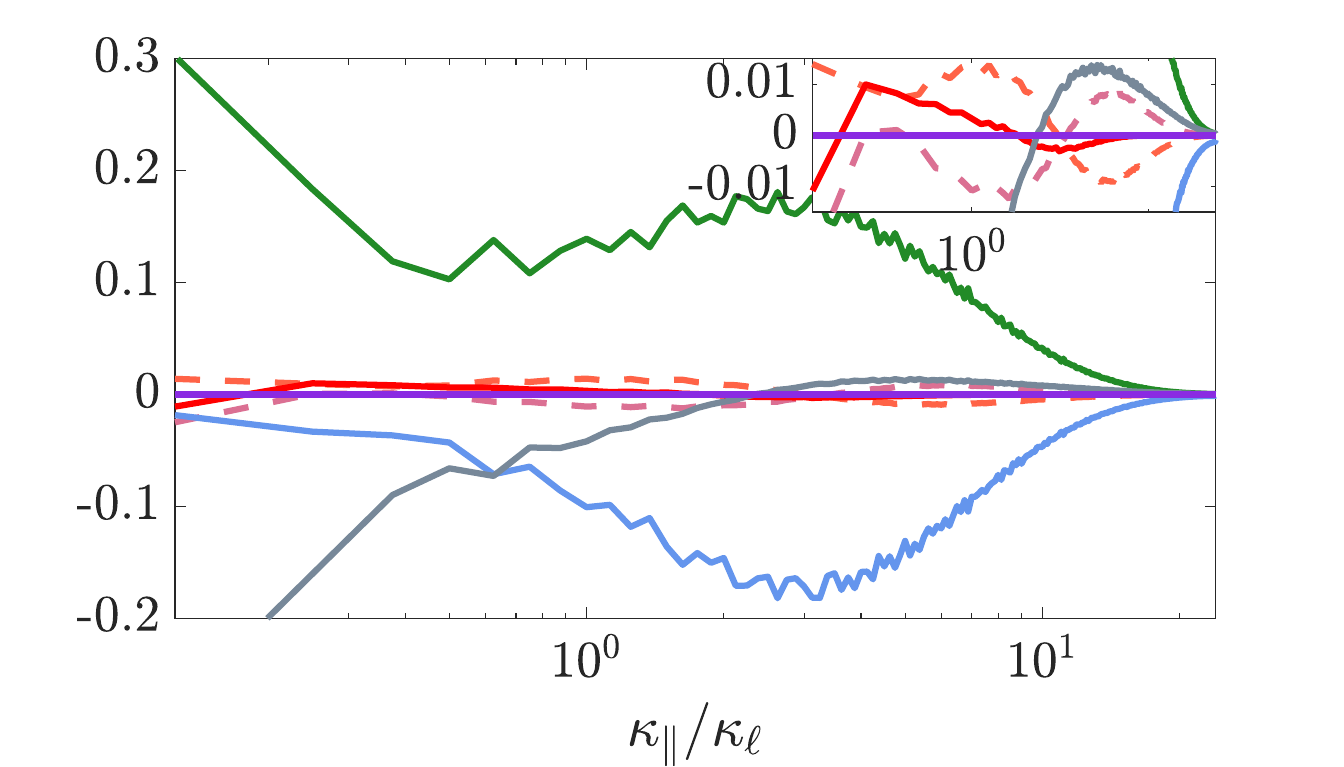}};
    \node at (7,-3.8) {\includegraphics[width=0.49\textwidth]{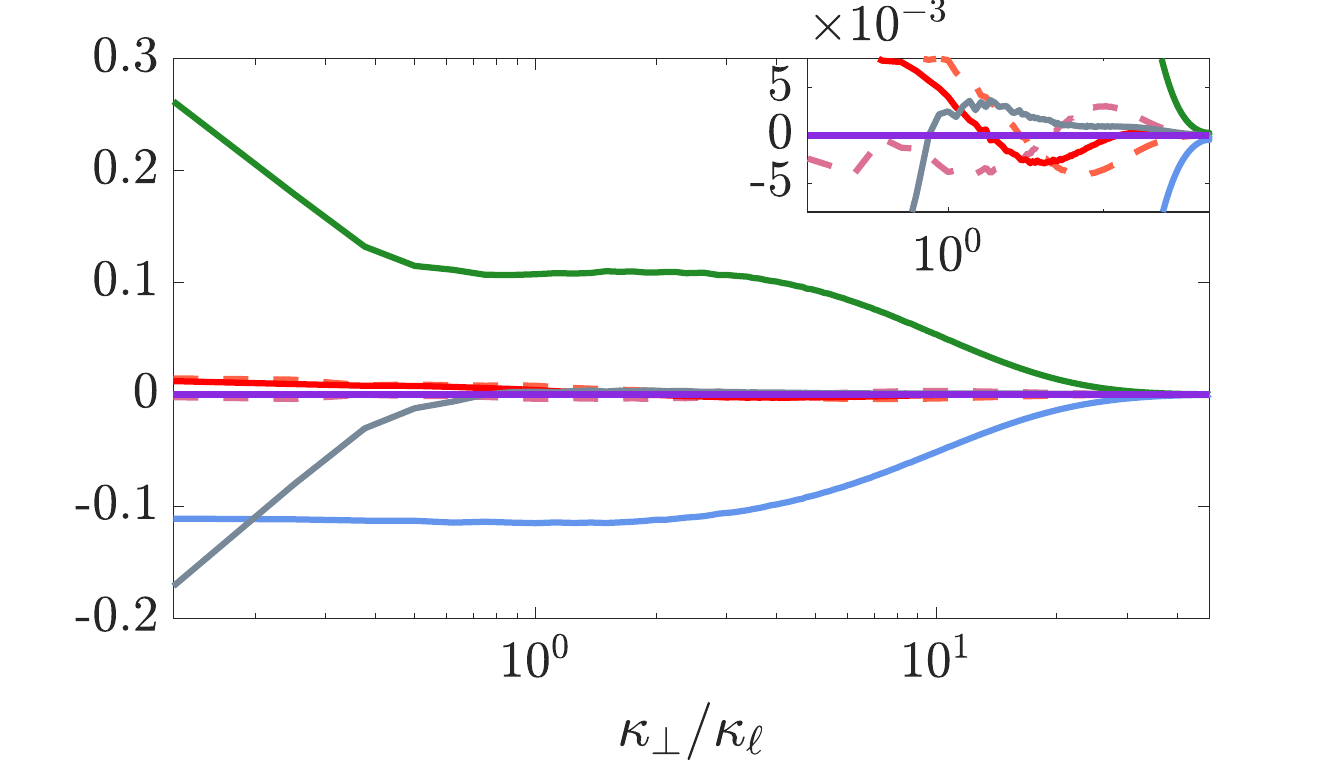}};     
    \node at (-3.2,1.6){$(a)$};
    \node at ( 3.8,1.6){$(b)$};
    \node at (-3.2,-2.2){$(c)$};
    \node at ( 3.8,-2.2){$(d)$};     
  \end{tikzpicture}    
  \caption{Scale-by-scale energy transfer balance for $n\ell_f^3 = 2.89$ and $Ga=180$. Top left: budget for $\mathcal{E}_{1c}^\parallel$. Top right: budget for $\mathcal{E}_{1c}^\perp$. Bottom left: $\mathcal{E}_{2c}^\parallel$. Bottom right: budget for $\mathcal{E}_{2c}^\perp$. All quantities are scaled with the averaged dissipation in the box $\langle \varepsilon \rangle$.}
  \label{fig:bud_Ga180}  
\end{figure}
\begin{figure}
  \centering
  \begin{tikzpicture}
    \node at (3,2.5) {\includegraphics[trim={0 7 0 0},clip,width=0.7\textwidth]{./fig/Budget/legend_bud-eps-converted-to.pdf}};  
    \node at (0,0) {\includegraphics[width=0.49\textwidth]{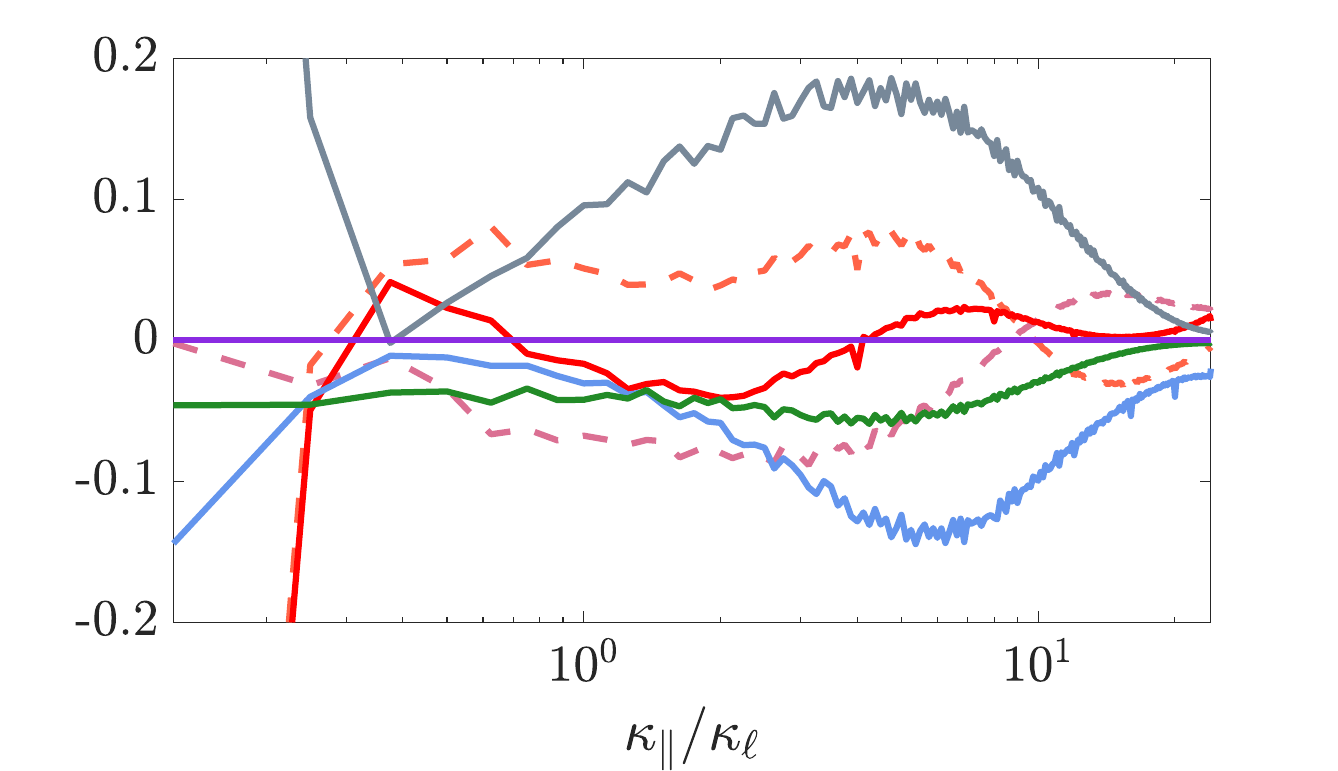}};
    \node at (7,0) {\includegraphics[width=0.49\textwidth]{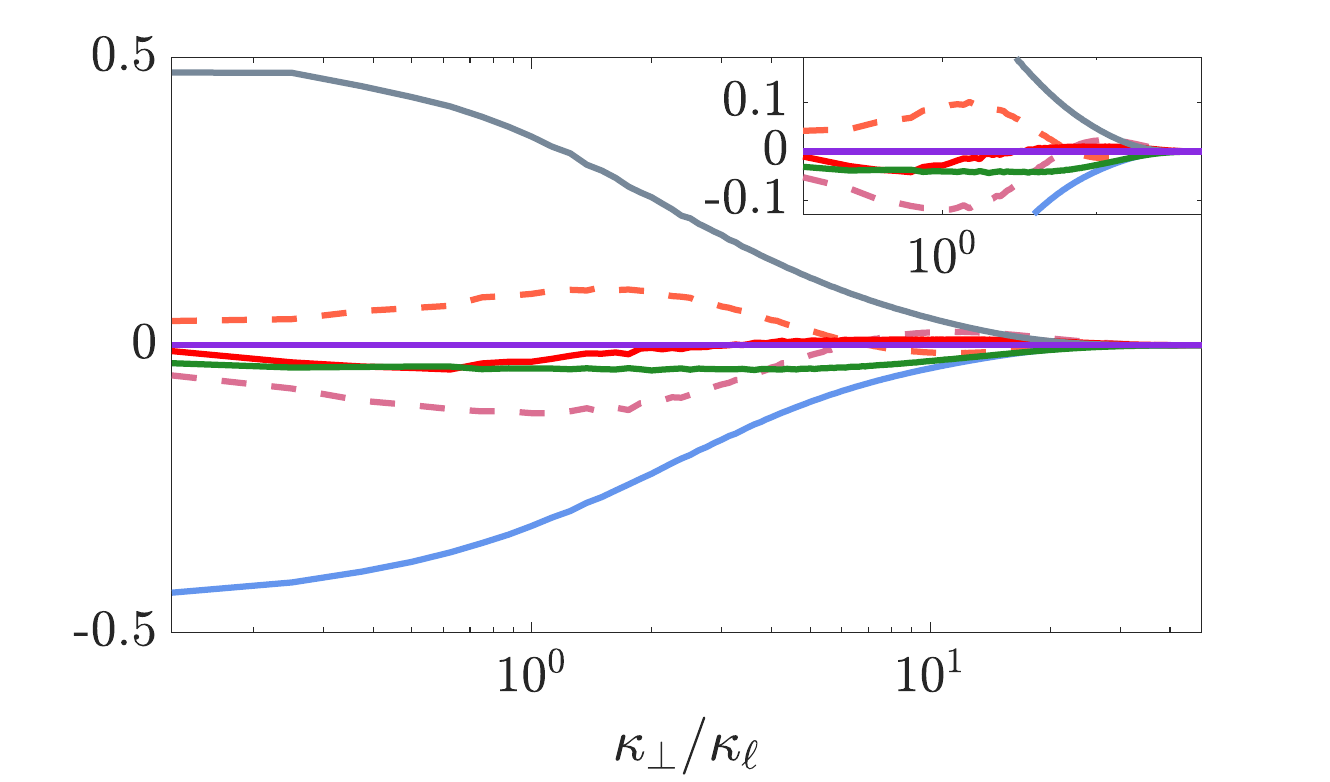}};
    \node at (0,-3.8) {\includegraphics[width=0.49\textwidth]{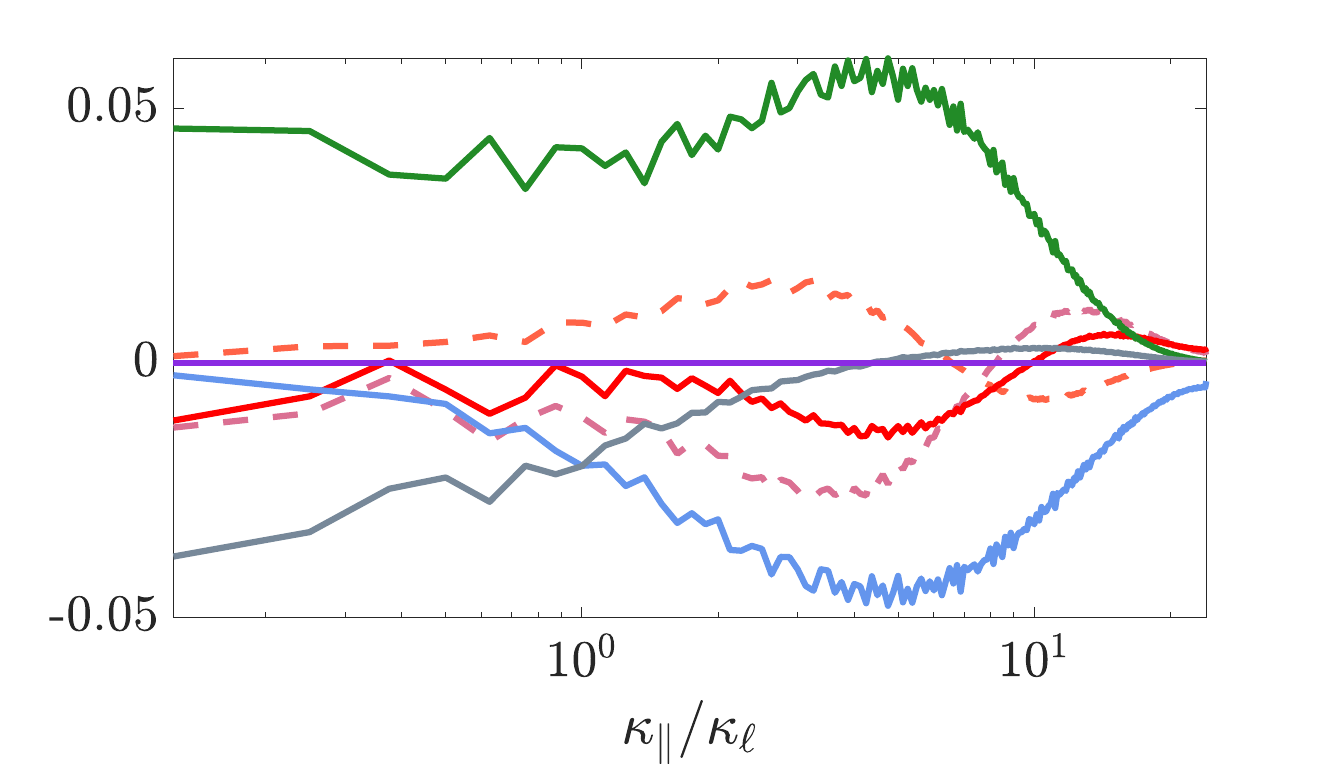}};
    \node at (7,-3.8) {\includegraphics[width=0.49\textwidth]{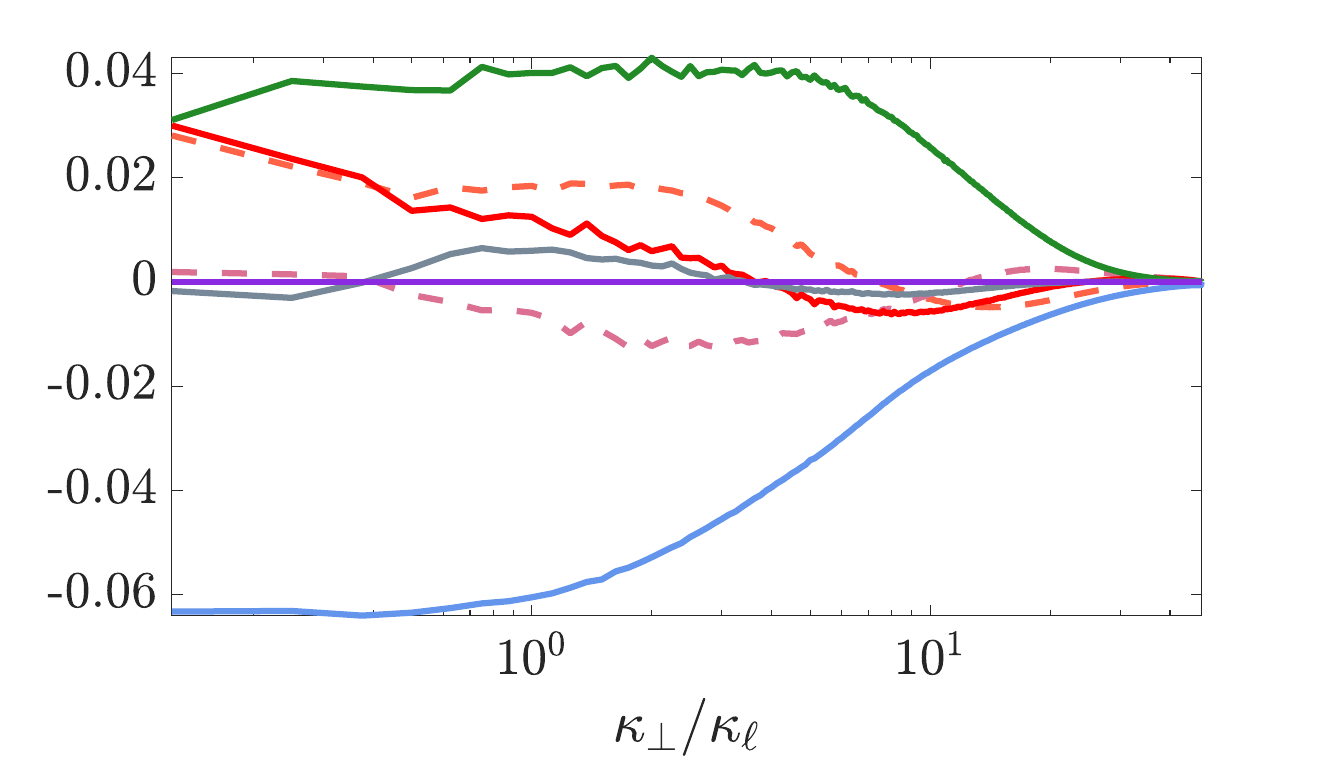}};     
    \node at (-3.2,1.6){$(a)$};
    \node at ( 3.8,1.6){$(b)$};
    \node at (-3.2,-2.2){$(c)$};
    \node at ( 3.8,-2.2){$(d)$};     
  \end{tikzpicture}    
  \caption{Scale-by-scale energy transfer balance for $n\ell_f^3 = 2.89$ and $Ga=900$. Top left: budget for $\mathcal{E}_{1c}^\parallel$. Top right: budget for $\mathcal{E}_{1c}^\perp$. Bottom left: $\mathcal{E}_{2c}^\parallel$. Bottom right: budget for $\mathcal{E}_{2c}^\perp$. All quantities are scaled with the averaged dissipation in the box $\langle \varepsilon \rangle$.}
  \label{fig:bud_Ga900}  
\end{figure}
Figures \ref{fig:bud_Ga180} and \ref{fig:bud_Ga900} present the budget terms for $n \ell_f^3 = 2.89$ at $Ga = \{180, 900\}$, representative of low and high fluctuation intensities. All quantities are normalised by the mean fluid dissipation. At fixed $Ga$, decreasing $n \ell_f^3$ enhances the velocity fluctuations (see figure \ref{fig:snapEner}), and the resulting energy budgets qualitatively mirror those obtained by increasing $Ga$.

We begin with the $\mathcal{E}_{1c}$ budget (vertical velocity fluctuations). As fibres settle under gravity, they convert potential energy into kinetic energy and transfer part of it to the surrounding fluid, thereby sustaining $w$ fluctuations across all scales. This is reflected in the positive fluid-solid coupling term $\mathcal{F}_{1c}^{\parallel,\perp}$, which exceeds the viscous dissipation across the scales, i.e., $|\mathcal{F}_{1c}^{\parallel,\perp}| > |\mathcal{D}_{1c}^{\parallel,\perp}|$. A portion of the injected energy is redistributed to the in-plane components via the pressure-strain term $\mathcal{P}_{1c}^{\parallel,\perp} < 0$, which promotes isotropisation. Thus, while vertical fluctuations are directly sustained by the fluid-solid interaction, the in-plane fluctuations are primarily maintained by pressure redistribution, a mechanism consistent with recent observations for spheroidal particles \citep{jiang-etal-2025}.

The $\mathcal{E}_{2c}$ budget (panels c and d) shows a distinct behaviour. Here, the fluid-solid coupling term $\mathcal{F}_{2c}^{\parallel,\perp}$ extracts energy at large scales ($\kappa_\parallel, \kappa_\perp < \kappa_p$) and injects it at smaller ones ($\kappa_\parallel, \kappa_\perp > \kappa_p$), where $\kappa_p \sim \kappa_\ell$ increases with $Ga$ and/or $n \ell_f^3$. This scale-dependent transfer, observed along both $\kappa_\perp$ and $\kappa_\parallel$, reflects the fibres' ability to locally deform the fluid, draining energy from the energy-containing scales and enhancing small-scale motions. As previously noted, pressure-strain ($\mathcal{P}_{2c}^{\parallel,\perp} > 0$) acts as the main energy source for in-plane fluctuations, balanced by $\mathcal{F}_{2c}^{\parallel,\perp} < 0$ at large scales ($\kappa_\parallel, \kappa_\perp < \kappa_p$), and by viscous dissipation $\mathcal{D}_{2c}^{\parallel,\perp} < 0$ at smaller scales ($\kappa_\parallel, \kappa_\perp \gtrapprox \kappa_p$).

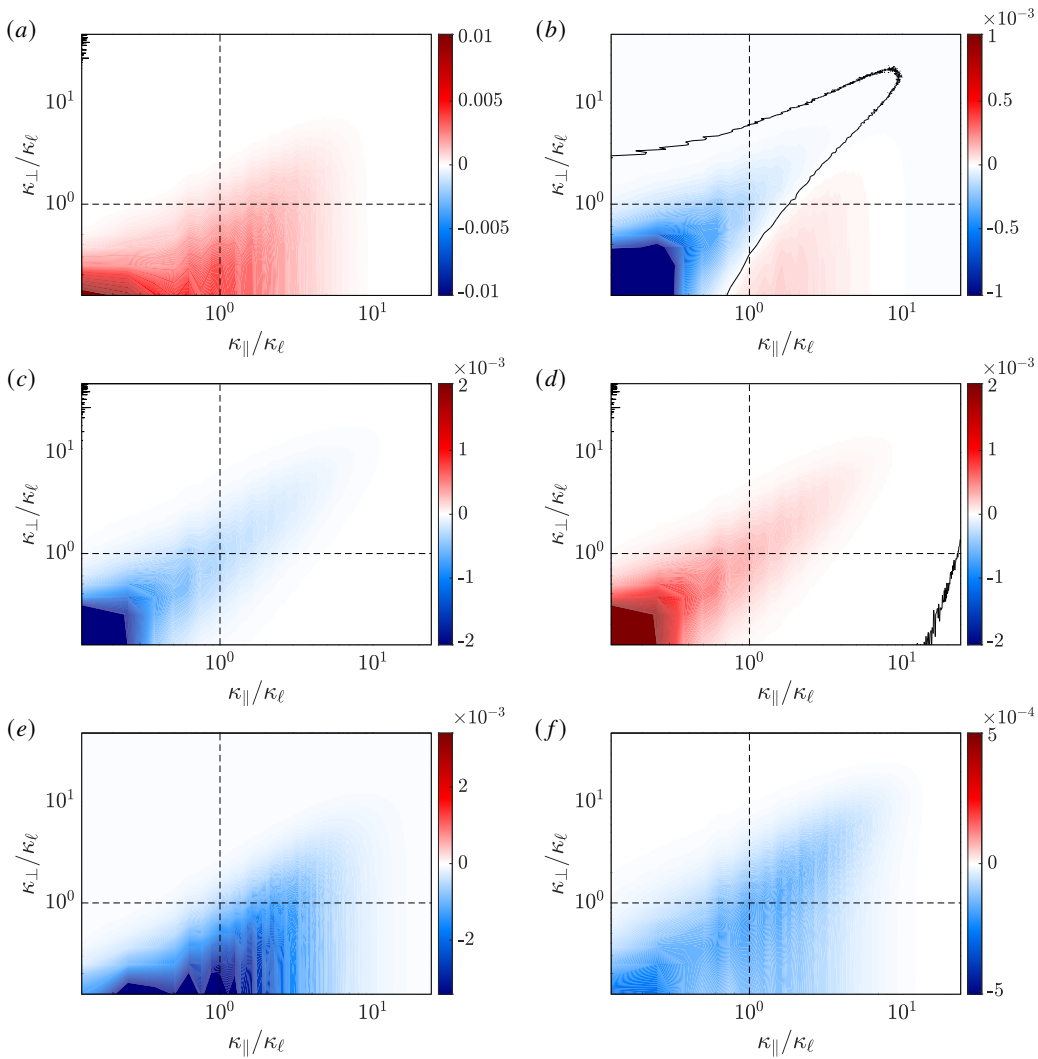
\begin{figure}
  \centering
  \begin{tikzpicture}
    \node at (0,0) {\includegraphics[width=0.49\textwidth]{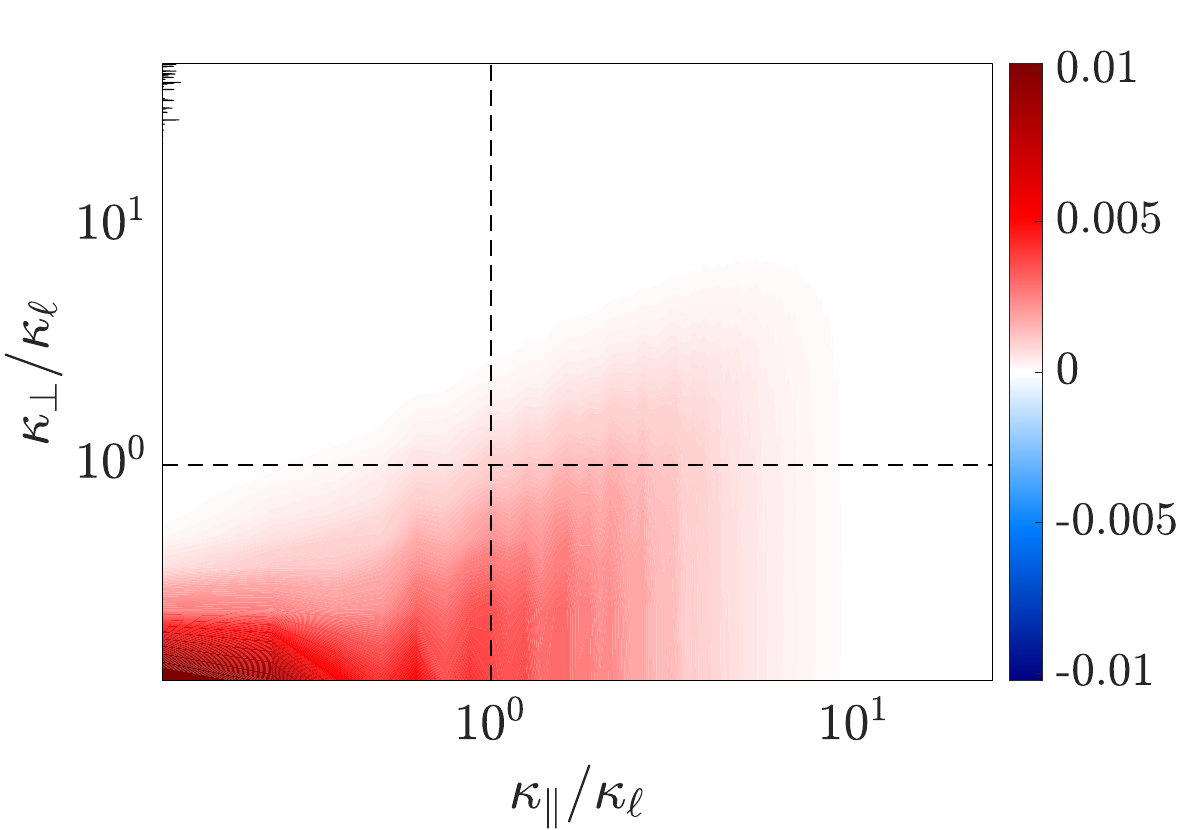}};
    \node at (7,0) {\includegraphics[width=0.49\textwidth]{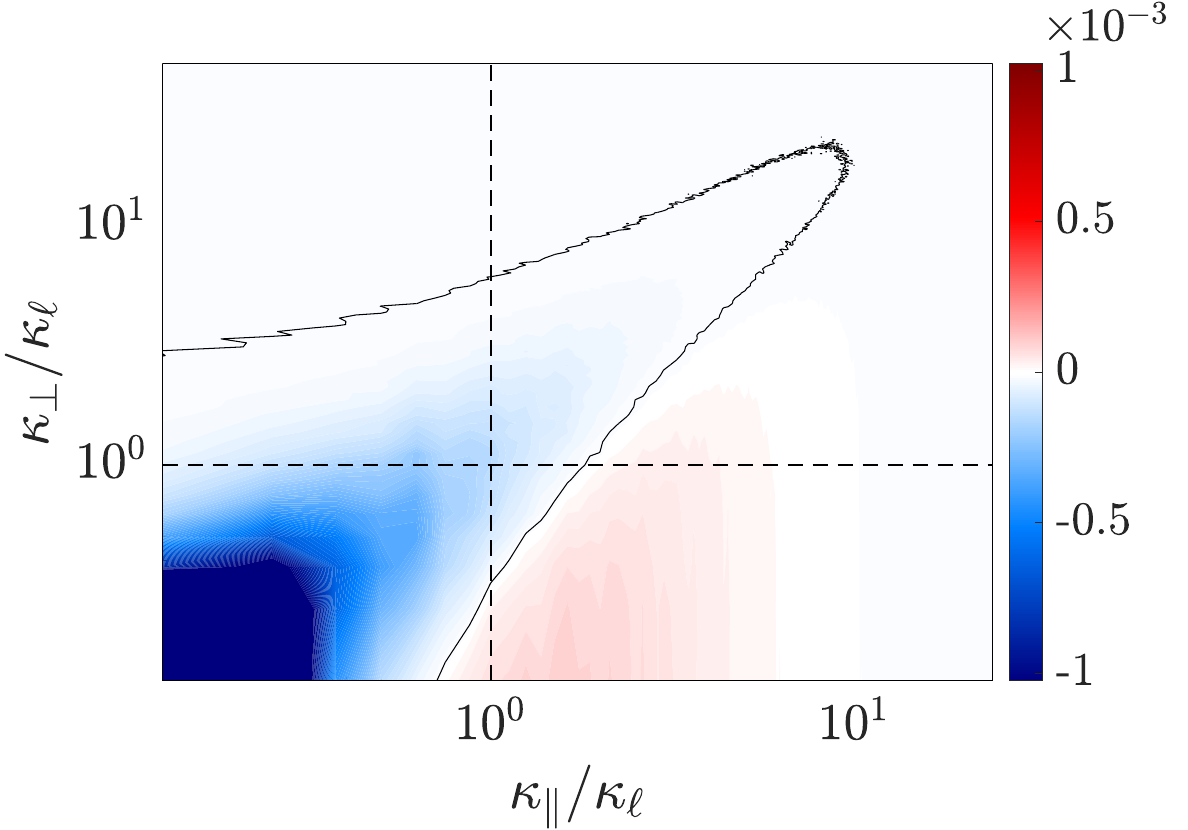}};
    \node at (0,-4.6) {\includegraphics[width=0.49\textwidth]{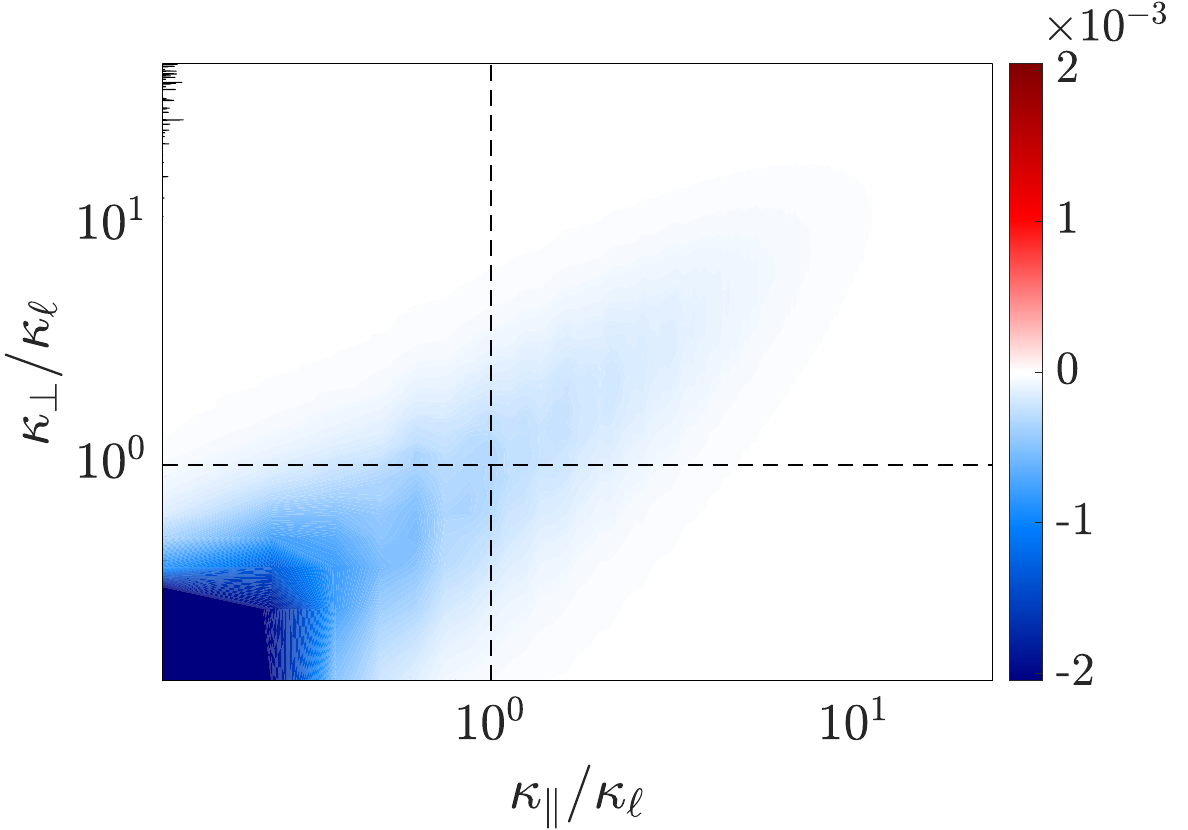}};
    \node at (7,-4.6) {\includegraphics[width=0.49\textwidth]{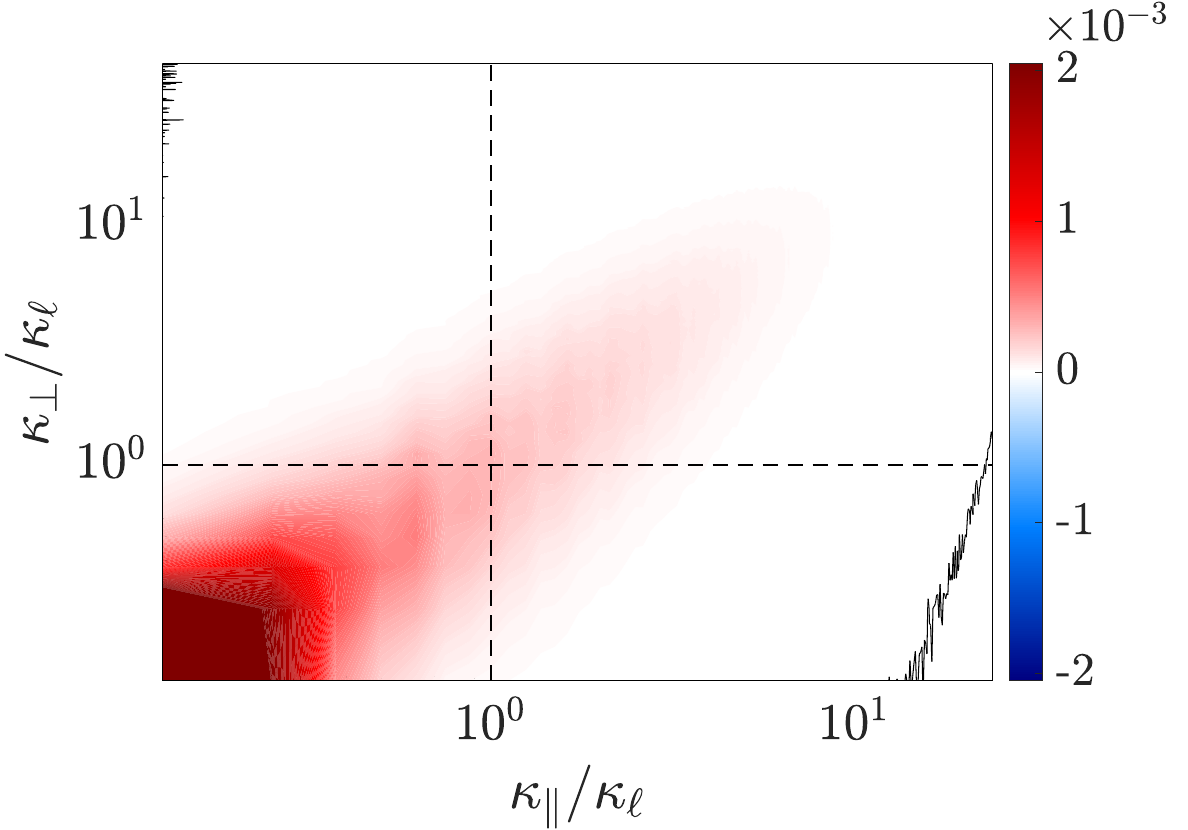}};
    \node at (0,-9.2) {\includegraphics[width=0.49\textwidth]{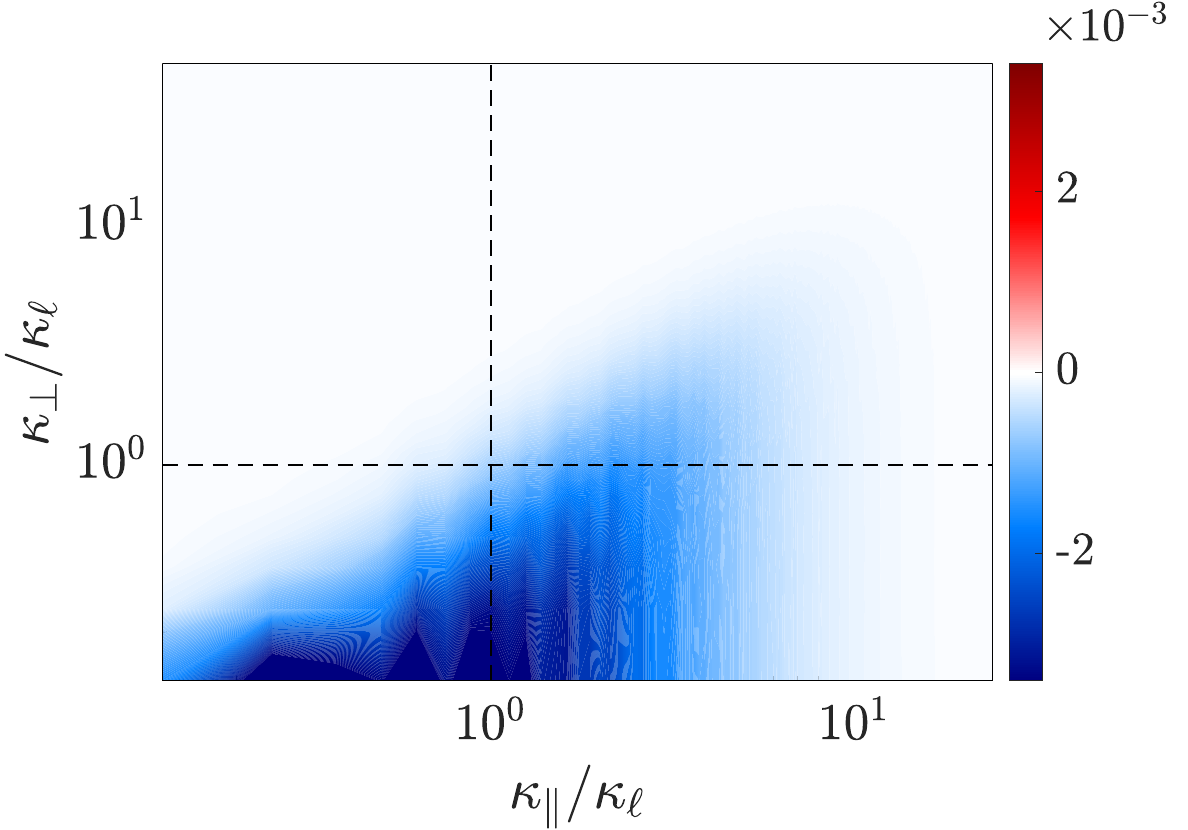}};
    \node at (7,-9.2) {\includegraphics[width=0.49\textwidth]{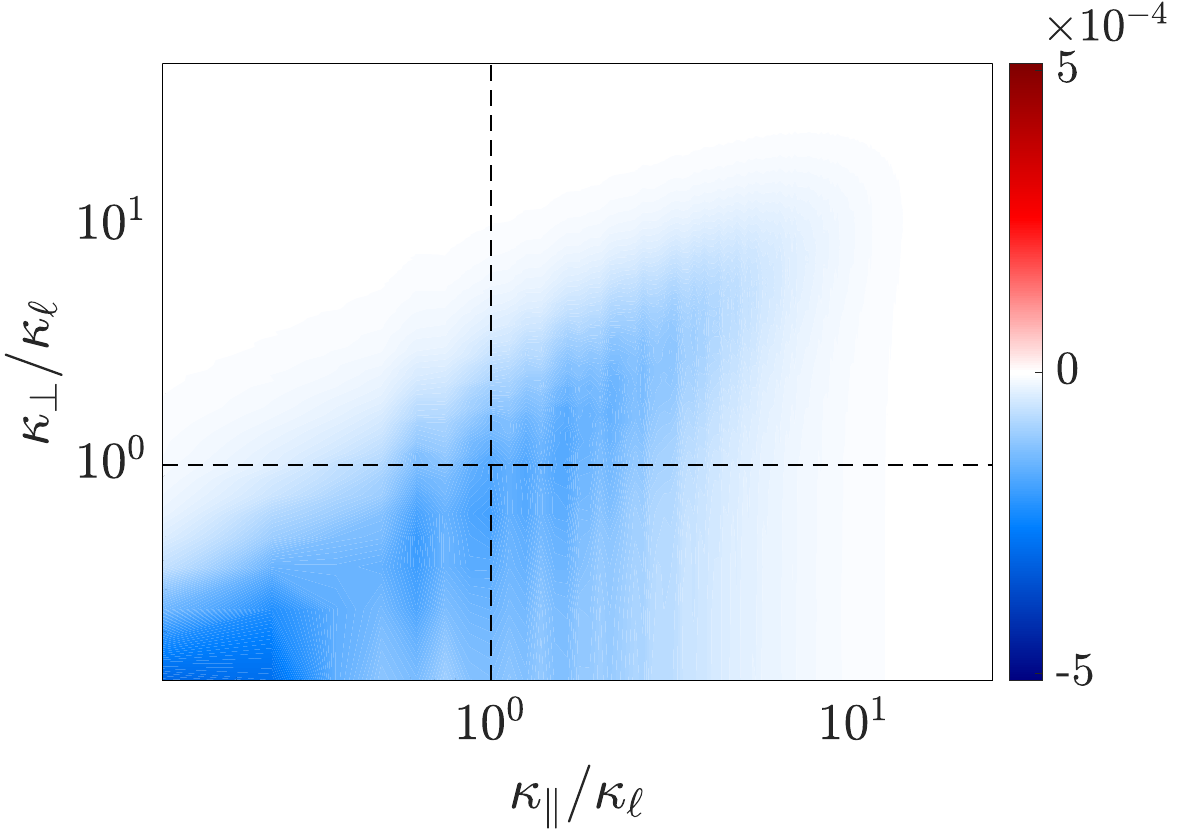}}; 
    \node at (-3.2,2){$(a)$};
    \node at ( 3.8,2){$(b)$};
    \node at (-3.2,-2.6){$(c)$};
    \node at ( 3.8,-2.6){$(d)$};
    \node at (-3.2,-7.2){$(e)$};
    \node at ( 3.8,-7.2){$(f)$}; 
  \end{tikzpicture}   
  \caption{Scale-by-scale energy transfer balance for $n \ell_f^3 = 2.89$ in the two-dimensional $(\kappa_\parallel,\kappa_\perp)$ space at $Ga = 180$: $(a)$ $\mathcal{F}_{1c}$, $(b)$ $\mathcal{F}_{2c}$, $(c)$ $\mathcal{P}_{1c}$, $(d)$ $\mathcal{P}_{2c}$, $(e)$ $\mathcal{D}_{1c}$, $(f)$ $\mathcal{D}_{2c}$. All quantities are scaled with the averaged dissipation in the box $\langle \varepsilon \rangle$. The black thick line denotes the zero value and separates sink and source regions.}
  \label{fig:Ga180_bud_FPD}
\end{figure}  
Figure \ref{fig:Ga180_bud_FPD} shows the two-dimensional distributions of the budget terms $\mathcal{F}_{ic}$, $\mathcal{P}_{ic}$, and $\mathcal{D}_{ic}$ ($i = 1, 2$) in the $(\kappa_\parallel, \kappa_\perp)$ space. For brevity, only the case with $Ga = 180$ and $n \ell_f^3 = 2.89$ is reported, though similar trends are observed across all cases.
The maps highlight pronounced flow anisotropy at large scales: all terms peak near $\kappa_\perp \to 0$ over a broad range of $\kappa_\parallel$, and decay more rapidly with increasing $\kappa_\perp$ than with $\kappa_\parallel$. This anisotropic distribution is consistent with the $\mathcal{E}_{1c}$ and $\mathcal{E}_{2c}$ spectra. At smaller scales, where $\kappa_\perp \approx \kappa_\parallel$, the distributions become more isotropic.
Focusing on the $\mathcal{E}_{1c}$ budget (left panels), the fluid-solid coupling term $\mathcal{F}_{1c}$ is positive across all wavenumbers, acting as a source of vertical velocity fluctuations. Its peak at low $\kappa_\parallel$ and $\kappa_\perp \to 0$ reflects the characteristic scales of the streamers. This input is primarily balanced by viscous dissipation $\mathcal{D}_{1c}$, except at low wavenumbers ($\kappa_\parallel, \kappa_\perp \lessapprox \kappa_\ell$), where the negative pressure-strain term $\mathcal{P}_{1c}$ also contributes significantly. Although smaller in magnitude, $\mathcal{P}_{1c} < 0$ persists down to the smallest scales, particularly where $\kappa_\perp \approx \kappa_\parallel$, indicating that pressure plays a consistent role in promoting isotropy across scales.
By contrast, the $\mathcal{E}_{2c}$ budget (right panels) reveals a different balance. Here, the in-plane velocity fluctuations are primarily sustained by the pressure-strain term $\mathcal{P}_{2c} = -\mathcal{P}_{1c} > 0$, which remains positive throughout. The coupling term $\mathcal{F}_{2c}$ is negative at large scales (especially where $\kappa_\perp \approx \kappa_\parallel$), suggesting that fibres extract energy from the in-plane motions. At smaller scales ($\kappa_\perp, \kappa_\parallel \gtrapprox \kappa_\ell$), however, $\mathcal{F}_{2c}$ becomes positive, indicating a reversal where fibres begin to energise the flow. The large-scale negative $\mathcal{F}_{2c}$ is balanced primarily by $\mathcal{P}_{2c}$, particularly at low wavenumbers and intermediate scales. Conversely, at small $\kappa_\perp$ and large $\kappa_\parallel$, the positive $\mathcal{F}_{2c}$ is balanced by viscous dissipation $\mathcal{D}_{2c}$.

We now move to the nonlinear term $\mathcal{N}_{ic}$. Figure \ref{fig:bud_Ga180} shows that when the intensity of the fluctuations is low, $\mathcal{N}_{ic}^{\parallel,\perp}$ is rather small meaning that the inter-scale energy transfer is substantially negligible and that the fluctuations are mainly sustained by local mechanisms. However, figure \ref{fig:bud_Ga900} shows that the nonlinear term becomes relevant as $Ga$ increases, in agreement with the increasing fluid inertia. Interestingly, the in-plane and out-of-plane terms $\mathcal{N}_{ic,\parallel}^{\parallel,\perp}$ and $\mathcal{N}_{ic,\perp}^{\parallel,\perp}$ behave differently. The $\mathcal{N}_{ic,\parallel}^{\parallel,\perp}$ term is a sink at small scales and a source at large scales, denoting an average energy transfer from smaller to larger scales. The $\mathcal{N}_{ic,\perp}^{\parallel,\perp}$ term, instead, exhibits a more classical behaviour. It is a sink at the large scales and a source at the smaller scales, denoting thus an average energy transfer from larger to smaller scales. Nonetheless, when examining the total scale-by-scale energy transfer rate $\mathcal{N}_{3c}^{\parallel,\perp}$, we observe that $|\mathcal{N}_{3c,\perp}^{\parallel,\perp}| > |\mathcal{N}_{3c,\parallel}^{\parallel,\perp}|$ across almost all wavenumbers. The overall energy transfer is directed from large to small scales, as $\mathcal{N}_{3c}^{\parallel,\perp} < 0$ at small $\kappa$ and $\mathcal{N}_{3c}^{\parallel,\perp} > 0$ at large $\kappa$.

\begin{figure}
\centering
  \begin{tikzpicture}
 \node at (3.5,1.7) {\includegraphics[trim={0 14 0 5},clip,width=0.9\textwidth]{./fig/legend_spectrum_Ga-eps-converted-to.pdf}};    
    \node at (0,0) {\includegraphics[width=0.49\textwidth]{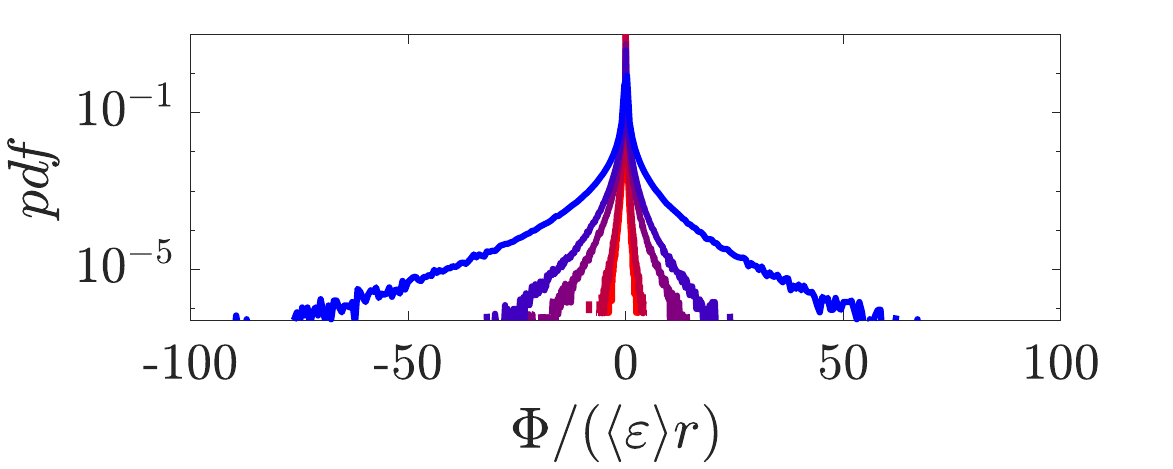}};
    \node at (7,0) {\includegraphics[width=0.49\textwidth]{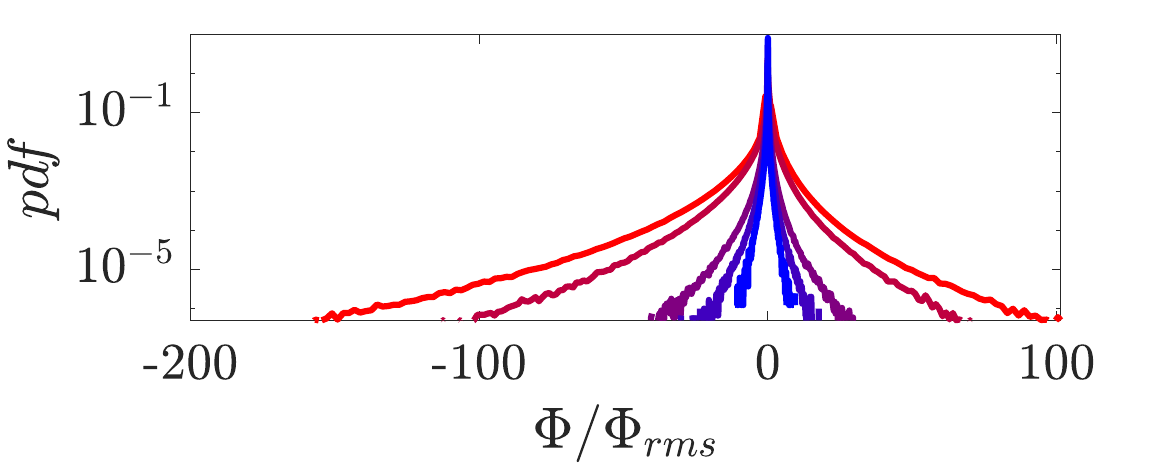}};
    \node at (0,-2.8) {\includegraphics[width=0.49\textwidth]{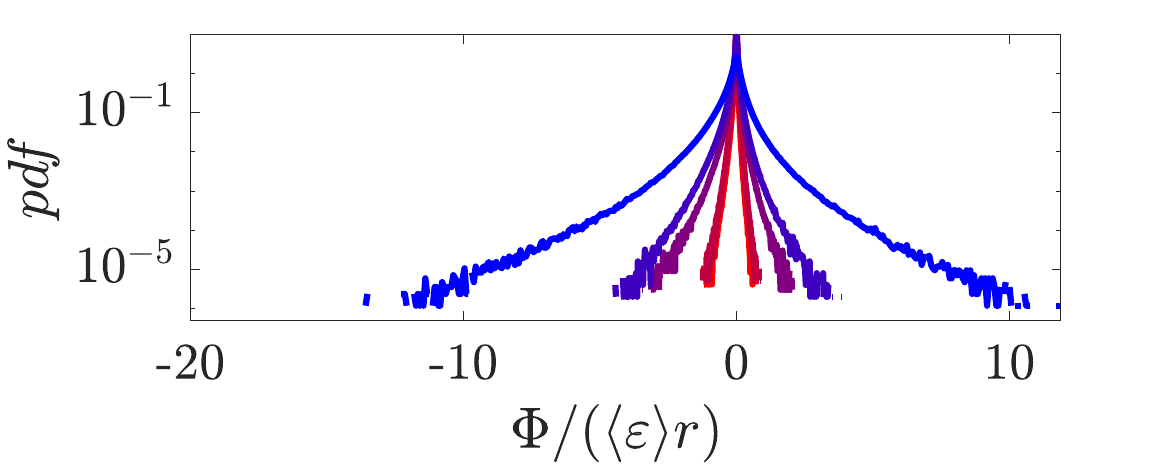}};
    \node at (7,-2.8) {\includegraphics[width=0.49\textwidth]{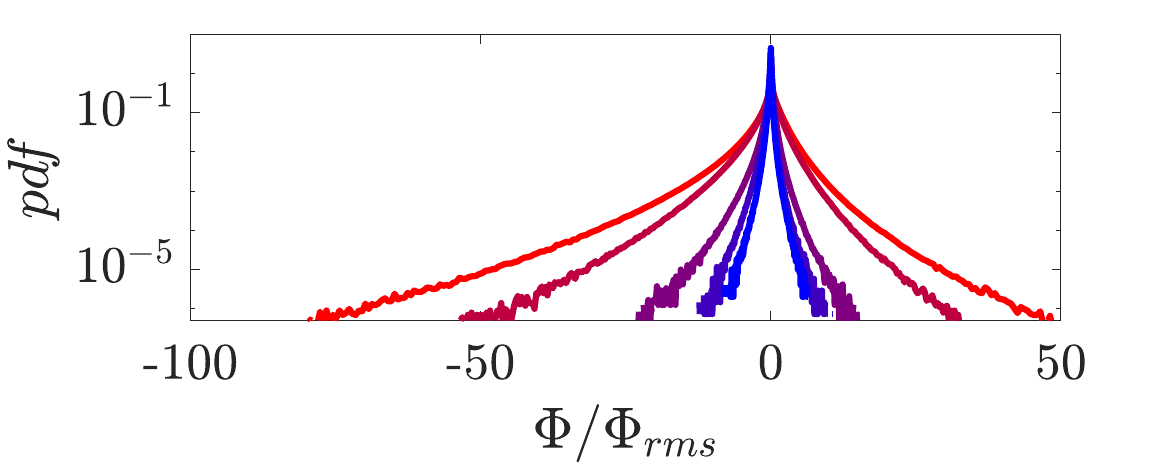}};
    \node at (0,-5.6) {\includegraphics[width=0.49\textwidth]{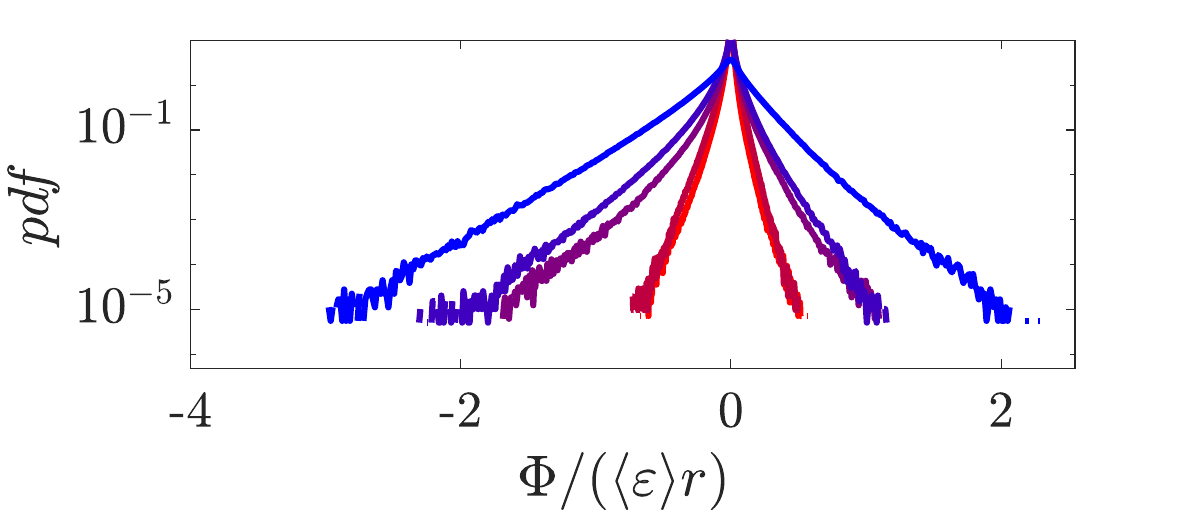}};
    \node at (7,-5.6) {\includegraphics[width=0.49\textwidth]{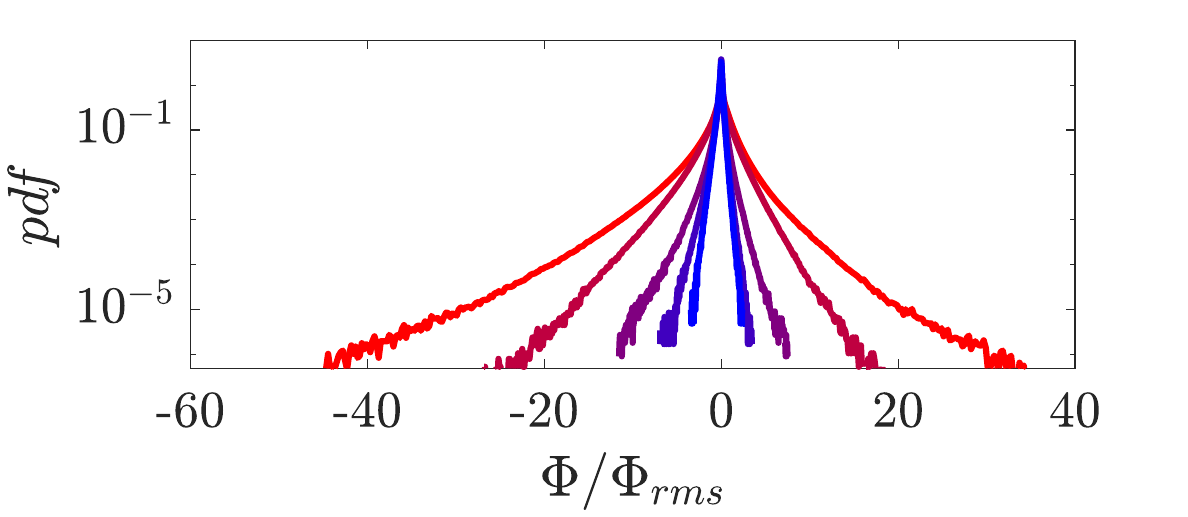}};
    \node at (-3.2,1.2){$(a)$};
    \node at ( 3.8,1.2){$(b)$};
    \node at (-3.2,-1.6){$(c)$};
    \node at ( 3.8,-1.6){$(d)$};
    \node at (-3.2,-4.4){$(e)$};
    \node at ( 3.8,-4.4){$(f)$};   
  \end{tikzpicture}      
\caption{Probability distribution function of the $\Phi$ fluxes for the $n \ell_f^3 = 2.89$ cases. Panels ($a,c,e$): Distribution of $\Phi/(\langle \varepsilon \rangle r)$. Panels ($b,d,f)$: Distribution of $\Phi/\Phi_{rms}$. From top to bottom: $r/\ell_f \approx 0.25$ ($a,b$), $0.76$ ($c,d$) and $1.5$ ($e,f$).}
\label{fig:fluxes}
\end{figure}

The average picture of the interscale exchanges is not the complete physical process: although the average picture may predict a transfer from larger to smaller scales, there may exist events localised in space and time where the opposite occurs. To address this aspect in the present context, we analyse the probability distribution functions of the energy transfer rate at different scales. We resort on the real-space counterpart of equation \eqref{eq:budEi}, i.e., the K\'{a}rm\'{a}n-Howarth-Monin-Hill (KHMH) equation, also known as the generalised Kolmogorov equation \citep{danaila-etal-2001,hill-2002}. The KHMH equation is the budget equation for the velocity structure function $\langle \delta q^2 \rangle = \langle \delta u_i \delta u_i \rangle$, where $\delta u_i$ is the velocity increment between two points $\bm{x}$ and $\bm{x}'$ identified using their mid-point $\bm{X} = (\bm{x} + \bm{x}')/2$ and their separation vector $\bm{r} = \bm{x}' - \bm{x}$, i.e. $\delta u_i(\bm{X},r,t) = u_i(\bm{X}+\bm{r}/2,t) - u_i(\bm{X}-\bm{r}/2,t)$. Since the present flow features three homogeneous directions and is statistically stationary, $\aver{\delta q^2}$ depends only on $\bm{r} = (r_x,r_y,r_z)$. To get rid of the directional information we also integrate in the $\bm{r}$ space over spherical shells of radius $r=|\bm{r}|$, with surface $S(r)$ and volume $V(r)$. In this homogeneous and isotropic framework, the KHMH equation reads:
\begin{equation}
\begin{gathered}
\underbrace{\frac{1}{S(r)} \int_{S(r)} \aver{ \delta u_j \delta q^2} n_j \text{d}\Sigma }_{\aver{\Phi}(r)} = \underbrace{\frac{1}{S(r)} \int_{S(r)} \frac{2}{Ga} \frac{\partial \aver{ \delta q^2 }}{\partial r_j } n_j \text{d}\Sigma }_{\aver{D}(r)}+ \\ - \frac{4}{3} \underbrace{ \frac{1}{Ga}\aver{ \frac{\partial u_i}{\partial x_j}\frac{\partial u_i}{\partial x_j} }}_{\aver{\varepsilon}}r + \underbrace{ \frac{1}{S(r)} \int_{V(r)} 2 \aver{\delta u_i \delta f_{fs,i} } \text{d}\Omega}_{\aver{F(r)}}.
\end{gathered} 
\end{equation}
Here, $D$ denotes the viscous diffusion, $\varepsilon$ the dissipation rate, and $F$ the forcing associated with the fluid--structure interaction, and $\Phi$, the quantity of interest, represents the nonlinear energy cascade rate at the considered scale $r$; $\bm{n}$ is the outward-pointing unit normal to the spherical surface. In this convention, $\Phi > 0$ corresponds to backscatter, i.e., energy transfer from smaller to larger scales, whereas $\Phi < 0$ denotes a forward cascade, with energy transferred from larger to smaller scales. For a full derivation of the governing equations and additional details, the reader is referred to \cite{gatti-etal-2020}.

Figure \ref{fig:fluxes} plots the probability distribution function of $\Phi(r)$ for $n \ell_f^3 = 2.89$ and different $Ga$. We consider three different separations: $r/\ell_f \approx 0.25, 0.75$, and $1.5$. For all scales, the distribution shows long positive and negative tails, hinting at an intermittent interscale energy transfer, similar to what has been found in sustained turbulent flows \citep{ishihara-gotoh-kaneda-2009}. Backscatter and forward transfers coexist, with extreme events occurring with a higher probability than for a normal distribution. However, for all separations, the distributions are asymmetric and exhibit a longer negative tail. Accordingly, $\langle \Phi \rangle$ is negative at all scales, and the average interscale energy transfer is from larger to smaller scales, like in classical homogeneous and isotropic turbulence \citep{frisch-1995}; this agrees with the Fourier-space investigation. In agreement with figures \ref{fig:bud_Ga180} and \ref{fig:bud_Ga900}, when $Ga$ increases and the fluid velocity fluctuations intensify, the tails of the distributions become wider (see the left panels) indicating that the inter-scale energy transfers become more relevant.

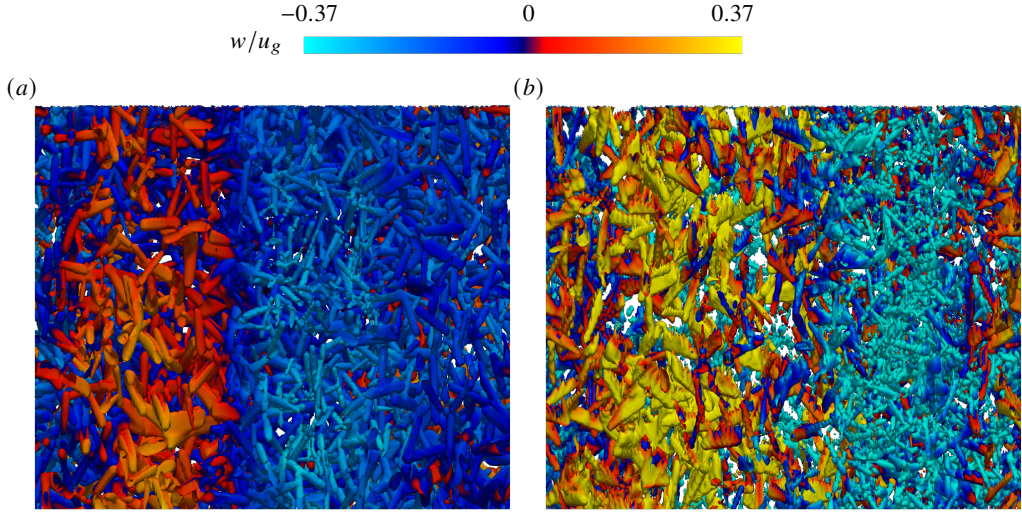
\begin{figure}
  \centering
\begin{tikzpicture}
    \node at (-0.1, 2.0) {\includegraphics[trim={0 1050 0 50},clip,width=1.0\textwidth]{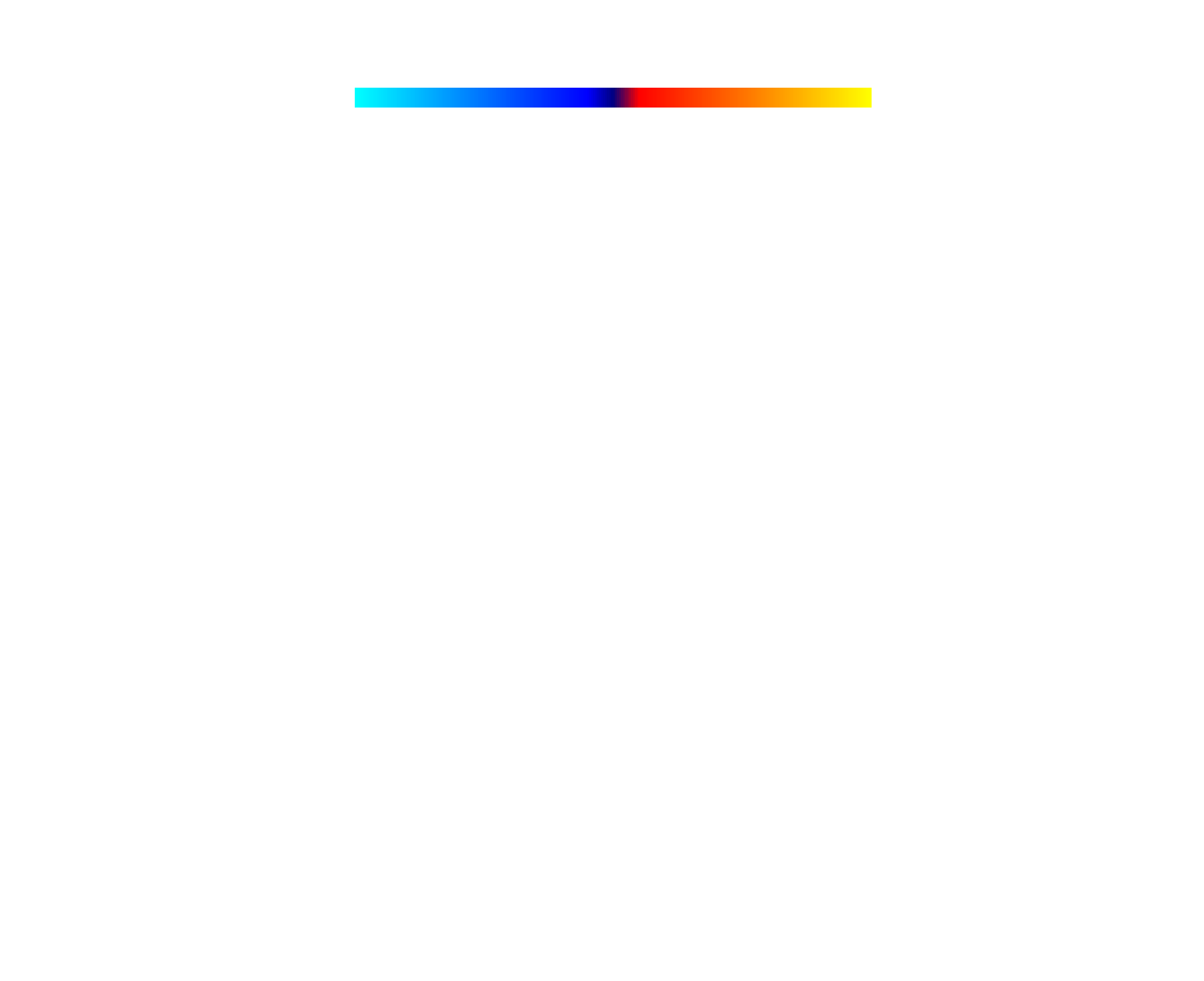}};
    \node at (0,-1.8) {\includegraphics[trim={0 150 0 190},clip,width=1.0\textwidth]{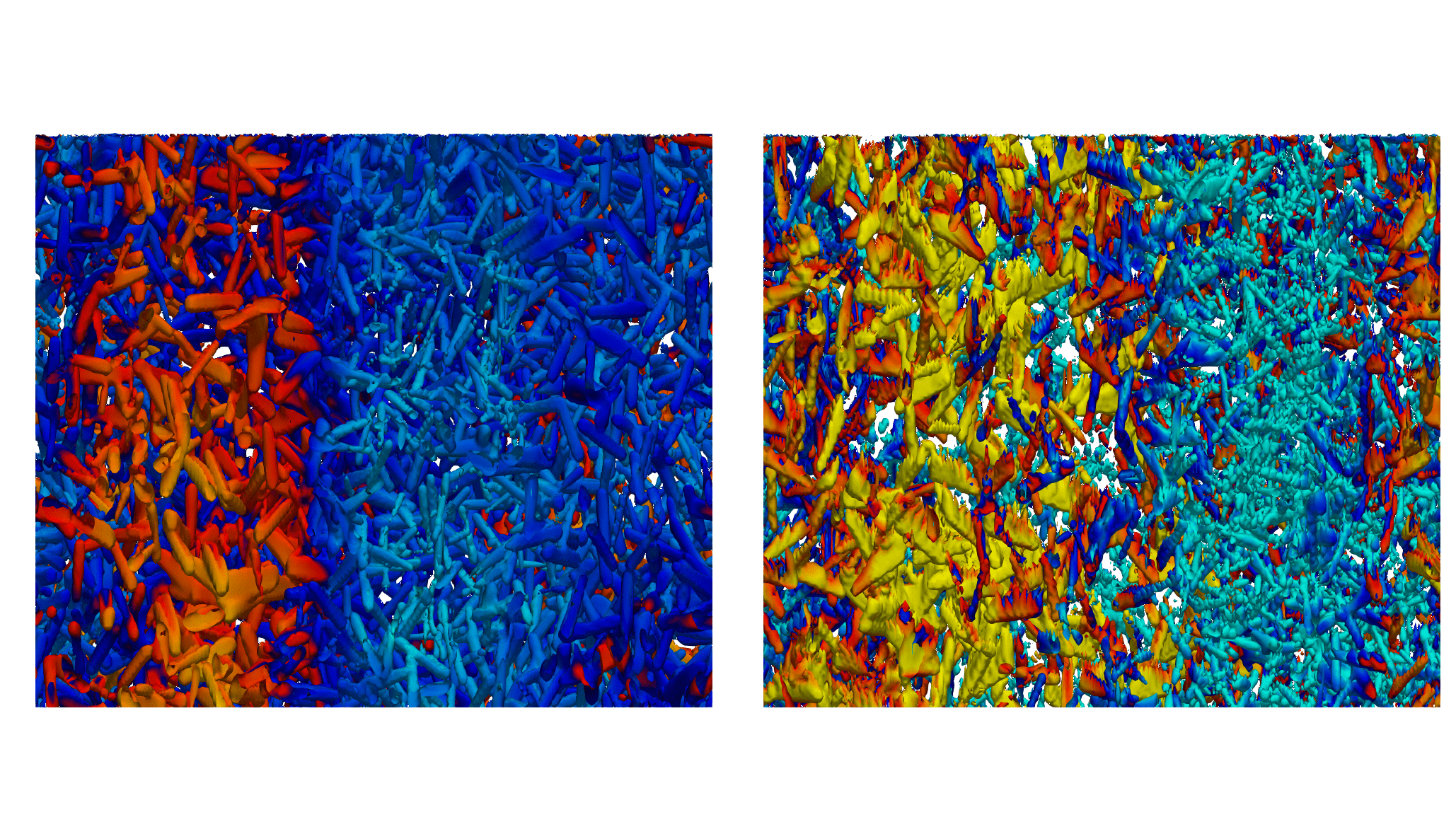}};  
\node at (0.1,2.2){$0$};
    \node at (-2.8,2.2){$-0.37$};
    \node at ( 2.8,2.2){$ 0.37$};  
    \node at (-3.5,1.8){$w/u_g$};
    \node at (-6.6,1.2){$(a)$};
    \node at ( 0.1,1.2){$(b)$};
    \end{tikzpicture}
  \caption{Isosurfaces of the dissipation, $0.03 \langle \varepsilon \rangle$ for $n \ell_f^3 = 2.89$ at ($a$) $Ga=180$ and ($b$) $Ga=900$. The colour is for the vertical velocity of the fluid phase, with blue colour denoting negative values and red colour positive values. For small $Ga$ the dissipation is confined in the near-fibre region, while for large $Ga$ large values of the dissipation are observed away from the fibres' surface.}
  \label{fig:diss}
\end{figure}

The fluxes normalised by their root-mean-square values (right panels of figure \ref{fig:fluxes}) highlight the influence of $Ga$ on the intermittency of the energy transfer. As $Ga$ increases, the tails of the normalised-flux distributions shrink across all separations, indicating a reduction in intermittency. This behaviour is linked to the nature of the velocity field. For lower $Ga$, the flow remains relatively smooth except in the immediate vicinity of the fibres, where sharp and intense fluctuations occur. This is illustrated in figure~\ref{fig:diss}, which presents isosurfaces of the dissipation field. At low $Ga$, regions of high dissipation are confined near the fibre surfaces, where flow deformation is most intense. In this regime, the energy injected at small scales is dissipated almost immediately and locally. In contrast, at higher $Ga$, the influence of the fibres extends farther into the fluid domain, and the injected energy is dissipated over a broader region. This results in a smoother, less spatially intermittent dissipation field.

 \section{The local structure of the fluctuations}
\label{sec:QR}

In this section, we characterise the local structure of the fluid-phase velocity fluctuations. We focus on the case with $n \ell_f^3 = 2.89$ and consider increasing values of $Ga$, thereby spanning conditions from weak to strong velocity fluctuations. When the fluid is sufficiently smooth in the neighbourhood of a point $(\bm{x}_0, t)$, and the spatial variation of the velocity gradient $a_{ij} = \partial u_j / \partial x_i$ is relatively small, the velocity field can be locally approximated by a first-order Taylor expansion: $u_i(\bm{x}, t) = u_i(\bm{x}_0, t) + a_{ij}(\bm{x}_0, t)(x_j - x_{0,j}) + \mathcal{O}(|\bm{x} - \bm{x}_0|^2)$. The velocity gradient tensor $a_{ij}$ encapsulates key information about the structure of the fluctuating field, as it: (i) reflects the dynamics at small scales \citep{meneveau-2011}, and (ii) is more sensitive to the non-Gaussian features of turbulence than the velocity field itself \citep{tsinober-2000}.

We decompose $a_{ij}$ into its symmetric and anti-symmetric parts, i.e., the strain-rate tensor $s_{ij} = (a_{ij} + a_{ji})/2$ and the rotation rate tensor $w_{ij} = (a_{ij}-a_{ji})/2$. The former is related to strain and dissipation, while the latter to vorticity and enstrophy. The velocity gradient field is completely characterised when knowing: (i) the three principal rates of strain $\alpha \ge \beta \ge \gamma$, i.e., the three eigenvalues of $s_{ij}$, (ii) the vorticity magnitude $\omega^2= \bm{\omega} \cdot \bm{\omega}$, i.e., the enstrophy, and (iii) the orientation of $\bm{\omega}$ relative to the three principal axes of strain, i.e., the eigenvectors of $s_{ij}$ ($\hat{\bm{e}}_\alpha$, $\hat{\bm{e}}_\beta$ and $\hat{\bm{e}}_\gamma$); see \cite{davidson-2004}. Due to the incompressibility constraint $\alpha + \beta + \gamma = 0$, meaning that $\alpha$ is always nonnegative, $\gamma$ is always nonpositive, while $\beta$ has a sign that determines the local straining state. The $s_{ij}$ and $w_{ij}$ tensors are strongly related to vortex stretching and self-amplification of the strain, which are manifested in the net production of enstrophy $\omega^2 = w_{ii}$ and strain $s^2 = s_{ii}$.

\subsection{Straining state and alignments}

\begin{figure}
  \centering
  \includegraphics[trim={0 14 0 5},clip,width=0.9\textwidth]{./fig/legend_spectrum_Ga-eps-converted-to.pdf}       
  \includegraphics[width=0.8\textwidth]{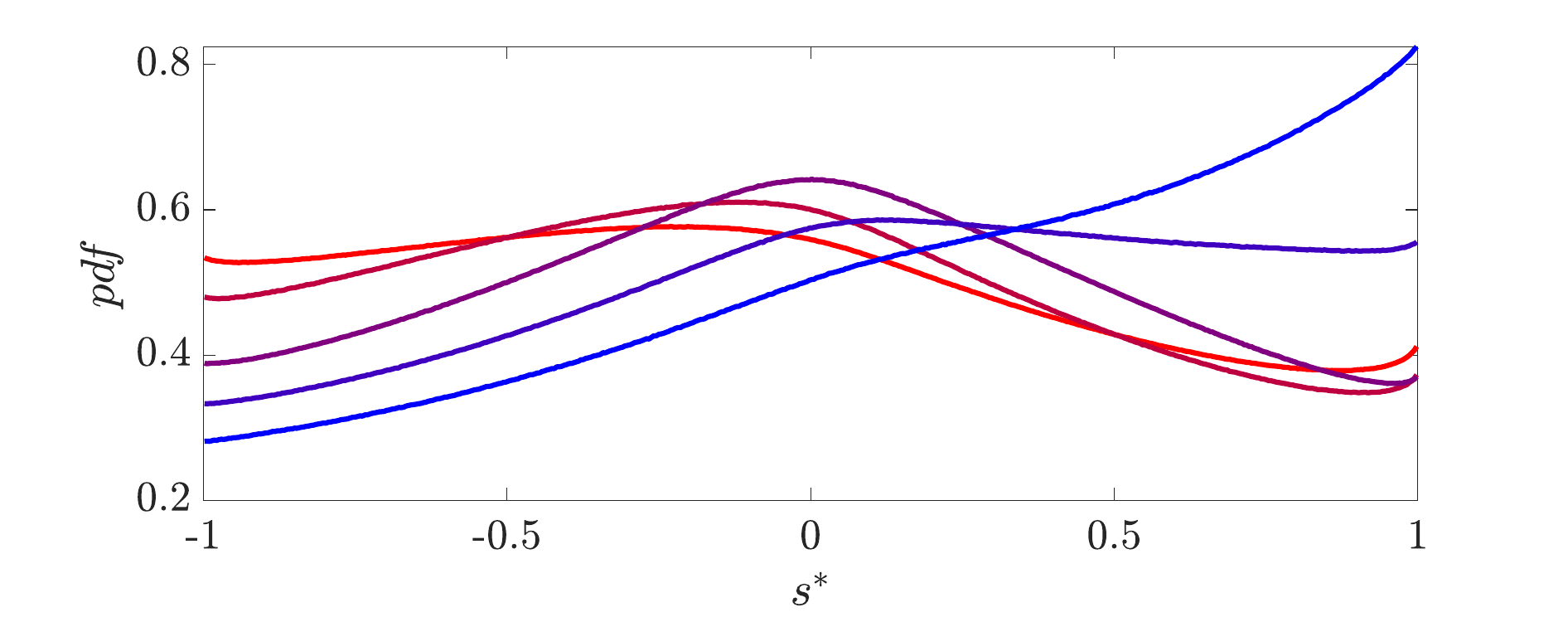}
\caption{Distribution of $s^*$ for $n \ell_f^3 = 2.89$ and different $Ga$.}
  \label{fig:sstar}
\end{figure}

We start characterising $\alpha$, $\beta$ and $\gamma$ (see figure \ref{fig:sstar}) by using $s^*$ \citep{lund-rogers-1994}, which is defined as:
\begin{equation}
  s^* = \frac{ 3 \sqrt{6} \alpha \beta \gamma }{ ( \alpha^2 + \beta^2 + \gamma^2 )^{3/2} }.
\end{equation}
For a random velocity gradient field with no preferred structure, all states are equally probable, and $s^*$ has a uniform distribution. When $s^* = 1$, we have that $\alpha = \beta = -\gamma/2$ meaning that the state of straining is an axisymmetric extension, in which a small element of fluid extends symmetrically in two directions and contracts in the third one. When $s^*=-1$, instead, $\gamma = \beta = - \alpha/2 <0$, and the state of straining corresponds to an axisymmetric compression, where the elements of fluid contract in two directions and extend in the third one. When $s^*=0$, we have that $\beta = 0$ meaning that the straining state is two-dimensional, which is typical in shear-dominated regions \citep{meneveau-2011}. 

Figure \ref{fig:sstar} shows that the local state of straining changes with $Ga$. For small $Ga$, events with $s^* \le 0$ are promoted and states with two directions of compression and one direction of extension are more frequent. At the intermediate $Ga =225$ and $Ga =450$, instead, the most likely value is $s^*=0$: two-dimensional straining states with one direction of compression and one direction of extension are mainly favoured. At these $Ga$, indeed, the fluid-phase velocity fluctuations are rather weak and mainly localised in the vicinity of the settling fibres (see figure \ref{fig:diss}), where the flow is shear-dominated. Similar results were observed in fully developed turbulent flows laden with particles or fibres \citep{schneiders-etal-2017,cannon-olivieri-rosti-2024,chiarini-tandurella-rosti-2025}. A further increase of $Ga$ leads to an enhancement of the $s^*>0$ events, and for $Ga=900$ the distribution peaks at $s^*=1$. At the large $Ga$, the fluid inertia is not negligible (see \S\ref{sec:bud}) and the local structure of the fluctuations is more similar to that of homogeneous and isotropic turbulence (HIT) \citep{lund-rogers-1994}, i.e., dominated by axisymmetric extensions. 

\begin{figure}
  \centering
    \begin{tikzpicture}
      \node at (0,0) {\includegraphics[width=0.49\textwidth]{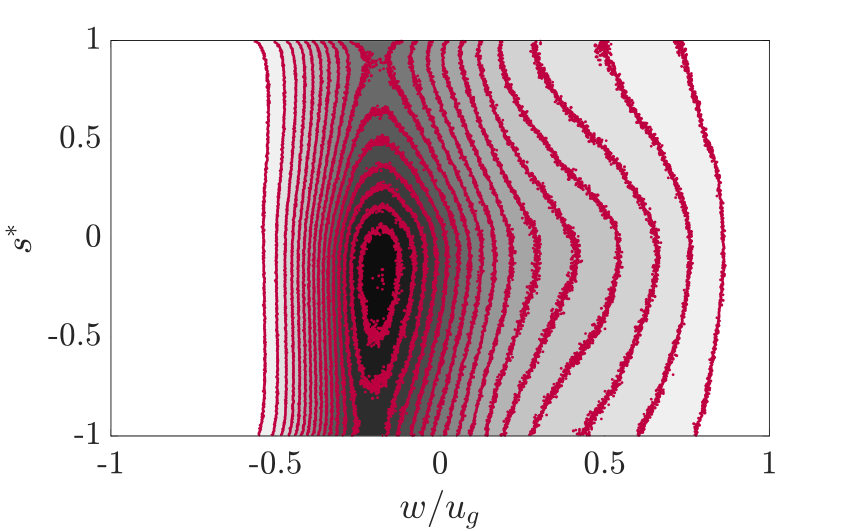}};
      \node at (7,0) {\includegraphics[width=0.49\textwidth]{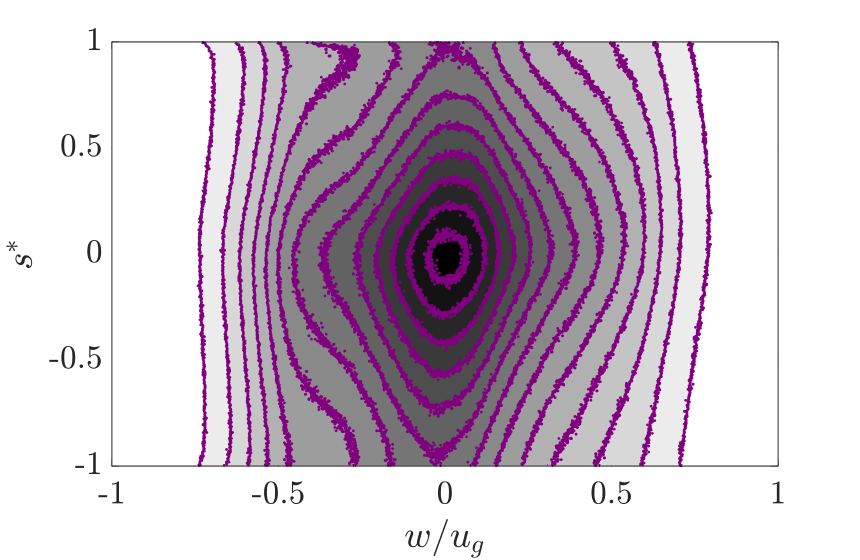}};
      \node at (0,-4.3) {\includegraphics[width=0.49\textwidth]{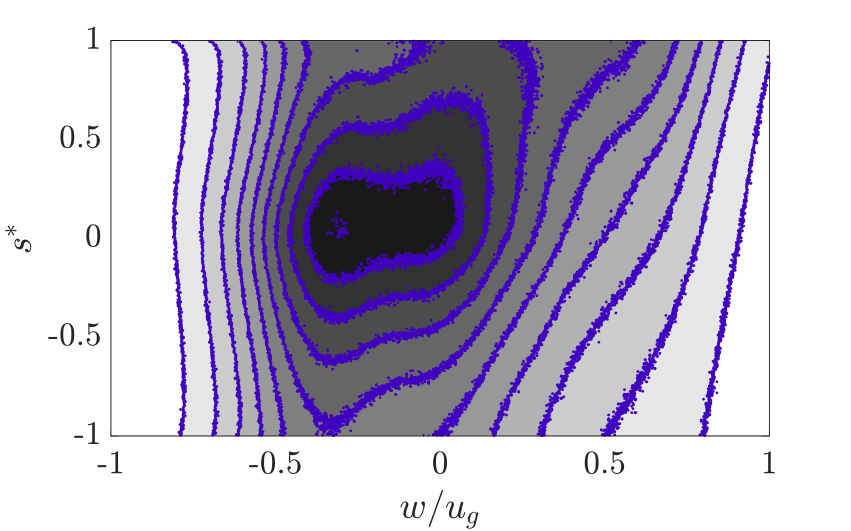}};
      \node at (7,-4.3) {\includegraphics[width=0.49\textwidth]{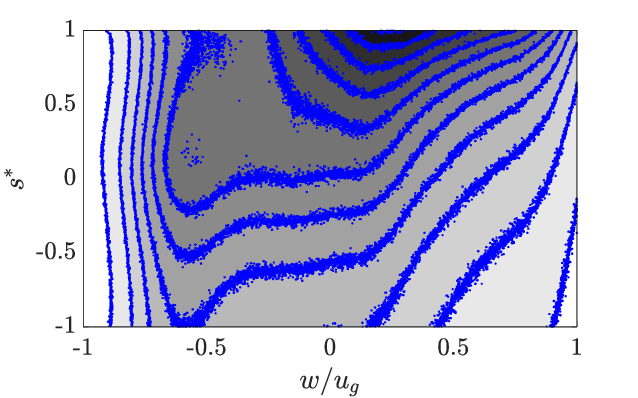}};
      \node at (-3.2,1.8){$(a)$};
      \node at ( 3.8,1.8){$(b)$};
      \node at (-3.2,-2.5){$(c)$};
      \node at ( 3.8,-2.5){$(d)$};
\end{tikzpicture}  
\caption{Joint probability distribution function of $s^*-w$ for $n \ell_f^3 = 2.89$ and ($a$) $Ga=225$, ($b$) $Ga=450$, ($c$) $Ga=675$ and ($d$) $Ga=900$. A linear spacing of $4.15 \times 10^{-3}$ is used between isolines.}
  \label{fig:swjpdf}
\end{figure}
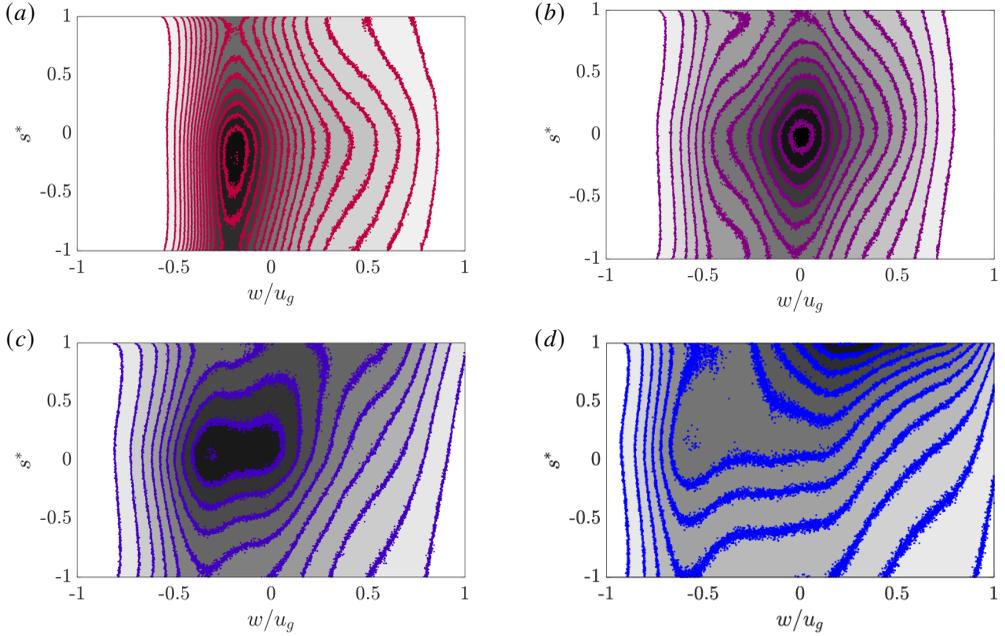
To account for the flow inhomogeneity induced by the presence of the streamers, figure \ref{fig:swjpdf} considers the $s^*-w$ JPDF. As mentioned above, $w<0$ locates the flow regions with a large concentration of fibres. For small $Ga$ ($Ga \le 450$), the most likely events have $w \approx 0$ and $s^* \le 0$. This is consistent with the observation that at these $Ga$ the fluid-phase velocity fluctuations are mainly associated with the shear layers that form in the vicinity of the fibres, which are dominated by a two-dimensional straining state. For larger $Ga$ the scenario changes due to the presence of the streamers, and the state of strain in the ascending ($w>0$) and descending ($w<0$) regions differs. In the ascending region the JPDF peaks at $s^*=1$, while in the descending one at $s^* \le 0$. The ascending stream is dominated by axisymmetric extensional events, reminiscent of HIT, while the descending stream is dominated by events characterised by a two-dimensional state of strain, like at lower $Ga$. At these values of $Ga$, two distinct mechanisms coexist in organising the fluid-phase velocity fluctuations (see \S\ref{sec:bud}). In the descending stream, where the fibre concentration is high, fluctuations are primarily driven by the fluid-solid interaction. In contrast, in the ascending stream, where the fibre concentration is low, the velocity fluctuations are also sustained by nonlinear mechanisms (see the following discussion).

\begin{figure}
  \centering
    \begin{tikzpicture}
 \node at (3.5,2) {\includegraphics[trim={0 14 0 5},clip,width=0.9\textwidth]{./fig/legend_spectrum_Ga-eps-converted-to.pdf}};     
    \node at (-1,0) {\includegraphics[width=0.32\textwidth]{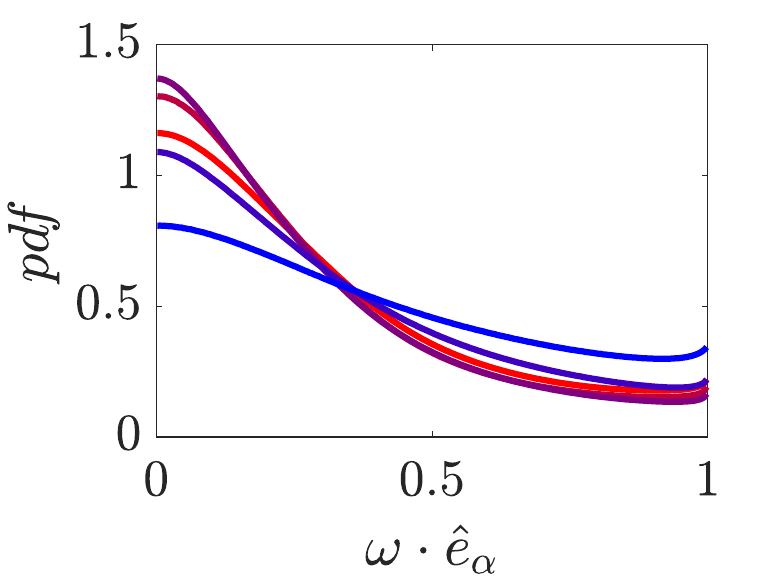}};
    \node at (3.25,0) {\includegraphics[width=0.32\textwidth]{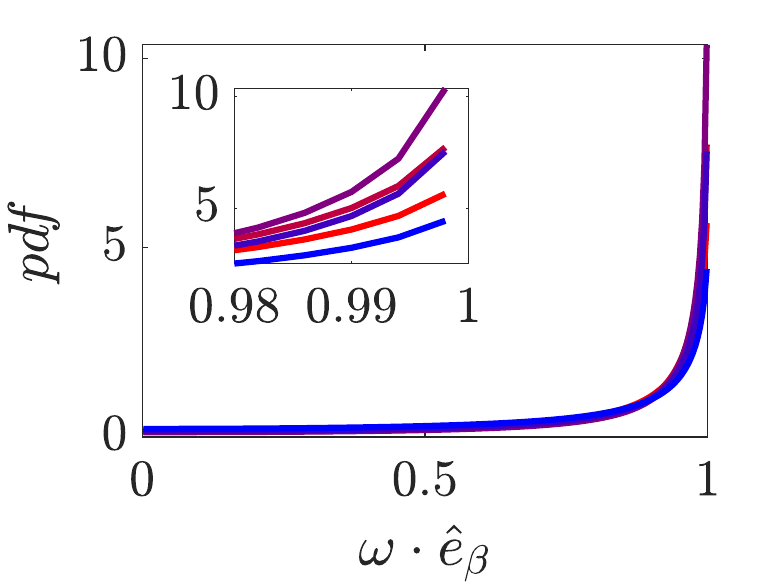}};
    \node at (7.5,0) {\includegraphics[width=0.32\textwidth]{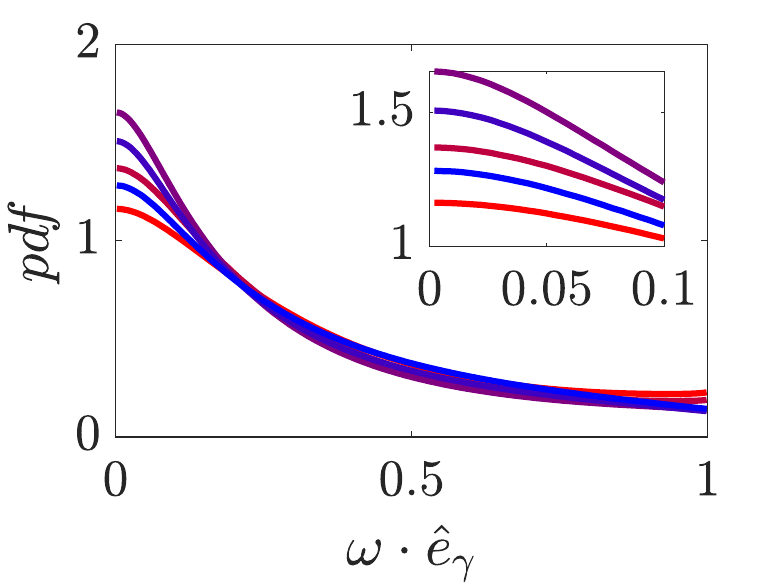}};
    \node at (-3,1.5){$(a)$};
    \node at ( 1.25,1.5){$(b)$};
    \node at ( 5.5,1.5){$(c)$};
    \end{tikzpicture}
  \caption{Distribution of the cosine of the angle between the vorticity and the eigenvectors of the strain-rate tensor for $n \ell_f^3 = 2.89$ and different $Ga$.}
  \label{fig:costh}
\end{figure}
Figure \ref{fig:costh} shows the alignment between $\bm{\omega}$ and the principal axes of strain. These quantities are strongly related to the dynamics of the fluctuations: mechanisms such as vortex stretching, that result from the interaction between the vorticity and the rate of strain tensor, do not depend only on the magnitude of $\omega^2$ and $s^2$, but also on the geometry of the velocity derivatives. For all cases $\bm{\omega}$ is mostly aligned with $\hat{e}_{\beta}$, similarly to HIT. This feature has been often associated with the mechanism of axial stretching of vortices \citep{ashurst-etal-1987}. Figure \ref{fig:costh} shows that the first eigenvector $\hat{\bm{e}}_\alpha$ shows very little correlation with $\hat{\bm{\omega}}$ (in particular at large $Ga$), as indicated by the almost flat distribution, while the last eigenvector is mostly perpendicular to the vorticity, with a peak at $|\hat{\bm{\omega}} \cdot \hat{\bm{e}}_\gamma | = 0$. The influence of $Ga$ is nonmonotonous and changes with the eigenvector. For $Ga \le 450$, an increase of $Ga$ promotes events with $|\hat{\bm{\omega}} \cdot \hat{\bm{e}}_{\alpha}|=0$, $|\hat{\bm{\omega}} \cdot \hat{\bm{e}}_\beta| = 1$ and $|\hat{\bm{\omega}} \cdot \hat{\bm{e}}_\gamma|=0$: the vorticity aligns more strongly with the intermediate eigenvector $\hat{\bm{e}}_\beta$ and becomes more perpendicular to the first and last eigenvectors $\hat{\bm{e}}_\alpha$ and $\hat{\bm{e}}_\gamma$. At these parameters, the nonlinear terms are subdominant and the fluid-phase fluctuations are mostly associated with the shear layers generated at the surface of the fibres, which become stronger as $Ga$ increases. In a purely shear flow, indeed, the compressional and extensional directions are in the plane of the shear flow, while the vorticity is aligned with the out-of-plane direction, leading thus to $|\hat{\bm{\omega}} \cdot \hat{\bm{e}}_\alpha | = 0$, $|\hat{\bm{\omega}} \cdot \hat{\bm{e}}_\beta | = 1$ and $|\hat{\bm{\omega}} \cdot \hat{\bm{e}}_\gamma | = 0$. However, when further increasing $Ga$ ($Ga>450)$ the opposite occurs and events with $|\hat{\bm{\omega}} \cdot \hat{\bm{e}}_{\alpha}|=1$, $|\hat{\bm{\omega}} \cdot \hat{\bm{e}}_\beta| = 0$ and $|\hat{\bm{\omega}} \cdot \hat{\bm{e}}_\gamma|=1$ are promoted, i.e. vorticity aligns more with the first and last eigenvectors and less with the second. For $Ga \ge 450$ the mechanisms associated with the nonlinear terms become relevant, and the geometry of the velocity derivative field changes accordingly.

\subsection{Vortex stretching and strain self-amplification}
The dynamics of the fluctuations is related to the local straining state of the flow and its geometrical properties. We briefly focus on the vortex stretching and on the self-amplification of the strain, which in three-dimensional flows results in a net production of enstrophy $\mathcal{P}_{\omega^2}>0$ and strain $\mathcal{P}_{s^2}>0$ \citep{tsinober-2000,davidson-2004}. Here,
\begin{equation}
  \mathcal{P}_{\omega^2} = \omega_i \omega_j s_{ij} = \bm{\omega} \cdot \bm{\mathsf{w}} = \underbrace{ \omega^2 \alpha \cos^2(\bm{\omega},\hat{\bm{e}}_\alpha ) }_{\mathcal{P}_{\omega^2}^\alpha} + \underbrace{ \omega^2 \beta \cos^2(\bm{\omega},\hat{\bm{e}}_\beta  ) }_{\mathcal{P}_{\omega^2}^\beta } + \underbrace{ \omega^2 \gamma \cos^2(\bm{\omega},\hat{\bm{e}}_\gamma ) }_{\mathcal{P}_{\omega^2}^\gamma},
  \label{eq:pvst}
\end{equation}
and
\begin{equation}
  \mathcal{P}_{s^2} = - s_{ij} s_{jk} s_{ki} = - (\alpha^3 + \beta^3 + \gamma^3 ) = - \alpha \beta \gamma,
\end{equation}
where $\mathsf{w}_i = \omega_j s_{ij}$ is the vortex-stretching vector, and after averaging $\langle \mathcal{P}_{\omega^2} \rangle = (4/3) \langle \mathcal{P}_{s^2} \rangle$.

\begin{figure}
\centering
  \begin{tikzpicture}
 \node at (3.5,2.2) {\includegraphics[trim={0 14 0 5},clip,width=0.9\textwidth]{./fig/legend_spectrum_Ga-eps-converted-to.pdf}};   
      \node at (0,-0.3) {\includegraphics[width=0.49\textwidth]{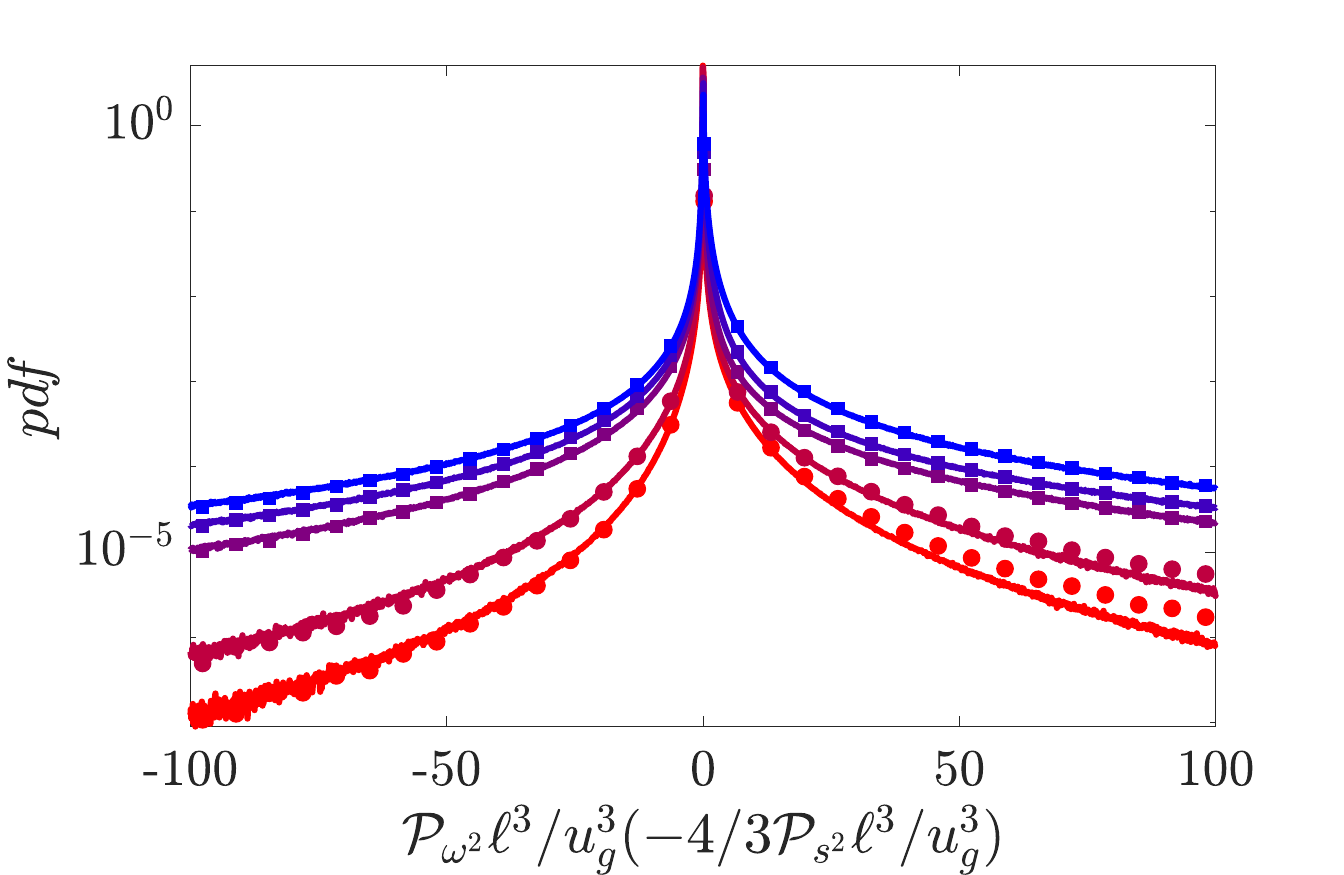}};
      \node at (7,-0.3) {\includegraphics[width=0.49\textwidth]{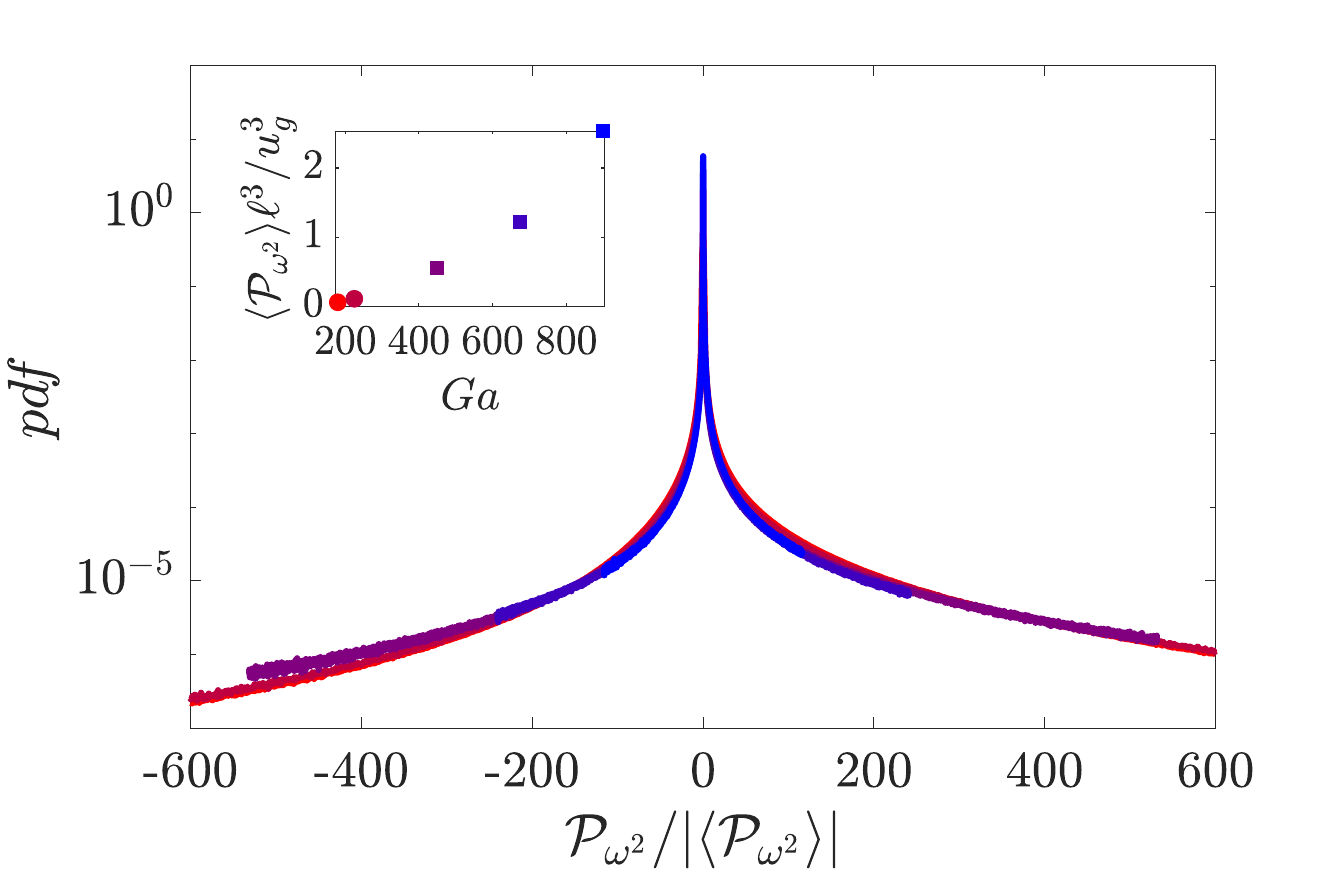}};
      \node at (0,-5.0) {\includegraphics[width=0.49\textwidth]{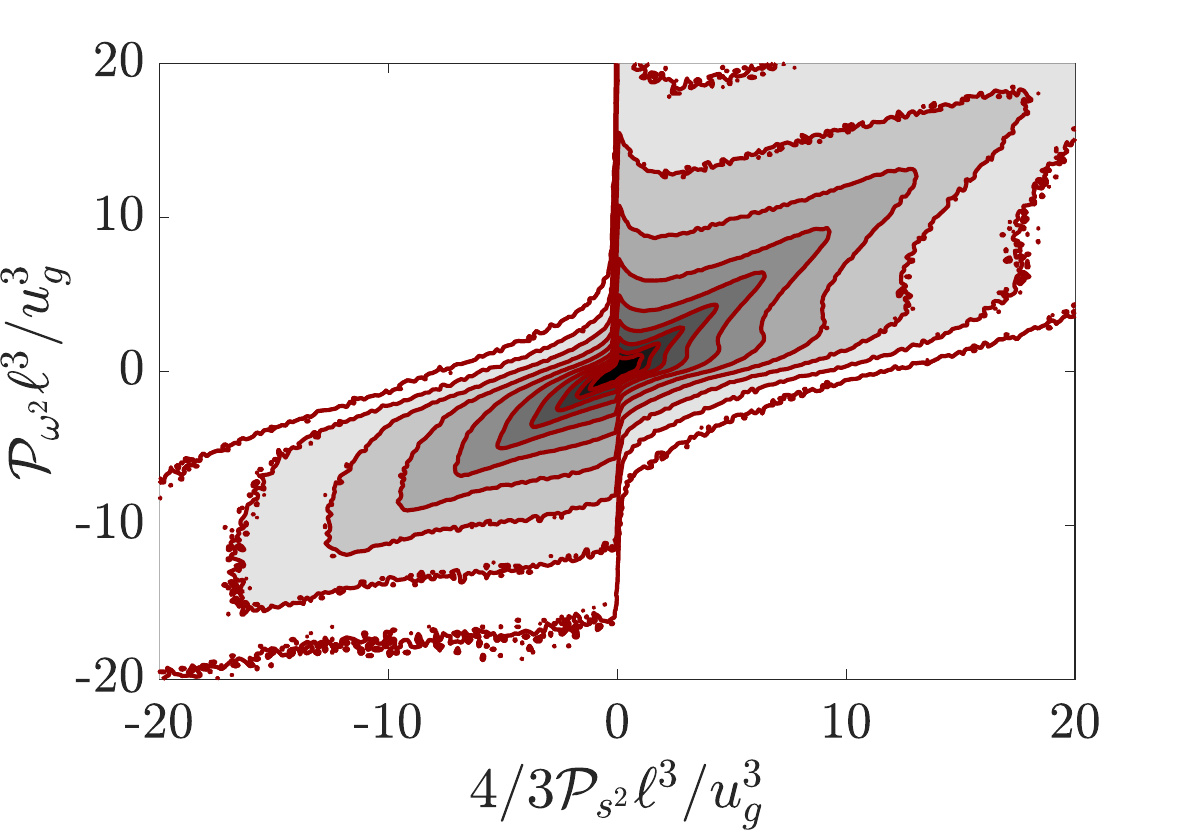}};
      \node at (7,-5.0) {\includegraphics[width=0.49\textwidth]{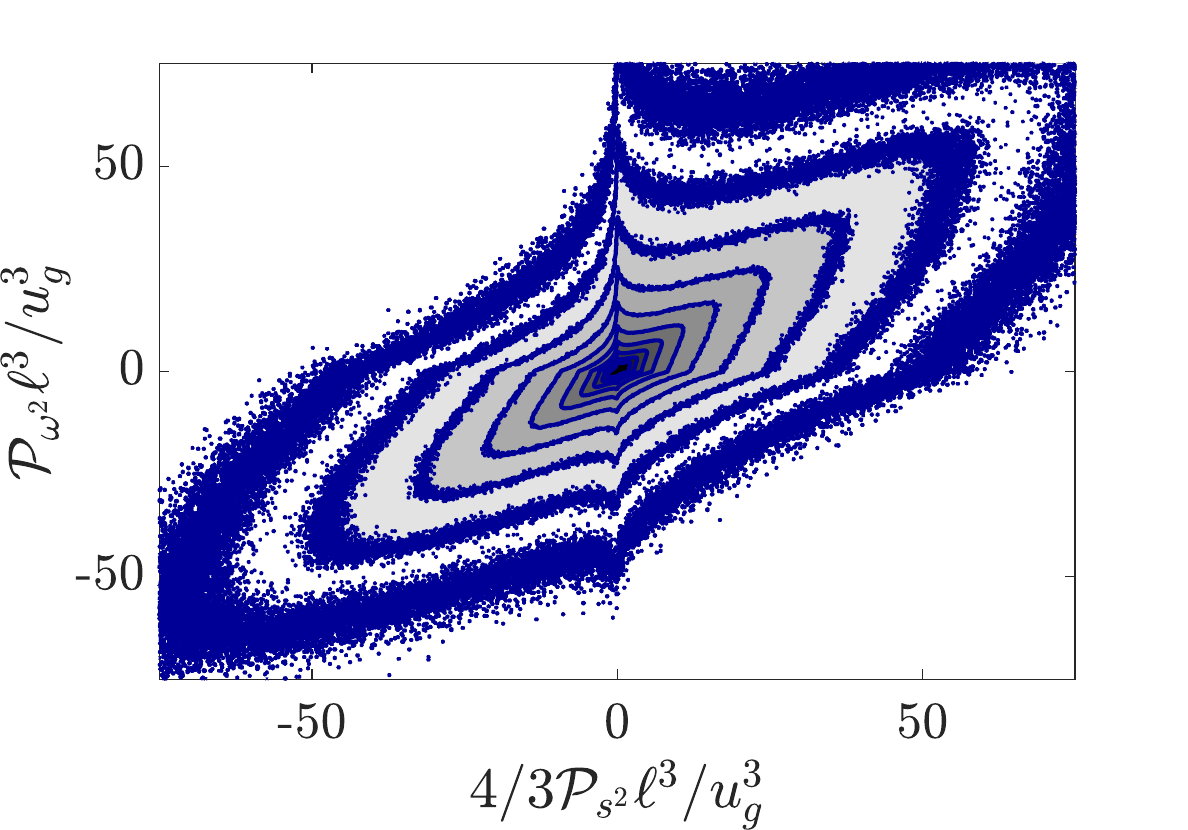}};
      \node at (-3.2, 1.5){$(a)$};
      \node at ( 3.8, 1.5){$(b)$};
      \node at (-3.2,-2.6){$(c)$};
      \node at ( 3.8,-2.6){$(d)$};
  \end{tikzpicture}
\caption{Influence of $Ga$ on $\mathcal{P}_{\omega^2}$ and $\mathcal{P}_{s^2}$ for $n \ell_f^3 = 2.89$. $(a)$: distribution of $\mathcal{P}_{\omega^2}$ (solid line) and $-(4/3) \mathcal{P}_{s^2}$ (circles). $(b)$: distribution of $\mathcal{P}_{\omega^2}/\aver{\mathcal{P}_{\omega^2}}$. In the inset the dependence of $\aver{\mathcal{P}_{\omega^2}}$ on $Ga$ is shown. Bottom panels: Joint probability distribution between $-(4/3) \mathcal{P}_{s^2}$ and $\mathcal{P}_{\omega^2}$, for $(c)$ $Ga=180$ and $(d)$ $Ga=900$. The $10$ isolines correspond to probability values between $1 \times 10^{-6}$ and $1 \times 10^{-2}$, distributed on a logarithmic scale.}
\label{fig:vst_ssa}
\end{figure}
\begin{table}
\centering
\begin{tabular}{ccccccc}
$Ga$      && $\langle \mathcal{P}_{\omega^2}^\alpha \rangle/ \langle \mathcal{P}_{\omega^2} \rangle$ &&
             $\langle \mathcal{P}_{\omega^2}^\beta  \rangle/ \langle \mathcal{P}_{\omega^2} \rangle$ &&
             $\langle \mathcal{P}_{\omega^2}^\gamma \rangle/ \langle \mathcal{P}_{\omega^2} \rangle$ \\
\hline
$180$     && $1.56$ && $0.28$  && $-0.84$ \\
$225$     && $1.45$ && $0.31$  && $-0.77$ \\
$450$     && $1.31$ && $0.54$  && $-0.85$ \\
$675$     && $1.35$ && $0.70$  && $-1.06$ \\
$900$     && $1.52$ && $0.75$  && $-1.28$
\end{tabular}
\caption{Contribution of $\aver{ \mathcal{P}_{\omega^2}^\alpha } \equiv \aver{ \omega^2 \alpha \cos^2(\bm{\omega} \hat{\bm{e}}_\alpha)}$, $\aver{ \mathcal{P}_{\omega^2}^\beta } \equiv \aver{\omega^2 \beta \cos^2(\bm{\omega} \hat{\bm{e}}_\beta)}$ and $\aver{ \mathcal{P}_{\omega^2}^\gamma } \equiv \aver{\omega^2 \gamma \cos^2(\bm{\omega} \hat{\bm{e}}_\gamma)}$ to the average vortex stretching $\aver{ \mathcal{P}_{\omega^2} }$ for $n \ell_f^3 = 2.89$ and different $Ga$.}
\label{tab:vst}
\end{table}
Figure~\ref{fig:vst_ssa} characterises the dependence of $\mathcal{P}_{\omega^2}$ and $\mathcal{P}_{s^2}$ on $Ga$. The top-left panel shows that for all considered $Ga$ the distributions of $\mathcal{P}_{\omega^2}$ and $\mathcal{P}_{s^2}$ are in very good agreement, similarly to what is observed in HIT \citep{tsinober-2000}. Nonetheless, these two quantities are rather different and their pointwise relation is strongly nonlocal~\citep{ohkitani-1994}. 
To investigate the correlation between the two quantities at different $Ga$, the bottom panels of figure \ref{fig:vst_ssa} show the $\mathcal{P}_{\omega^2}$--$\mathcal{P}_{s^2}$ JPDF for $Ga=180$ (left) and $Ga=900$ (right). Although the JPDFs exhibit a clear bias along the $\mathcal{P}_{\omega^2}=(4/3)\mathcal{P}_{s^2}$ line, the two quantities are weakly correlated, as there are frequent events where $\mathcal{P}_{\omega^2}$ is small and $\mathcal{P}_{s^2}$ is large, and vice versa. When increasing $Ga$, the correlation between negative (positive) $\mathcal{P}_{\omega^2}$ and positive (negative) $\mathcal{P}_{s^2}$ increases, and the overall distribution becomes closer to that observed in HIT~\citep{tsinober-2000}. 
The top panels of figure~\ref{fig:vst_ssa} show that the two distributions are right-skewed, with both positive and negative tails becoming longer as $Ga$ increases. The mean value $\langle \mathcal{P}_{\omega^2} \rangle$ (and $\langle \mathcal{P}_{s^2} \rangle$) increases more than linearly with $Ga$; see the inset in figure~\ref{fig:vst_ssa}$(b)$. This is consistent with the enhanced nonlinear energy transfer rate discussed in \S\ref{sec:bud}. Interestingly, figure~\ref{fig:vst_ssa}$(b)$ demonstrates that the distributions of $\mathcal{P}_{\omega^2}$ and $\mathcal{P}_{s^2}$ collapse nicely for all $Ga$, once the quantities are normalised with their mean value.

We now look at the contributions to $\mathcal{P}_{\omega^2}$ associated with $\alpha$, $\beta$ and $\gamma$; see equation \eqref{eq:pvst}. A positive $\mathcal{P}_{\omega^2}$ implies the alignment between $\bm{\omega}$ and $\bm{\mathsf{w}}$, which is obtained when (i) $\bm{\omega}$ is aligned with $\hat{\bm{e}}_\alpha$ or (ii) $\bm{\omega}$ is aligned with $\hat{\bm{e}}_2$ and $\beta>0$. We find that for all $Ga$ the largest contribution comes from the first eigenvalue, like in HIT \citep{ashurst-etal-1987}; see table \ref{tab:vst}. Indeed, although figure \ref{fig:costh} shows a strong alignment between $\bm{\omega}$ and $\hat{\bm{e}}_\beta$, $\beta$ can be both positive and negative, while $\alpha$ is always positive and generally much larger (see figure \ref{fig:sstar}). By looking at the influence of $Ga$, we find that the relative contribution of $\mathcal{P}_{\omega^2}^\beta$ increases with $Ga$, while that of $\mathcal{P}_{\omega^2}^\alpha$ and $\mathcal{P}_{\omega^2}^\gamma$ decreases and then increases. This nonmonotonous trend is inherited by the nonmonotonous dependence of the vorticity-eigenvector alignment on $Ga$ (see figure \ref{fig:costh} and the related discussion).

\begin{figure}
  \centering
    \begin{tikzpicture}
    \node at (0,0) {\includegraphics[width=0.49\textwidth]{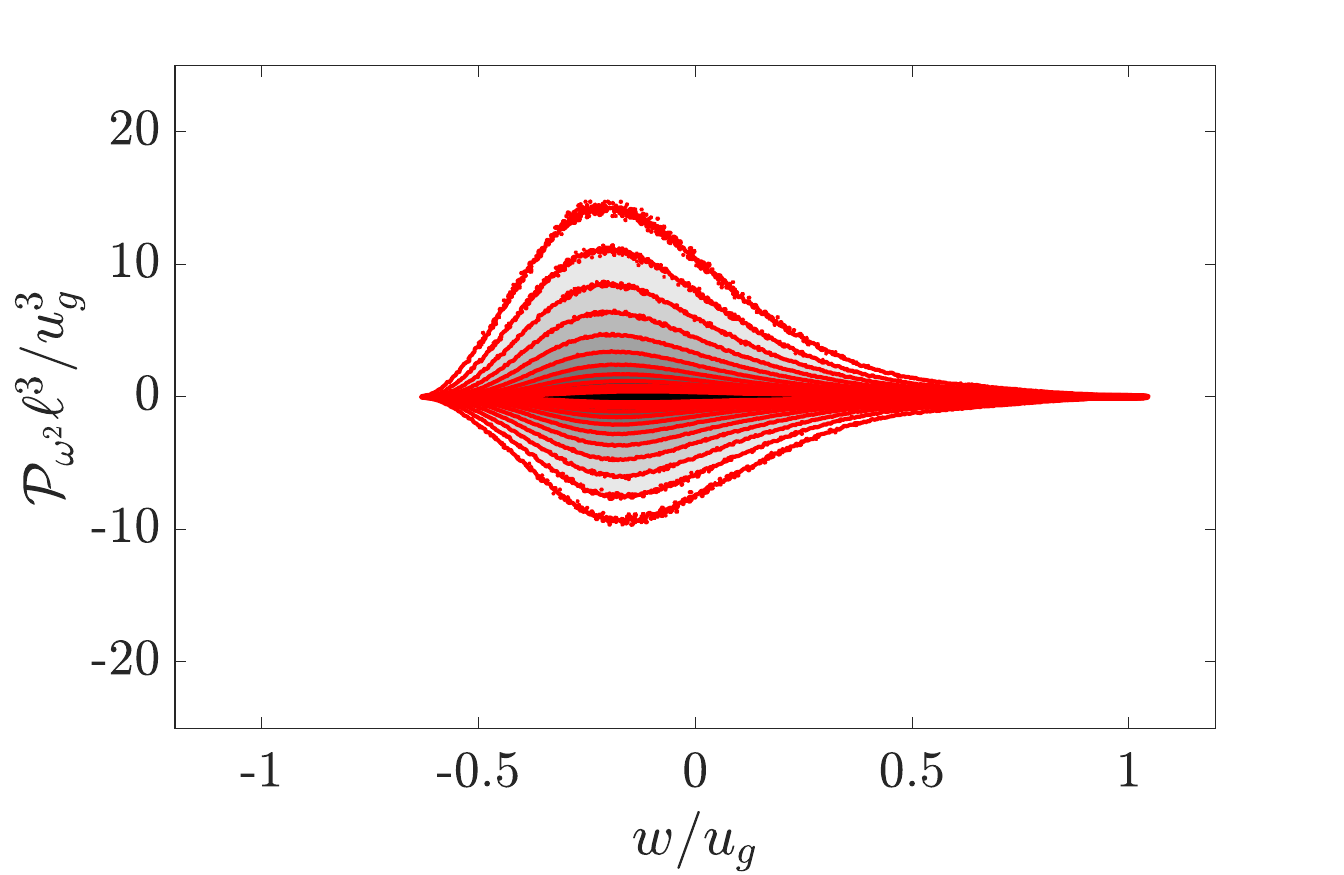}};
    \node at (7,0) {\includegraphics[width=0.49\textwidth]{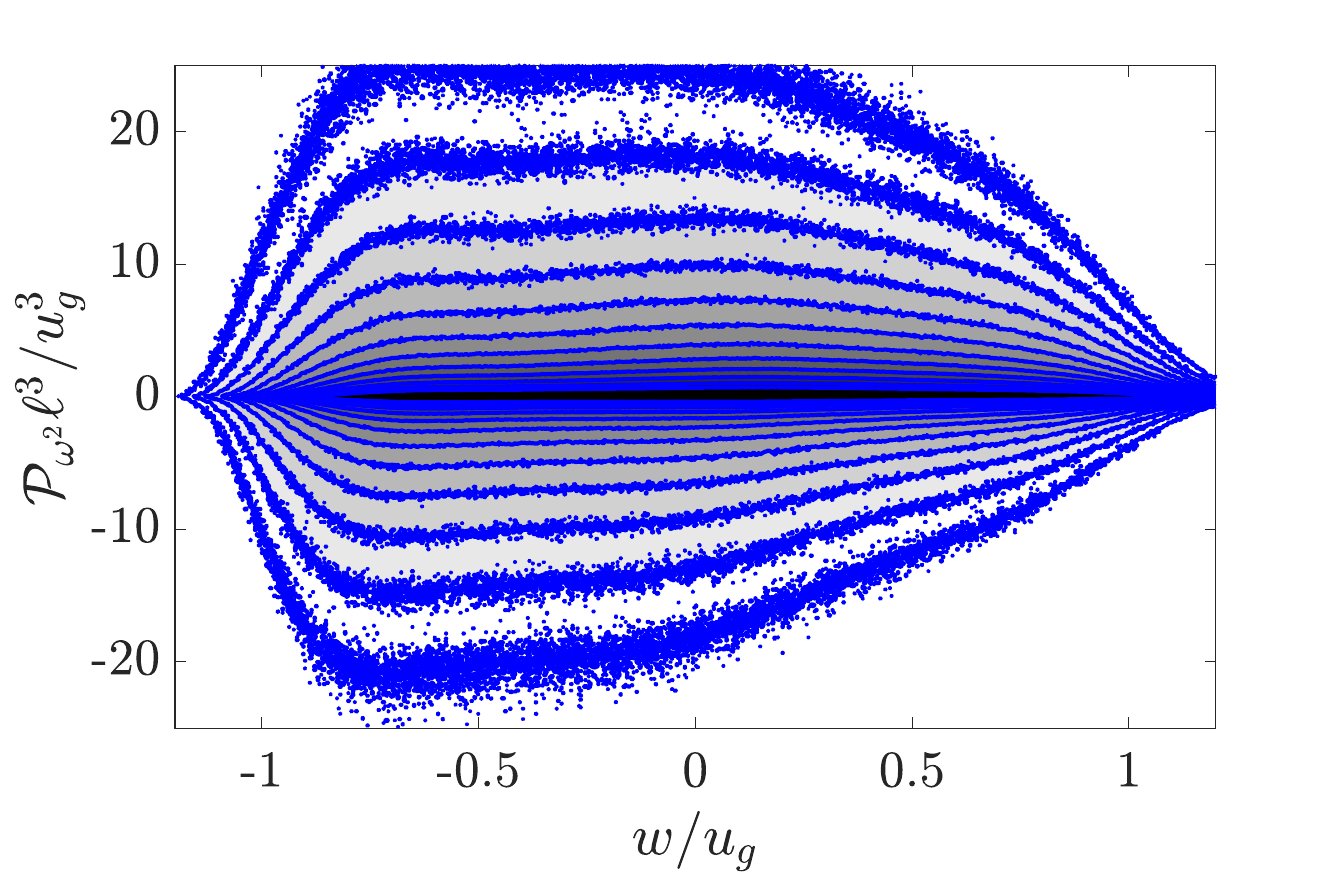}};
\node at (-3.2,1.8){$(a)$};
      \node at ( 3.8,1.8){$(b)$};
\end{tikzpicture}
\caption{ $\mathcal{P}_{\omega^2}-w$ JPDF for $n \ell_f^3 = 2.89$ and ($a$) $Ga=180$, ($b$) $Ga=900$. The $12$ isolines correspond to probability values between $3 \times 10^{-4}$ and $0.1$, distributed on a logarithmic scale.}
  \label{fig:vstwjpdf}
\end{figure}

We now account for the flow inhomogeneity and examine the $\mathcal{P}_{\omega^2}$--$w$ JPDF. Figure \ref{fig:vstwjpdf} shows that $\mathcal{P}_{\omega^2}>0$ and $\mathcal{P}_{\omega^2}<0$ events substantially balance at small $Ga$, where the average value $\langle \mathcal{P}_{\omega^2} \rangle$ is positive but rather small (see figure~\ref{fig:vst_ssa}). As expected, at these small $Ga$ values, events with large $|\mathcal{P}_{\omega^2}|$ are clustered in the $w<0$ region, where the fibres are primarily concentrated. 
As $Ga$ increases, large values of $|\mathcal{P}_{\omega^2}|$ become increasingly correlated with $w>0$ events. This observation is consistent with the distribution of $s^*$ (see figure~\ref{fig:swjpdf}) and confirms that the vortex-stretching mechanism becomes progressively more relevant, also in the ascending stream where the fibre concentration is low.

\subsection{Invariants of $a_{ij}$}

We close this section by looking at the invariants of the velocity gradient tensor $a_{ij}$ \citep{davidson-2004}. A second-order tensor possesses three invariants, $P$, $Q$ and $R$, which are directly related to its eigenvalues $\lambda$ by the characteristic polynomial function 
\begin{equation}
\lambda^3 + P \lambda^2 + Q \lambda + R = 0,
\end{equation}
where
\begin{equation}
\begin{gathered}
  P = \alpha + \beta + \gamma, \
  Q = -\frac{1}{2} \left( \alpha^2 + \beta^2 + \gamma^2 \right) + \frac{\omega^2}{4} \ \text{and} \
  R = - \frac{1}{3} \mathcal{P}_{s^2} - \frac{1}{3} \mathcal{P}_{\omega^2}.
\end{gathered}
\end{equation}
Here, $P=0$ is due to the incompressibility constraint. The second invariant $Q$ provides a measure of the relative intensity of the strain and vorticity, with large negative $Q$ indicating regions of strong strain and large positive $Q$ regions of intense vorticity. The third invariant $R$, instead, is a measure of the relative intensity of the production of vorticity/enstrophy, i.e. vortex stretching $(R<0)$ and production of strain/dissipation ($R>0$). The discriminant of the characteristic equation, i.e. $\Delta = 27 R^2/4 + Q^3$, is used to distinguish between regions where the motion is mainly vortical ($\Delta <0 $) and regions with a node-saddle streamline pattern ($\Delta > 0$). When $Q$ is large and positive, the strain is locally weak and $R \sim - \mathcal{P}_{\omega^2}$. In this case, $R <0$ implies vortex stretching, while $R>0$ implies vortex compression. When, instead, $Q$ is large and negative, then $R \sim \mathcal{P}_{s^2} = - \alpha \beta \gamma$. Here a negative $R$ implies a region of axial strain ($\alpha>0$, $\beta <0$ and $\gamma<0$), while a positive $R$ a region of biaxial strain ($\alpha>0$, $\beta>0$ and $\gamma<0$).

\begin{figure}
  \centering
  \begin{tikzpicture}
\node at (0,-4) {\includegraphics[width=0.49\textwidth]{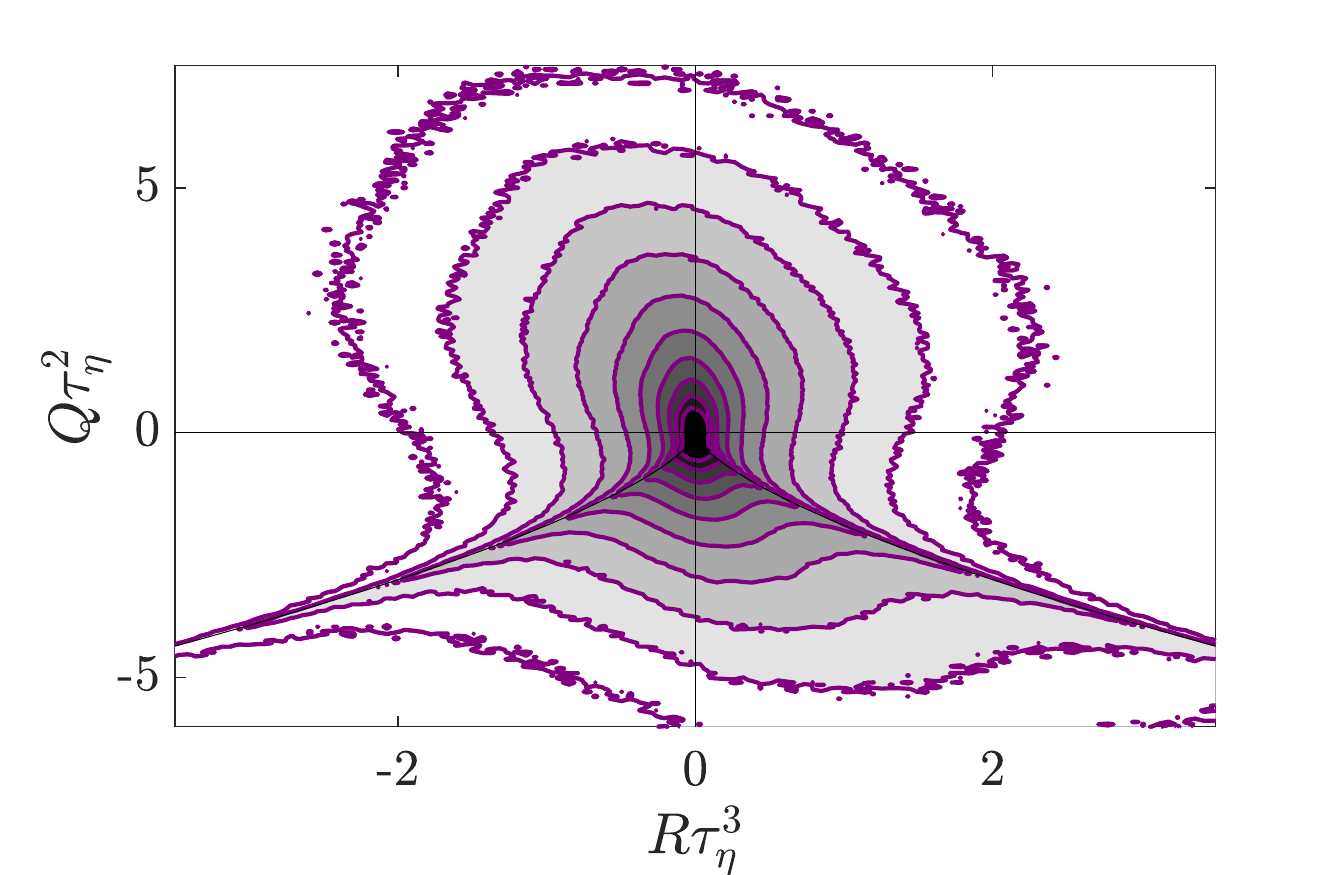}};
    \node at (7,-4) {\includegraphics[width=0.49\textwidth]{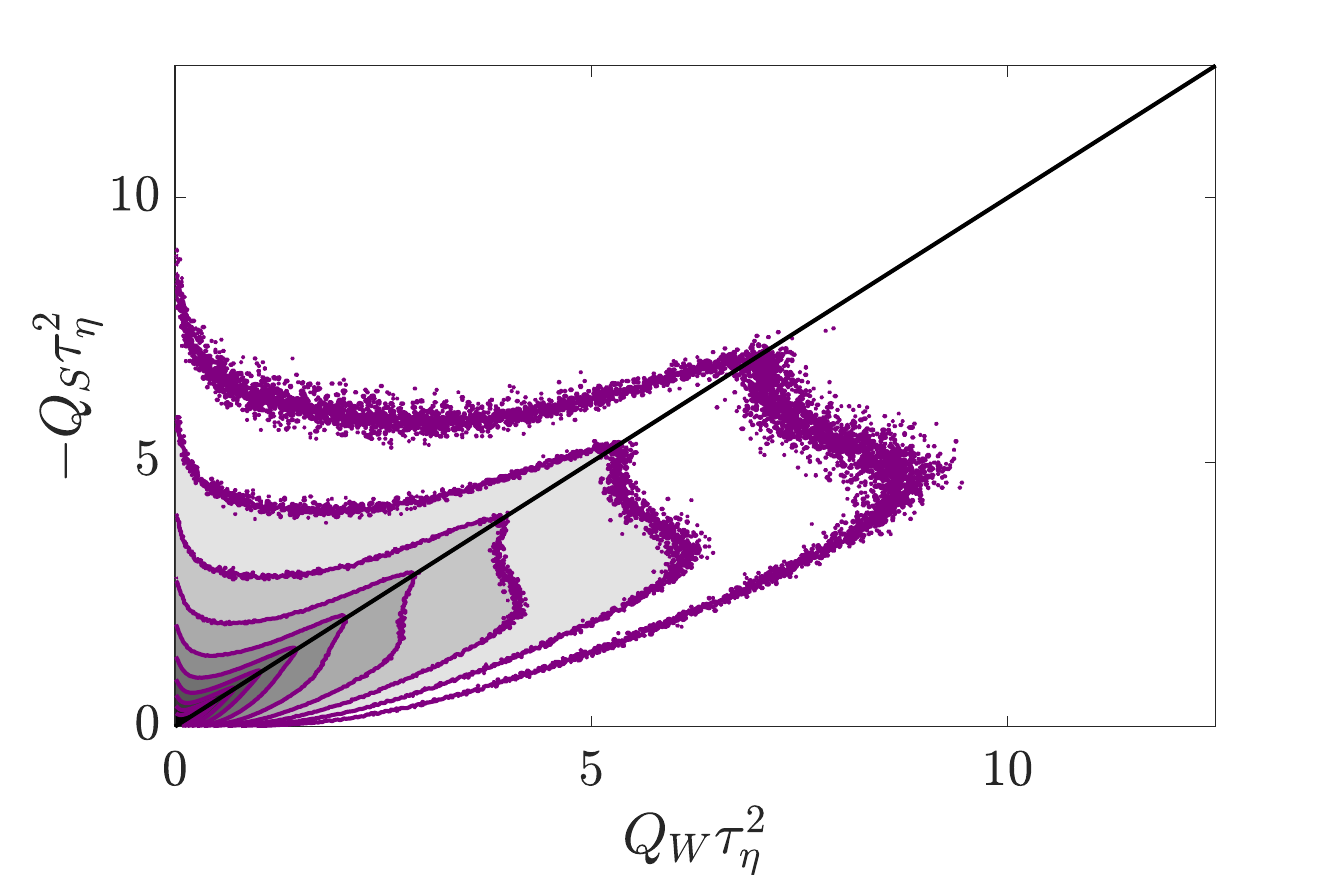}};
    \node at (0,-8) {\includegraphics[width=0.49\textwidth]{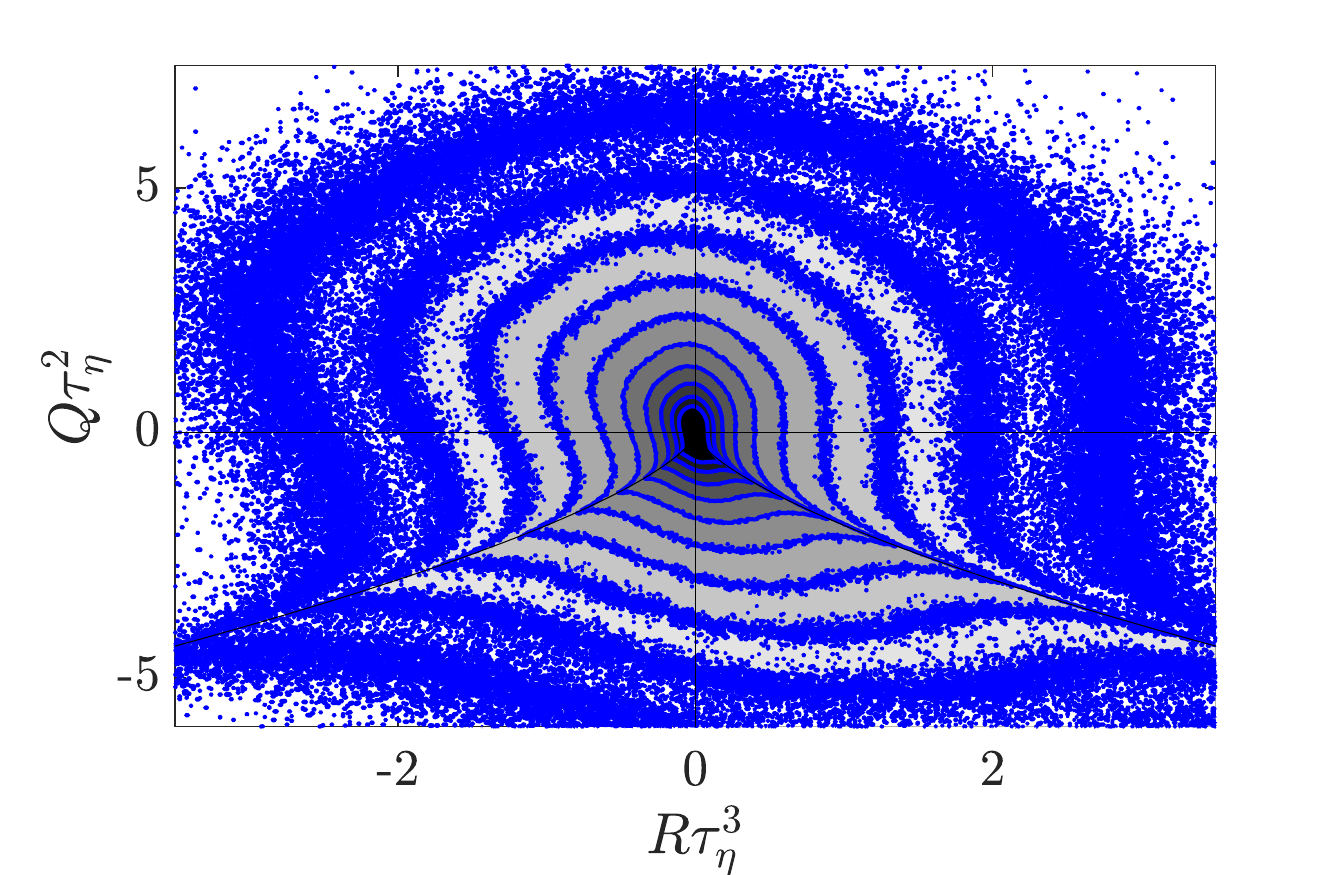}};
    \node at (7,-8) {\includegraphics[width=0.49\textwidth]{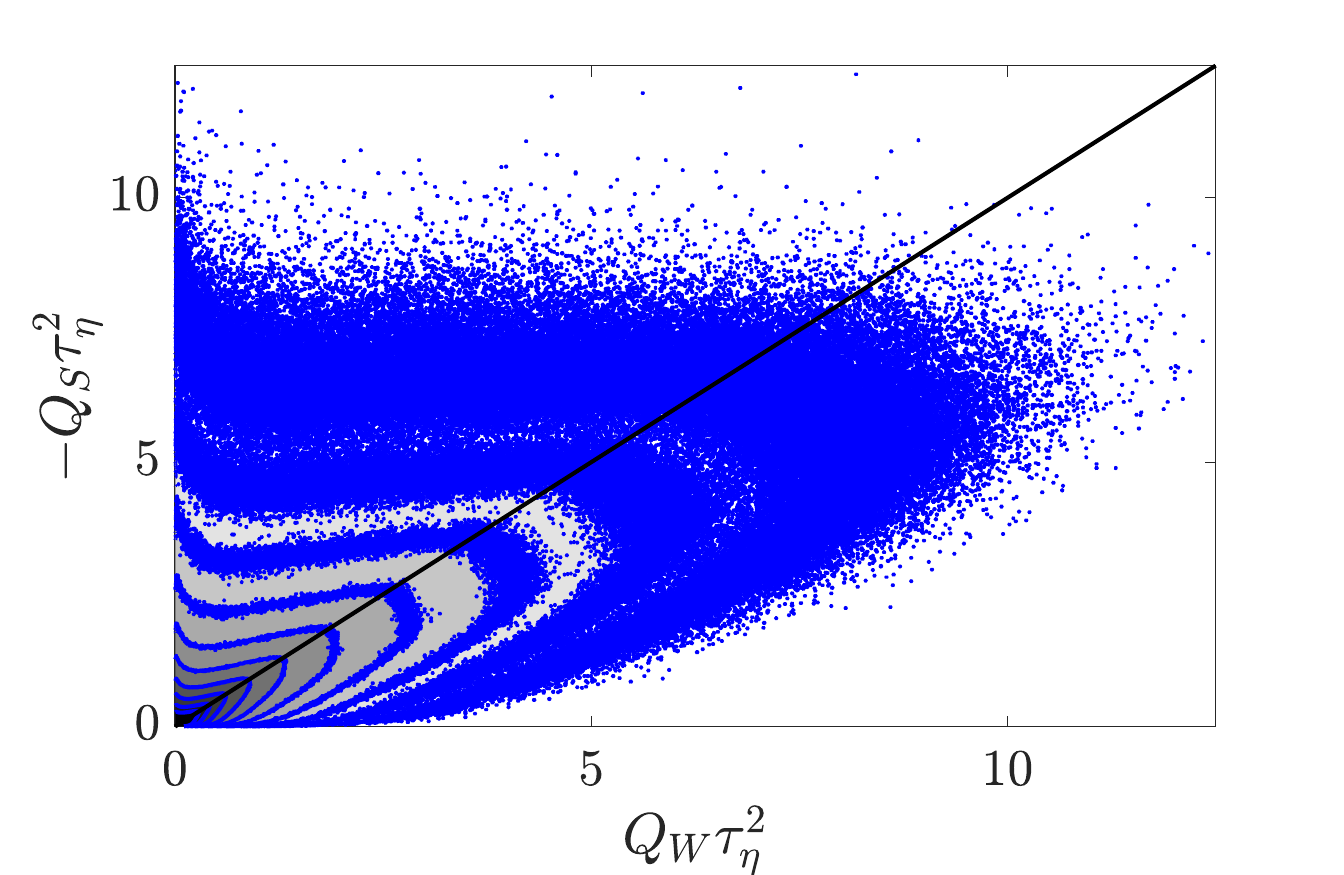} };
\node at (-3.2,-2.2){$(a)$};
    \node at ( 3.8,-2.2){$(b)$};
    \node at (-3.2,-6.2){$(c)$};
    \node at ( 3.8,-6.2){$(d)$}; 
\node[rotate=90] at (10.2,-3.8){$Ga=450$};
    \node[rotate=90] at (10.2,-7.8){$Ga=900$};    
  \end{tikzpicture}   
  \caption{Joint probability density function maps in the $Q-R$ (panels a,c) and $Q_S-Q_W$ (panels b,d) planes for $n \ell_f^3 = 2.89$, at $Ga = 450$ (a,b) and $Ga = 900$ (c,d). The 10 isolines in the left (right) panels correspond to probability values between $3.16 \times 10^{-5}$ and $10^2$ ($10^{-4}$ and $1$), distributed logarithmically. The Kolmogorov time scale is defined as $\tau_\eta = \sqrt{ \nu / \langle \epsilon \rangle }$. The case with $Ga = 450$ is representative of the behaviour observed at lower $Ga$.}
  \label{fig:QR_QsQw}
\end{figure}
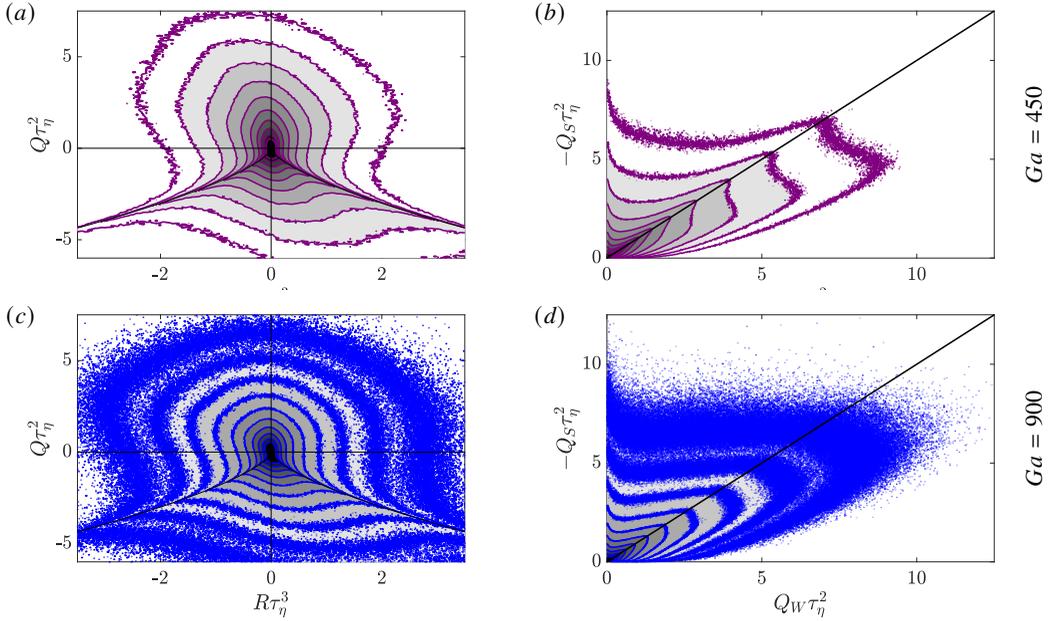
In HIT the shape of the $Q-R$ JPDF takes a tear-drop pattern, with a clear point at the right-Vieillefosse tail with $\Delta = 0$, $R>0$ and $Q <0$ \citep{meneveau-2011}. The largest probability is observed in the two quadrants with $Q<0$ and $R>0$, and $Q>0$ and $R<0$, meaning that there is a strong negative correlation between $Q$ and $R$, i.e. $\aver{QR}<0$. In other words, the most common states are vortex stretching $\mathcal{P}_{\omega^2}>0$ and biaxial strain $\mathcal{P}_{s^2}>0$ ($\alpha \beta \gamma<0$). In the present case, the $Q-R$ JPDFs show some different properties, as the presence of the fibres promotes shear-dominated events. The distribution of $Q$ is skewed towards negative values, and this is particularly true for the smaller $Ga$. The probability of events with negative $Q$ and small $R$ is relatively large, generating a second cluster close to the left Vieillefosse tail. A similar phenomenon was observed for heavy Kolmogorov-size particles in HIT and related to the deformation of the fluid in the proximity of the particles \citep{chiarini-tandurella-rosti-2025}. The left panels of figure \ref{fig:QR_QsQw} show that for smaller $Ga$ the probability of events with $Q>0$ and $R<0$ decreases, indicating a weakening of the vortex stretching events in agreement with the above discussion. 

Additional insights are provided in the right panels of figure \ref{fig:QR_QsQw}, where the JPDF of the second invariants of the $s_{ij}$ and $w_{ij}$ tensors, i.e.
\begin{equation}
  Q_S = -\frac{1}{2} \left( \alpha^2 + \beta^2 + \gamma^2 \right) \ \text{and} \
  Q_W = \frac{1}{4} \omega^2,
\end{equation}
are represented separately. These invariants are related to the fluid dissipation $\varepsilon = - 4 \nu Q_S$ and with the fluid enstrophy $\omega^2 = 4 Q_W$. Although the kinematic (i.e., it does not descend from the Navier--Stokes equations) relation $\langle \varepsilon \rangle  = \nu \langle \omega^2 \rangle$ holds, these two quantities are dynamically different and there is not a causal relationship between them. The $Q_S-Q_W$ JPDF shows whether the flow is dominated by fluid dissipation (extensional dominated regions with $Q_S>Q_W$) or by enstrophy (rigid rotation regions with $Q_W>Q_S$). In HIT, events with $Q_W>-Q_S$ are more frequent meaning that the flow is mainly dominated by rigid rotations \citep{meneveau-2011}. In the present case, the scenario is rather different. Indeed, events having large $-Q_S$ and small $Q_W$ are promoted as well as events with $-Q_S \approx Q_W$. The last effect is particularly prominent for the smaller $Ga$ ($Ga \le 450$) and is characteristic of shear-dominated flows \citep{soria-etal-1994}. This further confirms that, while settling, the fibres induce a chaotic motion of the fluid phase which (at least in the descending flow region) is mainly extensional dominated and characterised by a two-dimensional state of straining, consistently with the shear layers that generate at the fibres surface.
 \section{Conclusions}
\label{sec:conclusion}

We have investigated the fluid-solid interaction of suspensions of rigid fibres settling in a quiescent fluid using DNS coupled with an IBM. The main focus of the study is on the dynamics and organisation of the fluid-phase velocity fluctuations. Specifically, the objectives are: (i) to examine the role of the Galileo number and the fibre concentration on the scale-by-scale redistribution of fluid-phase kinetic energy; (ii) to characterise the flow anisotropy across scales; (iii) to elucidate the mechanisms sustaining fluid-velocity fluctuations; and (iv) to investigate the local structure of the fluid-phase velocity field under different parameter regimes.
We considered heavy fibres with a solid-to-fluid density ratio of $\rho_f/\rho = \mathcal{O}(100)$ and explored a broad range of the parameter space by varying the Galileo number in the range $Ga \in [180,900]$ and the concentration in the range $n \ell_f^3 \in [0.36,23.15]$.

The settling fibres generate a chaotic fluid motion with properties that vary as a function of $Ga$ and $n \ell_f^3$, and depend on the spatial organisation of the solid phase. At higher $Ga$ and lower $n \ell_f^3$, the fibres cluster and form streamers. At lower $Ga$ and/or higher $n\ell_f^3$, the fibres show a more homogeneous distribution and a lower level of clustering. At the present parameters, the fibres mostly settle aligned with the direction of gravity.

The fluid-phase kinetic energy increases with $Ga$ across the entire parameter space considered, and the complete energy spectrum $\mathcal{E}_{3c}$ shows that the excited range of scales broadens accordingly. At large $Ga$ and/or small $n \ell_f^3$, the relative amount of energy contained at the small scales ($\kappa \gtrapprox \kappa_\ell$) increases, while that at the larger scales ($\kappa \lessapprox \kappa_\ell$) decreases, due to the so-called spectral shortcut \citep{brandt-coletti-2022}. The anisotropy of the flow across scales is analysed by separately considering the out-of-plane ($\mathcal{E}_{1c}$) and in-plane ($\mathcal{E}_{2c}$) contributions to the total energy spectrum in the full $(\kappa_\perp,\kappa_\parallel)$ wavenumber space. We observe that motions with in-plane length scales larger than those in the vertical direction ($\kappa_\parallel < \kappa_\perp$) are dominated by $u$ and $v$ fluctuations, whereas vertically elongated motions ($\kappa_\parallel > \kappa_\perp$) are primarily associated with $w$ fluctuations.

Scale-by-scale energy budgets in Fourier space reveal that the out-of-plane ($\mathcal{E}_{1c}$) and in-plane ($\mathcal{E}_{2c}$) velocity fluctuations are sustained by distinct physical mechanisms. The vertical fluctuations are primarily driven by the fluid-solid coupling term, $\mathcal{F}_{1c}$, which acts as a source for $\mathcal{E}_{1c}$ across all scales. In contrast, the pressure-strain term, $\mathcal{P}_{2c}$, acts to energise the in-plane components by redistributing energy from the vertical direction, thereby promoting isotropy in the flow. This term serves as a sink in the $\mathcal{E}_{1c}$ budget and a corresponding source in the $\mathcal{E}_{2c}$ budget, consistently across all scales.
In addition, the fluid-solid coupling term in the in-plane budget, $\mathcal{F}_{2c}$, exhibits a scale-dependent behaviour: it extracts energy from the large scales ($\kappa < \kappa_\ell$) and injects it into the small scales ($\kappa > \kappa_\ell$). The influence of nonlinear transfer becomes increasingly prominent with $Ga$, in line with the growing role of fluid inertia. At high $Ga$ and/or low fibre concentrations ($n \ell_f^3$), this results in an average energy transfer from larger to smaller scales.
However, by employing the K\'{a}rm\'{a}n--Howarth--Monin--Hill (KHMH) equation \citep{hill-2002}, which enables a rigorous definition of the scale-by-scale energy cascade rate, we also identify intense localised backscatter events. These correspond to regions where energy is transferred from smaller to larger scales and are present at both small and large $Ga$.

The influence of the governing parameters on the local structure of the fluctuating velocity field has been analysed through the velocity gradient tensor $a_{ij} = \partial u_i / \partial x_j$ and its invariants \citep{meneveau-2011}. At low $Ga$, the flow is dominated by axisymmetric compression and two-dimensional straining motions, typically associated with shear layers forming at the fibre surfaces. In contrast, increasing $Ga$, the local strain state becomes largely spatially dependent due to the presence of streamers. In particular, descending regions (characterised by a high local fibre concentration) remain dominated by two-dimensional strain states, similar to the low-$Ga$ regime, whereas ascending regions (where the fibre concentration is lower) are dominated by axisymmetric extensions, more reminiscent of homogeneous isotropic turbulence.
The joint probability density function of the second and third invariants of $a_{ij}$ reveals that events involving axial strain and vortex compression occur with significantly higher frequency than in HIT \citep{ashurst-etal-1987}. In agreement with the enhanced nonlinear energy transfer, at higher $Ga$ the vortex stretching and strain self-amplification mechanisms are amplified, leading to increased production of both vorticity/enstrophy and strain/dissipation. Finally, an inspection of the JPDF of the second invariants of the symmetric and antisymmetric parts of $a_{ij}$ indicates that extensional events are more prevalent than rotational ones, highlighting the dominant strain-driven nature of the fluctuations in these flows.
 
These findings open several avenues for future investigation. In particular, it would be valuable to examine how the scale-by-scale organisation of fluid velocity fluctuations is modified by the presence of flexible fibres with varying lengths and bending stiffnesses. Fibre geometry and flexibility are known to influence their settling configurations \citep{banaei-etal-2020, marchioli-rosti-verhille-2025}, and flexible fibres, in particular, interact strongly with fluid motions at comparable scales \citep{rosti-etal-2018}. Another promising direction is the study of fibre settling in non-Newtonian fluids, which have recently been shown to alter settling behaviour significantly \citep{alghalibi-etal-2020}. Understanding these effects would extend the current framework and offer further insight into the complex dynamics of fibre-laden flows.

\backsection[Acknowledgements]{
The research was supported by the Okinawa Institute of Science and Technology Graduate University (OIST) with subsidy funding to M.E.R. from the Cabinet Office, Government of Japan. M.E.R. also acknowledges funding from the Japan Society for the Promotion of Science (JSPS), grant 24K17210 and 24K00810. The authors acknowledge the computer time provided by the Scientific Computing \& Data Analysis section of the Core Facilities at OIST, and by HPCI, under the Research Project grants hp220402, hp240006, hp250035. E.G. is thankful to OIST for the support received during his visit.
}

\backsection[Declaration of interests]{The authors report no conflict of interest.}

\bibliographystyle{jfm}     

\bibliography{./Wallturb}

\end{document}